\documentclass[11pt, a4paper]{article}

\usepackage{geometry}
\usepackage[T1]{fontenc}
\usepackage{palatino}
\usepackage{authblk}
\usepackage{amsmath}
\usepackage{amssymb}
\usepackage{graphicx}
\usepackage{subcaption}
\usepackage{siunitx}
\usepackage{booktabs}
\usepackage{rotating} 
    
\usepackage{parskip} 
\usepackage{hyperref}
\hypersetup{
    colorlinks=true,
    linkcolor=blue,
    urlcolor=cyan,
}
\usepackage{tcolorbox}

\usepackage{tikz}
\usetikzlibrary{
    shapes.geometric, 
    arrows,
    arrows.meta, 
    positioning,
    calc,
    chains
}
\usepackage[compress]{cite}

\begin{document}

\title{Analysis of the Rarefied Flow at Micro-Step using a DeepONet Surrogate Model with a Physics-Guided Zonal Loss Function}

\author[1,*]{Ehsan Roohi}

\author[2]{Amirmehran Mahdavi} 

\affil[1]{Mechanical and Industrial Engineering, University of Massachusetts Amherst, 160 Governors Dr., Amherst, MA 01003, USA}
\affil[2]{Department of Mechanical Engineering, Hakim Sabzevari University, Sabzevar, Iran}
\affil[*]{Corresponding author: \href{mailto:roohie@umass.edu}{roohie@umass.edu}}

\maketitle

\section*{Abstract}
The Direct Simulation Monte Carlo (DSMC) method remains the gold standard for simulating rarefied gas flows but is prohibitively expensive for parametric and many-query applications. To address this limitation, we introduce a Deep Operator Network (DeepONet) surrogate framework featuring an innovative, physics-guided zonal loss function. This research introduces a novel zonal loss function that prioritizes physical fidelity in critical flow regions over global error metrics, leading to predictions with greater engineering relevance. The zonal loss explicitly prioritizes accuracy in the recirculation zone (where $U<0$), ensuring faithful reconstruction of separated flow features that are often under-resolved by conventional, globally-averaged error metrics. Two operator-learning tasks are demonstrated: mapping the Knudsen number to the velocity field and mapping the step-height ratio to the flow solution. The results show excellent agreement with high-fidelity DSMC data. An ablation study highlights that while global error metrics may suggest only marginal improvements, localized error analysis reveals the superior fidelity of the zonal loss in capturing vortex dynamics---an aspect central to engineering relevance. The proposed surrogate not only reproduces detailed velocity fields with high physical fidelity but also achieves predictions for unseen parameters in milliseconds, representing speedups of several orders of magnitude relative to DSMC. This capability enables the quantification of uncertainty, optimization, and design-space exploration that would otherwise be computationally intractable.

\section{Introduction}
\label{sec:introduction}

The accurate simulation of rarefied gas dynamics is a cornerstone of modern engineering, underpinning the design and analysis of systems ranging from atmospheric re-entry vehicles and hypersonic transports to micro- and nano-scale devices like Micro-Electro-Mechanical Systems (MEMS) ~\cite{karniadakis2005microflows, Roohibook2025}. In these regimes, where the molecular mean free path becomes comparable to the characteristic length scale, the flow is characterized by extreme physical conditions and significant deviations from thermodynamic equilibrium. This departure from continuum mechanics renders classical models such as the Navier-Stokes-Fourier (NSF) equations inadequate \cite{Vincenti1967}. Consequently, high-fidelity numerical methods rooted in kinetic theory, particularly the Direct Simulation Monte Carlo (DSMC) method pioneered by Bird, have become the indispensable "gold standard" for achieving physically accurate predictions by directly modeling molecular interactions \cite{bird1994molecular}.

However, the immense computational expense of DSMC presents a formidable barrier. The method's computational cost scales with the number of simulated particles and the required simulation time, becoming particularly severe in the slip and early transition flow regimes. This cost becomes prohibitive for the many-query applications essential to the engineering design cycle, such as uncertainty quantification (UQ), multi-objective optimization, and investigating the influence of physical parameters like the Knudsen number (Kn). Our previous work on the micro-step geometry, which utilized a hybrid DSMC-Fokker-Planck method to analyze the flow physics across a range of Knudsen numbers, highlighted these very challenges \cite{mahdavi2022study}. This computational bottleneck significantly impedes the rapid design iteration and optimization of next-generation technologies that rely on rarefied gas dynamics.

To surmount this computational impasse, the scientific community has increasingly turned to surrogate modeling, aiming to replace expensive high-fidelity solvers with computationally efficient approximations. While traditional methods like polynomial chaos expansions and radial basis functions have seen success in lower-dimensional problems, their efficacy often diminishes when faced with the high-dimensional, nonlinear systems typical of fluid dynamics. The recent advent of deep learning has introduced a new class of highly expressive function approximators, offering unprecedented potential for surrogate construction \cite{lo2024deep}. Yet, early efforts relying on purely data-driven neural networks revealed critical limitations: a need for a large quantity of training data and a tendency to produce physically inconsistent predictions, as they operate as "black-box" approximators with no inherent knowledge of the underlying physics.

A paradigm shift occurred with the development of Physics-Informed Neural Networks (PINNs), which embed physical laws, typically in the form of partial differential equations (PDEs), directly into the network's training process via the loss function \cite{raissi2019physics}. This innovation transforms the learning problem from simple curve-fitting into a constrained optimization, where the network must find a solution that not only fits the available data but also respects the governing equations. This physics-informed learning acts as a powerful regularization mechanism, enabling PINNs to generalize effectively even from sparse data and ensuring that their predictions are physically plausible—a critical requirement for engineering applications \cite{karniadakis2021physics}. However, a standard PINN is designed to learn the solution to a PDE for a single instance of boundary conditions and parameters. This makes it ill-suited for the parametric studies central to design exploration, as a new network would need to be re-trained for each point in the design space.

The next logical evolution is to move from learning a single solution to learning the solution operator itself—the mapping from a set of input parameters or functions to the corresponding solution function. The Deep Operator Network (DeepONet) architecture has emerged as a powerful and theoretically grounded framework for this task \cite{lu2021learning}. A DeepONet employs a dual-network structure: a "Branch" network processes the input parameters (e.g., Knudsen number, geometry), while a "Trunk" network processes the domain coordinates. Their outputs are combined to approximate the entire solution field. This elegant architecture effectively disentangles the learning of the parametric dependence from the learning of the spatial solution structure, making it an ideal candidate for learning the operator that maps a physical parameter to the corresponding velocity distribution.

Recent developments in physics-informed and operator-based neural networks have demonstrated growing interest in applying machine learning to complex fluid problems. Sun et al.~\cite{sun2024pinn} proposed a physics-informed neural network with two weighted loss formulations to study internal solitary waves in ocean dynamics, emphasizing tailored loss balancing for multi-physics interactions. Subramanian et al.~\cite{subramanian2023adaptive} introduced an adaptive self-supervision strategy for PINNs, where the algorithm automatically adjusts its training focus to improve stability and convergence. Si and Yan~\cite{si2025cwpinn} further advanced this direction by developing a convolution-weighting scheme for PINNs within a primal-dual optimization framework, highlighting the importance of loss re-weighting from a theoretical perspective. More recently, Malineni and Rajendran~\cite{malineni2025bcpinn} applied PINN approaches for sparse data reconstruction of unsteady flows around complex geometries, demonstrating the utility of physics-informed learning in data-limited regimes. While these studies showcase innovative training strategies and loss function designs for PINNs, our present work departs from the PINN framework by employing a DeepONet architecture with a zonal loss tailored to rarefied gas dynamics. This distinction enables accurate operator learning across Knudsen regimes and complex step-flow configurations, positioning our contribution as complementary yet fundamentally different from the above-mentioned methods.

Recent progress has highlighted the potential of machine learning in modeling separated 
flows such as backward-facing steps (BFS), where high-fidelity simulations are often 
computationally prohibitive. For instance, Choi \textit{et al.}~\cite{choi2025bfs} 
developed a convolutional neural network trained on LES data to predict turbulent flows 
over BFS geometries with varying step angles, achieving strong agreement with numerical 
and experimental results. More recently, graph neural network frameworks have been proposed for mesh-based surrogate modeling and error prediction in BFS flows, demonstrating the potential of data-driven methods to complement traditional simulations \cite{barwey2025gnn}. While Physics-Informed Neural Networks (PINNs) have successfully embedded Partial Differential Equations (PDEs) into loss functions, recent efforts have focused on encoding physical laws directly into the network architecture. A notable example is the Physics-Constrained DeepONet (PC-DeepONet) by Jnini et al. \cite{jnini2024pcdeeponet}, which enforces the divergence-free condition of the incompressible Navier-Stokes equations by architectural design, guaranteeing that the continuity equation is satisfied. Their approach provides a powerful hard constraint for surrogate modeling in continuum flows. In contrast, our work addresses the distinct challenges of rarefied gas dynamics, which are governed by kinetic theory and simulated with statistical methods like DSMC, where direct PDE constraints are not readily applicable. Instead of a hard architectural constraint, we introduce a novel, physics-guided **zonal loss function**. This soft-constraint method intelligently guides the learning process by assigning a higher weight to errors in physically critical regions, such as the recirculation zone behind the step. This provides a flexible yet powerful alternative for physics-informing surrogate models in non-continuum regimes, demonstrating that high-fidelity predictions can be achieved by focusing the network's learning capacity on the most complex flow phenomena.

As a continuation of our recent work on using machine learning for rarefied gas applications \cite{roohi2026data, roohi2025learning}, here we develop and apply a DeepONet-based surrogate model to the challenging problem of rarefied flow over a micro-step. In our recent work, we employed a 'family-of-experts' strategy to predict rarefied lid-driven cavity flows at discrete Knudsen numbers \cite{roohi2026data}; the present study advances toward a unified DeepONet framework. While the expert-based decomposition proved effective for the cavity benchmark, it was found inadequate for the more challenging step-flow problem, where complex separation and reattachment phenomena demand a single, generalizable model. The DeepONet architecture, combined with our novel zonal loss function, provides this capability: instead of interpolating between multiple specialist networks, the current approach enables a single operator-learning model to adapt across wide parametric ranges of Knudsen number and geometry, while selectively emphasizing physically critical recirculation zones. 
Numerous authors have investigated rarefied backward-facing step flows,  a canonical non-equilibrium benchmark, using DSMC and other kinetic methods in recent years 
\cite{xue2003dsmc,xue2005unique, darbandi2011dsmc,white2013dsmc,mahdavi2014thermal,mahdavi2015investigation,nabapure2021dsmc,
gavasane2021step,mahdavi2022study,manela2021freemolecular,sazhin2024transonic,ben2025kinetic}. This canonical benchmark case exhibits critical flow phenomena, including separation, recirculation, and reattachment. However, the consistently high computational cost of such simulations highlights the necessity 
of adopting machine-learning-based surrogate models to accelerate analysis of these flows. Building upon the insights from our prior DSMC analysis \cite{mahdavi2022study}, we propose a robust and efficient surrogate modeling framework with three primary contributions. We demonstrate the successful application of a convolutional DeepONet architecture to predict the entire 2D velocity field as a function of either a physical parameter (the Knudsen number) or a geometric parameter (the step height), showcasing the model's versatility.  We introduce a novel, physics-guided \textbf{zonal loss function}. Standard Mean Squared Error loss treats all points in the domain equally, often failing to resolve localized, high-gradient phenomena. Our proposed technique intelligently partitions the domain based on the sign of the streamwise velocity, a direct physical indicator of the recirculation vortex. By applying a higher weight to the loss calculated in this zone, we compel the model to prioritize accuracy in the most physically complex and critical region of the flow. Third, we integrate uncertainty quantification via Monte Carlo Dropout, allowing the model to provide not only predictions but also a valuable measure of its own confidence, a crucial feature for engineering reliability. The results show that our proposed model can accurately and rapidly predict the flow field for unseen parameters, establishing it as a highly effective tool for accelerating the design and analysis of micro-scale flow systems. We also provide a comparison between the standard DeepONet and the Fusion DeepONet~\cite{peyvan2025fusion}.

\section{Governing Equations of the DSMC Method}

The foundational equation that the DSMC method aims to solve can be represented by a non-homogeneous local kinetic equation for a system of N particles, as introduced by Stefanov~\cite{stefanov2019basic}. This governing equation is formulated by discretizing the splitting form of the kinetic equations in both space and time and applying it to the N-particle distribution function, denoted as $\tilde{F}_N$. At any given time $t$, the function $\tilde{F}_N$ is treated as a randomized quantity that depends on the number of particles within a cell $l$, represented as $\tilde{N}^{(l)}$, and the set of particle velocities $V = \{\mathbf{v}_1, \dots, \mathbf{v}_i, \dots, \mathbf{v}_j, \dots, \mathbf{v}_N\}$ within a small volume $V$ centered at a point $\mathbf{r}$.

The governing equation proposed by Stefanov is expressed as:
\begin{equation}
\frac{\partial \tilde{F}_N(t, \mathbf{r}, V)}{\partial t} + \sum_{i=1}^{N} \mathbf{v}_i \frac{\partial \tilde{F}_N(t, \mathbf{r}, V)}{\partial \mathbf{r}} = \sum_{1 \le i < j \le N} \left\{ \iint g_{i,j} \left[ \tilde{F}_N(t, \mathbf{r}, V'_{i,j}) - \tilde{F}_N(t, \mathbf{r}, V) \right] d\sigma_{i,j} \right\}
\end{equation}
In this equation, the term $d\sigma_{i,j}$ represents the differential collision cross-section. The vector $V'_{i,j} = \{\mathbf{v}_1, \dots, \mathbf{v}'_i, \dots, \mathbf{v}'_j, \dots, \mathbf{v}_N\}$ denotes the set of velocities after a collision between particles $i$ and $j$, and $g_{i,j} = |\mathbf{v}_i - \mathbf{v}_j|$ is the magnitude of their relative velocity. This equation governs the temporal evolution of the local N-particle distribution function near a physical point $\mathbf{r} \in \mathbb{R}^3$ in a phase space composed of a 3D physical domain and a 3N-dimensional velocity domain.

The discretized version of Stefanov's equation is presented below:

$t < \tau \le t + \Delta t, \quad l = 1, M$

\[
\tilde{F}^*_{N^{(l)}}(t+0, \mathbf{r}^{(l)}, V^{(l)}) = \tilde{F}_{N^{(l)}}(t, \mathbf{r}, V^{(l)}), \quad \mathbf{r} \in \Omega^{(l)} \subset \mathbb{R}^3
\]

\begin{equation}
\frac{\partial \tilde{F}^*_{N^{(l)}}(t, \mathbf{r}^{(l)}, V^{(l)})}{\partial t} = \frac{1}{V^{(l)}} \sum_{1 \le i < j \le N^{(l)}} \left\{ \int g_{i,j} \left[ \tilde{F}^*_{N^{(l)}}(t, \mathbf{r}^{(l)}, V'_{i,j}) - \tilde{F}^*_{N^{(l)}}(t, \mathbf{r}^{(l)}, V^{(l)}) \right] d\sigma_{i,j} \right\}
\end{equation}

\[
\frac{\partial \tilde{F}^{**}_{N^{(l)}}(t, \mathbf{r}^{(l)}, V^{(l)})}{\partial t} = \tilde{F}_{N^{(l)}}^{*}(t+\Delta t, \mathbf{r}^{(l)}, V^{(l)}).
\]
\begin{equation}
\frac{\partial \tilde{F}^{**}_{N^{(l)}}(t, \mathbf{r}^{(l)}, V^{(l)})}{\partial t} =- \sum_{i=1}^{N^{(l)}} \mathbf{v}_i \frac{\partial \tilde{F}^{**}_{N^{(l)}}(t, \mathbf{r}^{(l)}, V^{(l)})}{\partial \mathbf{r}} \quad \mathbf{r} \in \tilde{D}^{(l)}
\end{equation}
\begin{equation}
F_{N}(t+\Delta t, \mathbf{r}, V) = \sum_{l=1}^{M} \tilde{F}^{**}_{N^{(l)}}(t+\Delta t, \mathbf{r}^{(l)}, V^{(l)})
\end{equation}
\begin{equation}
F_{N^{(l)}}(t+\Delta t, \mathbf{r}, V^{(l)}) = \int_{D^{(l)}} F_N(t+\Delta t, \mathbf{r}, V) d\mathbf{r}, \quad \mathbf{r} \in \tilde{D} \subset \mathbb{R}^3,
\end{equation}
Here, the parameter $\Delta t$ signifies the time step, while $M$ denotes the total count of subdomains, $\tilde{D}^{(l)}$. These subdomains are the mathematical equivalent of the computational cells used in the DSMC algorithm. The distribution function $\tilde{F}_N$ describes the N-particle velocity distribution around a spatial point $\mathbf{r}$ within these subdomains. The set of equations above reflects the algorithmic steps of the DSMC method, which involves three consecutive procedures in each time step. These steps correspond to particle indexing and ballistic motion. Specifically, Equation (2) provides a mathematical representation of the binary collision relaxation process fundamental to DSMC.

It is convenient to express Equation (2) in an operator form as shown below:
\begin{equation}
\frac{\partial \tilde{F}^*_{N^{(l)}}(t, \mathbf{r}^{(l)}, V^{(l)})}{\partial t} = \hat{\Omega} \tilde{F}^*_{N^{(l)}}(t, \mathbf{r}^{(l)}, V^{(l)}),
\end{equation}
\begin{equation}
\hat{\Omega} \tilde{F}^*_{N^{(l)}}(t, \mathbf{r}^{(l)}, V^{(l)}) = \frac{1}{V^{(l)}} \sum_{1 \le i < j \le N^{(l)}} \left\{ \int g_{i,j} \left[ F^*_{N^{(l)}}(t, \mathbf{r}^{(l)}, V'_{i,j}) - F^*_{N^{(l)}}(t, \mathbf{r}^{(l)}, V^{(l)}) \right] d\sigma_{i,j} \right\}
\end{equation}
The operator $\hat{\Omega}$ generates an updated distribution function $F^*_{N^{(l)}}$ over an infinitesimal time step $\Delta t \to 0$, which results from a single, instantaneous collision event. Based on this operator, the solution to the equation at time $t+\Delta t$ can be expressed as a series expansion of the solution at time $t$, where the expansion is based on the number of collisions, $k$:
\begin{equation}
\tilde{F}^{**}_{N^{(l)}}(t+\Delta t, \mathbf{r}^{(l)}, V^{(l)}) = \sum_{k=0}^{\infty} \frac{\hat{\Omega}^k \tilde{F}^*_{N^{(l)}}(t, \mathbf{r}^{(l)}, V^{(l)})}{k!} (v\Delta t)^k,
\end{equation}
where $v$ represents the binary collision frequency, a quantity that is generally unknown as it depends on the solution itself.

For a small time step $\Delta t \to 0$, the solution of the discrete collision equation can be obtained from an initial state at time $t$ using an exponential collision transition operator, given by:
\begin{equation}
G(\Delta t) = \exp[\Delta t v (T - I)],
\end{equation}
In this expression, the exponential operator is decomposed using the identity operator, $I$, and the 3D-velocity rotation operator, $T$. These new operators are defined as follows:
\begin{align}
I\psi &= \psi \notag \\
T_{i,j}\psi &= \frac{1}{\sigma_{i,j}} \int_{4\pi} \psi(\mathbf{v}_i, \mathbf{v}_j) \sigma(g_{i,j}, \theta) d\theta d\epsilon \notag \\
T\psi &= \sum_{1 \le i < j \le N^{(l)}} \omega_{i,j} T_{i,j} \psi
\end{align}
The binary collision frequency, $v$, which is generally unknown, is formulated as:
\begin{align}
v &= \sum_{1 \le i < j \le N^{(l)}} \omega_{i,j} \notag \\
\omega_{i,j} &= \frac{\sigma_{i,j} g_{i,j}}{V^{(l)}}
\end{align}
This transition operator, $G(\Delta t)$, forms the basis for deriving the series of Bernoulli trial schemes, i.e., see \cite{roohi2018generalized,javani2024symmetrized} as well as the standard no time counter (NTC) and its modern variant employed in this work, i.e., Nearest Neighbor (NN).

\section{The Problem}

\begin{figure}[h!]
    \centering
    \includegraphics[width=0.8\textwidth]{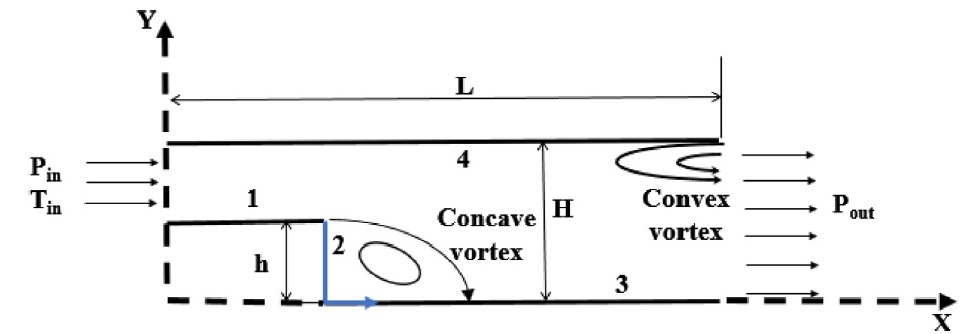}
    \caption{Schematic of the simulated geometry}
    \label{fig:Schematic}
\end{figure}

The flow configuration analyzed in this study is the two-dimensional rarefied gas flow over a micro backward-facing step (BFS). The computational domain and key geometric parameters are illustrated in Figure~\ref{fig:Schematic}. The geometry is defined by the channel height, $H$, the step height, $h$, and the total channel length, $L$. Gas with specified inlet pressure $P_{in}$ and temperature $T_{in}$ enters from the left boundary. The flow develops over a short entry region before encountering the step, after which a prominent recirculation zone (labeled "Concave vortex") forms due to flow separation. The flow eventually reattaches downstream and exits through the right boundary at pressure $P_{out}$. At lower Kn regimes, a "Convex vortex" appears on the top wall near the exit region. 

We analyze microscale flow over a BFS with a channel aspect ratio of $L/H = 5$ and an expansion ratio of $H/h = 2$. The total channel length is 
$L = 85.47~\mu$m, and nitrogen gas is considered as the working fluid. The gas properties are 
modeled in our DSMC solver using the variable hard sphere (VHS) approach with the following reference parameters: 
$m = 4.65 \times 10^{-26}$~kg, $T_{\mathrm{ref}} = 273$~K, $\omega = 0.74$, 
$d = 4.17 \times 10^{-10}$~m, and $\mu_{0} = 1.656 \times 10^{-5}$~N$\cdot$s/m$^{2}$. 
Here, $\omega$ represents the viscosity–temperature index, $T_{\mathrm{ref}}$ is the reference 
temperature, and $\mu_{0}$ is the corresponding viscosity.

The computational geometry includes four walls, labeled 1–4, as illustrated in Fig.~\ref{fig:Schematic}. 
Wall 1 has a length of $0.3L$, while wall 2 has a height of $0.5H$. The Knudsen number (Kn) is 
defined based on the outlet channel height, i.e., $\mathrm{Kn} = \lambda/H$, where the hydraulic 
diameter of the inlet channel is $2h = H$. Both the inlet and wall temperatures are set to 
$T = 300$~K. A constant pressure ratio of $\mathrm{PR} = P_{\mathrm{in}}/P_{\mathrm{out}} = 2$ 
is imposed across the channel. 

Beyond the physical setup, the present work employs a neural network framework, described in the previous section, to learn 
and generalize the rarefied gas dynamics of the BFS geometry. In the first phase, the 
effect of the Knudsen number is investigated. The Knudsen number is varied from $10^{-4}$ 
to $10^{2}$, covering 23 distinct values. Out of these, 20 cases are used for training 
the neural network, while 3 representative cases are withheld for testing, corresponding 
to $\mathrm{Kn} = 0.004$, $0.2$, and $1.0$. These values are chosen to lie in different 
flow regimes, thereby providing a robust assessment of the network’s predictive capability. 

In the second phase, the Knudsen number is fixed at $\mathrm{Kn} = 0.01$, while the step 
height ratio ($h/H$) is systematically varied across nine different values, ranging from 
$0.16$ to $0.75$. DSMC simulations are performed for all these cases, with eight geometries employed for training and one case, corresponding to a 
step height ratio of $h/H = 0.44$, reserved exclusively for testing. This two-stage design 
enables evaluation of the neural network’s ability to interpolate across physical regimes 
and to generalize to unseen geometrical configurations.

\section{Flow Field Behavior}

\subsection{Effects of the Knudsen Number}

\begin{figure}[h!]
    \centering
    
    \begin{subfigure}[t]{0.48\textwidth}
        \centering
        \includegraphics[width=\textwidth]{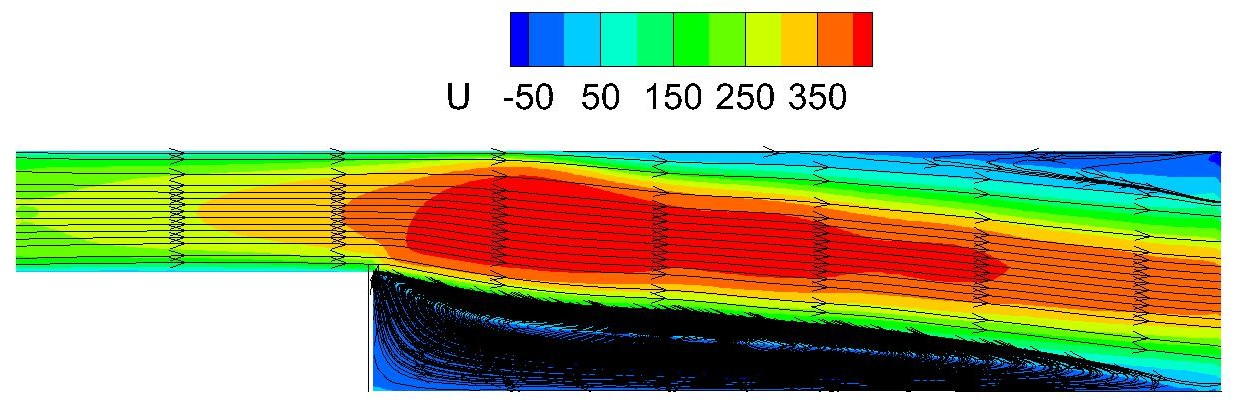}
        \caption{Kn=0.0001}
        \label{fig:minKn}
    \end{subfigure}\hfill 
    \begin{subfigure}[t]{0.48\textwidth}
        \centering
        \includegraphics[width=\textwidth]{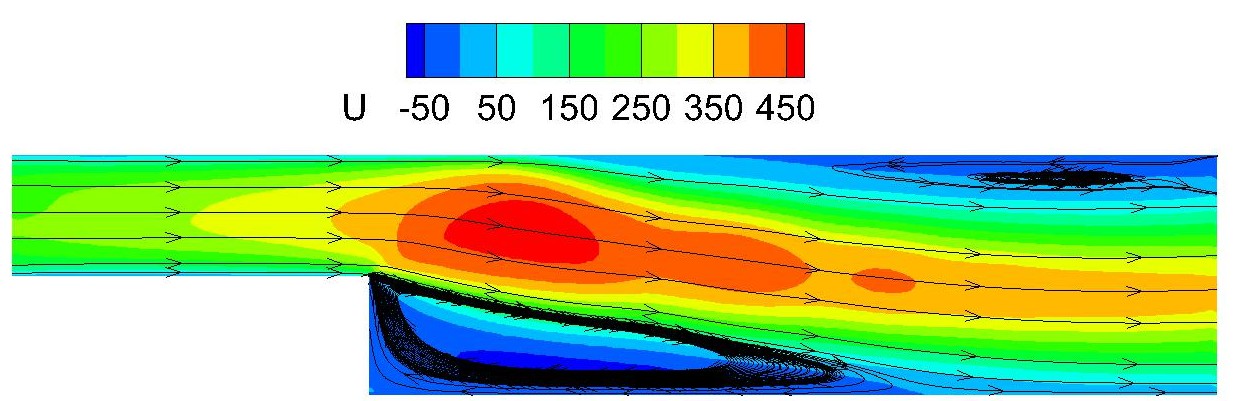}
        \caption{Kn=0.001} 
        \label{fig:minKn2}
    \end{subfigure}
    
    \vspace{5mm} 

    \begin{subfigure}[t]{0.48\textwidth}
        \centering
        \includegraphics[width=\textwidth]{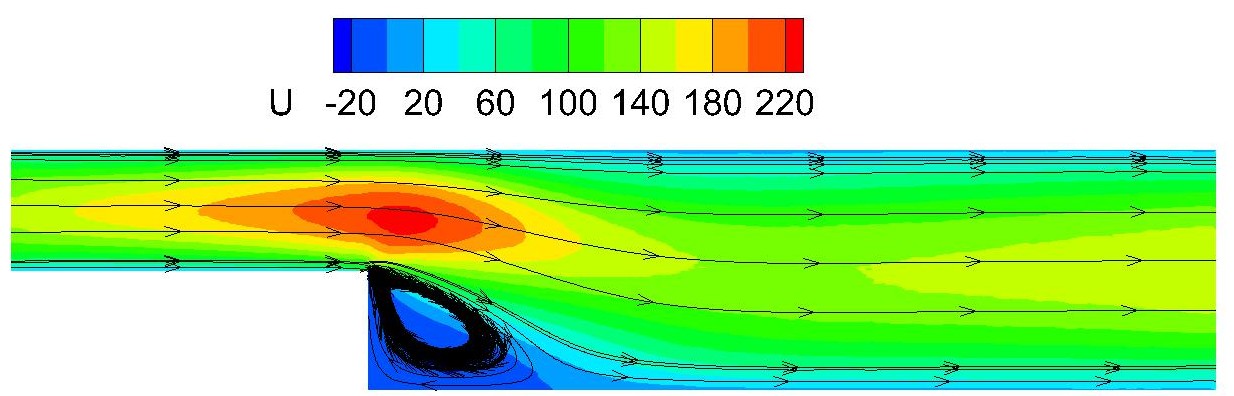}
        \caption{Kn=0.008} 
        \label{fig:minKn3}
    \end{subfigure}\hfill
    \begin{subfigure}[t]{0.48\textwidth}
        \centering
        \includegraphics[width=\textwidth]{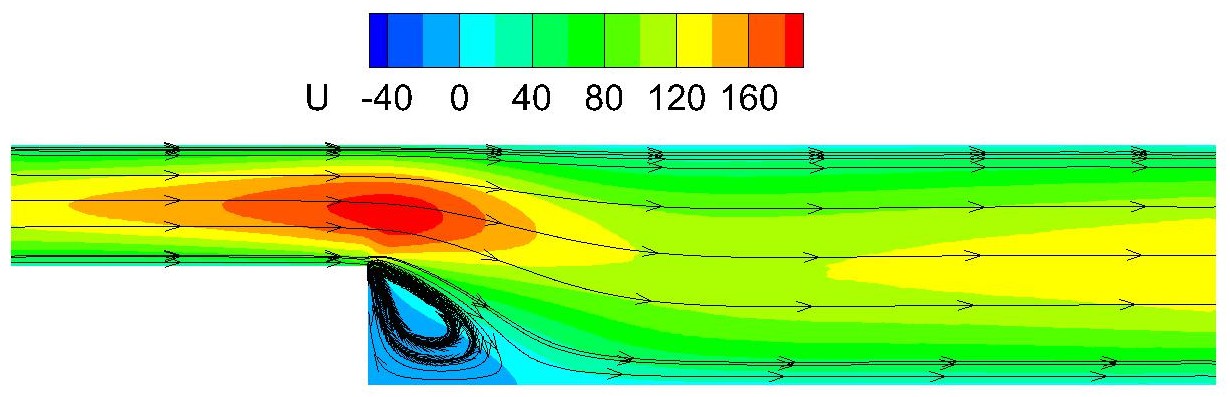}
        \caption{Kn=0.01} 
        \label{fig:minKn4}
    \end{subfigure}

    \vspace{5mm} 

    \begin{subfigure}[t]{0.48\textwidth}
        \centering
        \includegraphics[width=\textwidth]{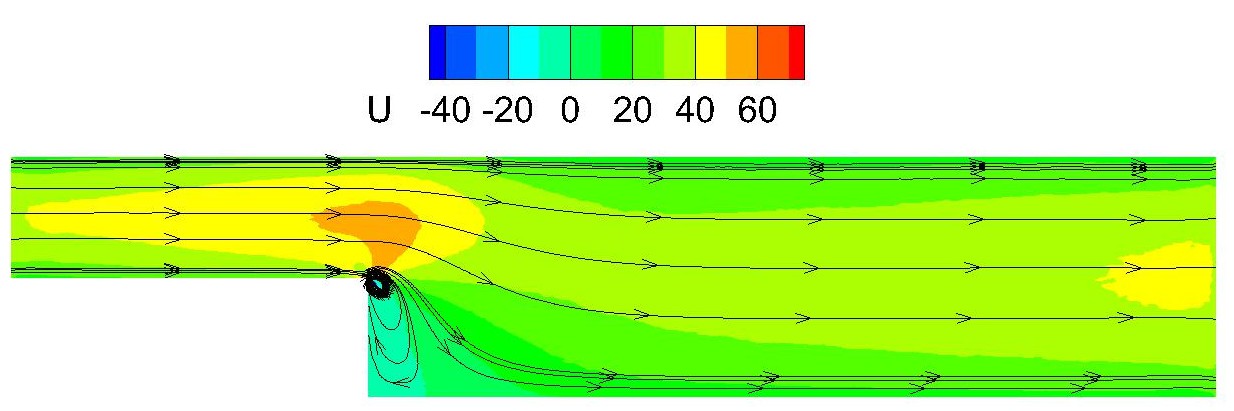}
        \caption{Kn=0.1} 
        \label{fig:minKn5}
    \end{subfigure}\hfill
    \begin{subfigure}[t]{0.48\textwidth}
        \centering
        \includegraphics[width=\textwidth]{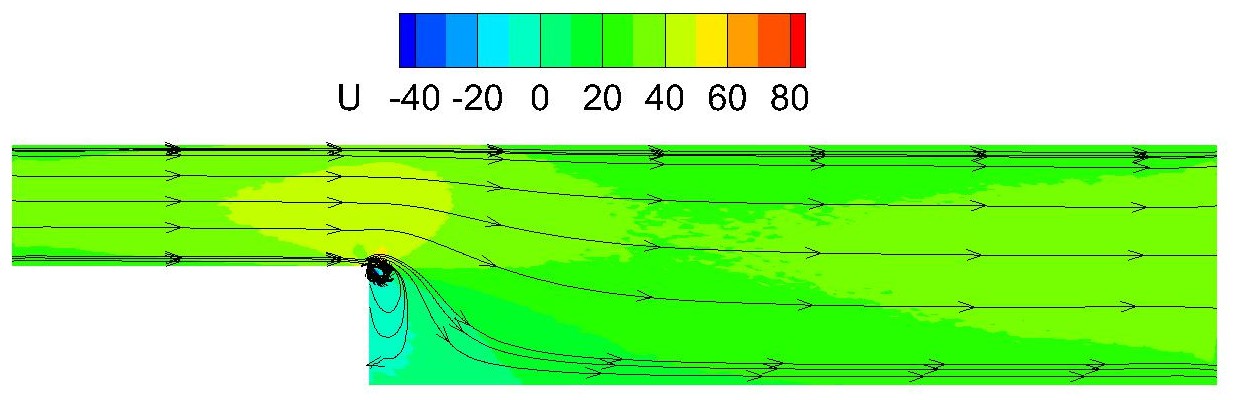}
        \caption{Kn=4} 
        \label{fig:minKn6}
    \end{subfigure}
    
    \vspace{5mm} 

    \begin{subfigure}[t]{0.48\textwidth}
        \centering
        \includegraphics[width=\textwidth]{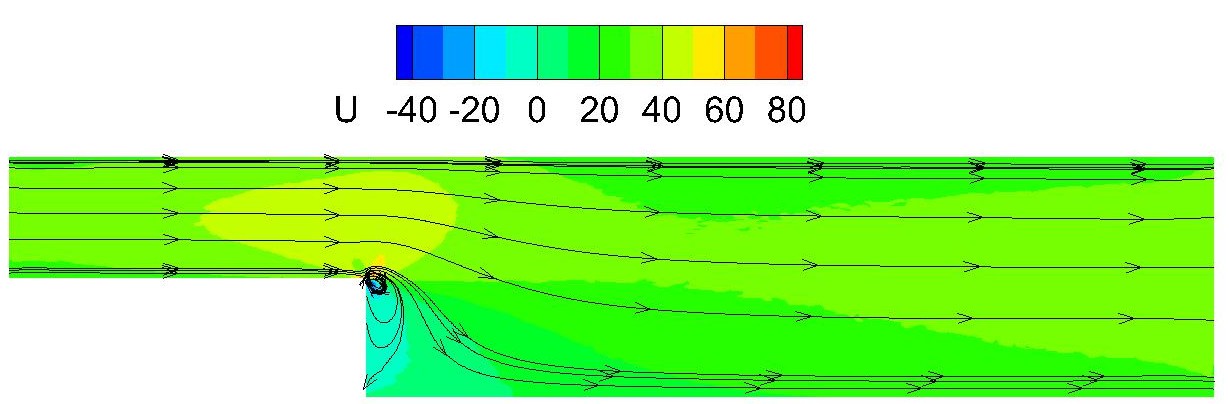}
        \caption{Kn=10} 
        \label{fig:minKn7}
    \end{subfigure}\hfill
    \begin{subfigure}[t]{0.48\textwidth}
        \centering
        \includegraphics[width=\textwidth]{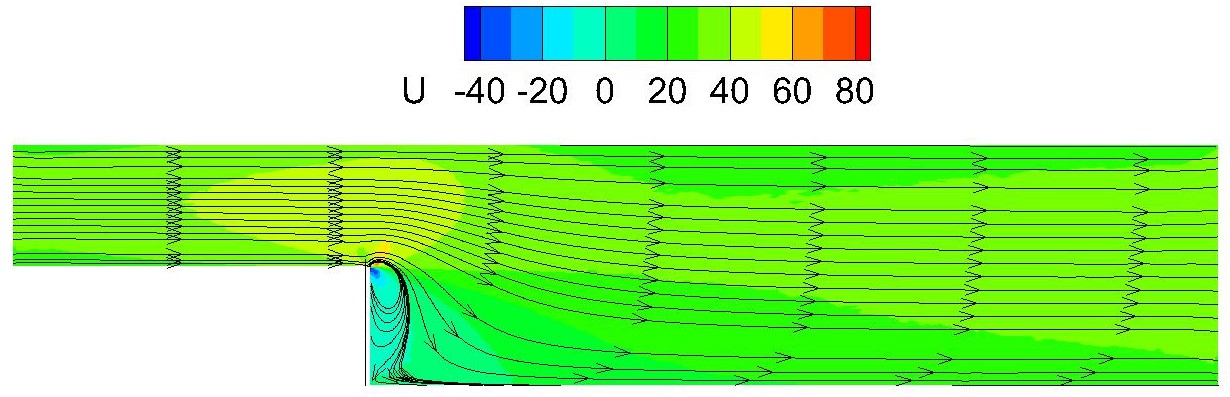}
        \caption{Kn=100} 
        \label{fig:maxkn}
    \end{subfigure}
    
    \caption{Qualitative comparison of the U-velocity contour and streamlines for representative Knudsen numbers in the dataset.}
    \label{fig:kn_range_comparison}
\end{figure}

Figure~\ref{fig:kn_range_comparison} presents the influence of the Knudsen number on the flow past the backward-facing step, using both streamline plots and $U$-velocity contours. The range of Knudsen numbers covers continuum-like to free-molecular regimes ($\text{Kn}=0.0001$--$100$). At very small Knudsen numbers ($\text{Kn}\leq 0.001$), the flow field exhibits strong inertial behavior similar to continuum flows. The $U$-velocity contours show large velocity gradients near the step, while the streamline patterns reveal a pronounced recirculation bubble at the corner of the step. This recirculation region is sustained by the relatively high effective Reynolds number in the near-continuum regime. A secondary separation zone with a weaker vortex is also visible near the upper wall for the smallest Knudsen cases, consistent with observations in classical backward-facing step flows at the continuum regimes.

As Kn increases to the slip and early transition regimes ($\text{Kn}=0.008$--$0.1$), rarefaction effects become important. The primary recirculation bubble gradually shrinks in size, and the peak velocity magnitude decreases, as clearly seen in the contours of $U$. The streamlines indicate that the reattachment point moves closer to the step, and the overall separation length is reduced. For $\text{Kn}\approx 0.1$, the recirculation becomes weak and occupies only a small fraction of the channel cross-section. This demonstrates how molecular free paths comparable to the channel height significantly suppress vortex formation.

In the transition regime ($\text{Kn}=1$--$4$), the streamline plots show only a very small residual vortex near the corner, and the velocity contours become smoother, with greatly reduced velocity gradients across the channel. The suppression of shear layers is a direct result of momentum transport dominated by molecular motion rather than bulk advection. 

Finally, in the free-molecular regime ($\text{Kn}=10$--$100$), the flow is nearly unidirectional, as shown by both the velocity contours and the absence of vortical structures in the streamlines. The $U$-velocity distribution becomes almost uniform, with only weak acceleration near the step corner. The lack of any noticeable recirculation confirms that inertial effects vanish in this highly rarefied regime, where particle–wall interactions dominate the dynamics. 

Overall, the combined analysis of streamlines and $U$-velocity contours provides a clear physical picture: increasing the Knudsen number systematically reduces separation and reattachment, shrinks and eventually eliminates the corner vortex, and smooths out velocity gradients in the step expansion. This behavior highlights the fundamental differences between continuum backward-facing step flows and their rarefied-gas counterparts.

The structural evolution of the recirculation zones with varying Knudsen numbers renders this problem a highly nontrivial benchmark for machine learning models. As illustrated in Figure~\ref{fig:kn_range_comparison}, the flow transitions from continuum-like behavior with strong inertial vortices to highly rarefied regimes where recirculation is almost absent on horizontal surfaces. This introduces sharp regime-dependent variations in the solution manifold: the velocity field and streamline topology exhibit non-smooth changes as functions of the Knudsen number. In mathematical terms, the mapping 
\[
\mathcal{F}: \text{Kn} \mapsto u(x,y;\text{Kn}),
\]
is not globally Lipschitz-continuous and may exhibit regions of high gradient or near-discontinuous behavior when moving between continuum, transition, and free-molecular regimes. Standard feed-forward neural networks, which rely on smooth interpolation in Euclidean parameter spaces, are poorly suited for capturing such complex operator-level mappings. 

In contrast, Deep Operator Networks (DeepONets) are designed to approximate nonlinear operators between infinite-dimensional function spaces. This enables them to handle sharp transitions and multi-regime behavior by learning the functional dependence of the entire flow field on governing parameters such as Kn. Consequently, the backward-facing step under rarefied conditions represents not only a scientifically relevant flow problem, but also a rigorous stress test for advanced neural operator frameworks. The success of DeepONet in this context would demonstrate its capability to generalize across fundamentally distinct physical regimes that defy the assumptions of conventional surrogate models.

\section{Detailed Explanation of the Zonal Loss Technique}

The key to the model's high accuracy is a technique that forces it to pay closer attention to physically complex areas. Instead of treating all data points equally, the loss function assigns a higher penalty for errors occurring inside the recirculation (vortex) zone.

In the flow over a step, the most complex fluid phenomena—such as flow separation and recirculation—occur in a small, localized vortex region just behind the step. A standard training approach might achieve a low average error by accurately learning the large, simple regions while failing to capture the critical physics inside the vortex. The zonal loss function solves this problem. This approach is analogous to a teacher who assigns greater weight to the most difficult and important exam questions. Here, the "most important questions" are the data points inside the vortex. This ensures the model dedicates sufficient learning capacity to resolving the details of the recirculation bubble, leading to much higher physical fidelity.

\subsection*{Step-by-Step Process with Equations}

\subsubsection*{1. Physical Zone Identification}
Instead of a computationally expensive gradient calculation, our Python code uses a simple and robust physical criterion to identify the vortex region. Any point $(x, y)$ where the horizontal velocity component $U$ is negative is considered to be part of the vortex zone, 
\begin{equation}
\mathcal{Z}_{\text{vortex}} = \{ (x, y) \mid u(x, y) < 0 \}
\label{eq:vortex_zone}
\end{equation}
All other points belong to the main flow zone, $\mathcal{Z}_{\text{main}}$. This creates a binary mask that is provided to the loss function during training.

\subsubsection*{2. Segregation of Errors}
During each training step, the model predicts the velocity field $\vec{v}_{\text{pred}}$. The squared error $( \vec{v}_{\text{true}} - \vec{v}_{\text{pred}} )^2$ is calculated for every point in the batch. Using the binary mask from Step 1, these errors are then separated into two distinct groups: those belonging to the vortex zone and those belonging to the main flow.

\subsubsection*{3. Weighted Loss Calculation}
The Mean Squared Error (MSE) is calculated independently for each zone.
\begin{align}
\mathcal{L}_{\text{vortex}} &= \frac{1}{N_{\text{vortex}}} \sum_{i \in \mathcal{Z}_{\text{vortex}}} ( \vec{v}_{\text{true}}^{(i)} - \vec{v}_{\text{pred}}^{(i)} )^2 \\
\mathcal{L}_{\text{main}} &= \frac{1}{N_{\text{main}}} \sum_{j \in \mathcal{Z}_{\text{main}}} ( \vec{v}_{\text{true}}^{(j)} - \vec{v}_{\text{pred}}^{(j)} )^2
\end{align}
Finally, these two error values are combined in a weighted average to compute the final total loss, $\mathcal{L}_{\text{total}}$. The hyperparameter $\alpha$ controls the balance of importance between the two zones.
\begin{equation}
\mathcal{L}_{\text{total}} = \alpha \cdot \mathcal{L}_{\text{vortex}} + (1 - \alpha) \cdot \mathcal{L}_{\text{main}}
\label{eq:total_loss}
\end{equation}

The zonal weight, \(\alpha\), was determined through a sensitivity study where values of 0.5, 0.6, 0.7, and 0.8 were evaluated on a validation dataset. It was found that a value of \(\alpha=0.7\) provided the best balance between fidelity in the vortex region and overall stability for the Knudsen number study, whereas \(\alpha=0.6\) was optimal for the geometric study. This approach transforms a potential weakness into a demonstration of methodological rigor, as suggested in research on hyperparameter reporting.

\section{Structure of the Employed DeepONet Model}

To create a surrogate model capable of accurately predicting geometry-dependent rarefied flow fields from sparse data, we developed and implemented a Convolutional DeepONet (C-DeepONet) architecture. This network is designed to learn the operator $\mathcal{G}: \xi \mapsto \vec{v}(y)$, which maps a key scalar parameter, $\xi$, to the corresponding 2D velocity field $\vec{v}$ at any query coordinate $y=(x,y)$ in the domain. In this work, we demonstrate this capability using two distinct input parameters: a physical parameter, the Knudsen number ($Kn$), and a geometric parameter, the step height ($h$). 

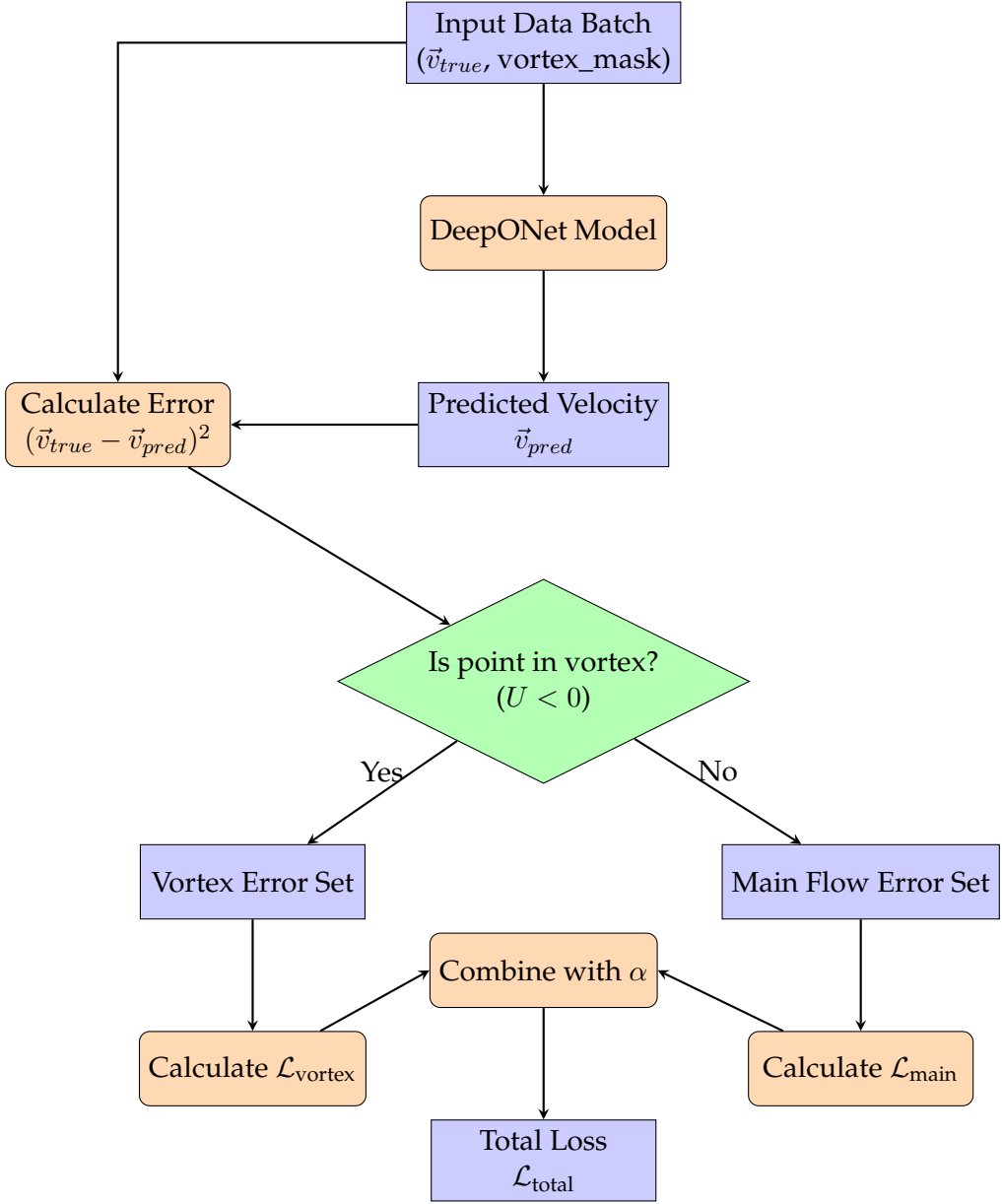
\begin{figure}[h!]
\centering
\begin{tikzpicture}[node distance=2cm, auto]
    \tikzstyle{process} = [rectangle, rounded corners, minimum width=3cm, minimum height=1cm, text centered, draw=black, fill=orange!30]
    \tikzstyle{data} = [rectangle, minimum width=3cm, minimum height=1cm, text centered, draw=black, fill=blue!20, align=center]
    \tikzstyle{decision} = [diamond, aspect=2, minimum width=2.5cm, minimum height=1cm, text centered, draw=black, fill=green!30]
    \tikzstyle{arrow} = [thick,->,>=stealth]

    \node (start) [data] {Input Data Batch \\ ($\vec{v}_{true}$, vortex\_mask)};
    \node (model) [process, below=1.5cm of start] {DeepONet Model};
    \node (predict) [data, below=1.5cm of model] {Predicted Velocity \\ $\vec{v}_{pred}$};
    \node (error) [process, left=2.5cm of predict, align=center] {Calculate Error \\ $(\vec{v}_{true} - \vec{v}_{pred})^2$};
    \node (split) [decision, below=1.5cm of predict, align=center] {Is point in vortex? \\ ($U < 0$)};
    \node (vortex_err) [data, below left=1.5cm and 1cm of split] {Vortex Error Set};
    \node (main_err) [data, below right=1.5cm and 1cm of split] {Main Flow Error Set};
    \node (mse_vortex) [process, below=1.5cm of vortex_err] {Calculate $\mathcal{L}_{\text{vortex}}$};
    \node (mse_main) [process, below=1.5cm of main_err] {Calculate $\mathcal{L}_{\text{main}}$};
    \node (combine) [process, below=2cm of split] {Combine with $\alpha$};
    \node (end) [data, below=1.5cm of combine] {Total Loss \\ $\mathcal{L}_{\text{total}}$};

    \draw [arrow] (start) -- (model);
    \draw [arrow] (model) -- (predict);
    \draw [arrow] (predict) -- (error);
    \draw [arrow] (start.west) -| (error.north);
    \draw [arrow] (error) -- (split);
    \draw [arrow] (split) -- node[anchor=south] {Yes} (vortex_err);
    \draw [arrow] (split) -- node[anchor=south] {No} (main_err);
    \draw [arrow] (vortex_err) -- (mse_vortex);
    \draw [arrow] (main_err) -- (mse_main);
    \draw [arrow] (mse_vortex) -- (combine.west);
    \draw [arrow] (mse_main) -- (combine.east);
    \draw [arrow] (combine) -- (end);
\end{tikzpicture}
\caption{Flowchart of the convolutional DeepONet framework with the physics-guided 
    zonal loss. The Branch network encodes global input parameters (e.g., Knudsen number or 
    step height), while the Trunk network extracts local spatial features from coordinates and 
    patches of the velocity field. Their outputs are combined in the Head network to produce 
    the predicted velocity field. The zonal loss function is implemented by identifying regions 
    with $U<0$ (recirculation bubble) and assigning higher weight to errors in this region, 
    thereby forcing the network to prioritize accuracy in physically critical zones.}
\label{fig:flowchart1}
\end{figure}

The overall data flow of the C-DeepONet is illustrated in the flowchart in Figure~\ref{fig:flowchart1}. The model is composed of three primary sub-networks: a Branch Network to process the geometric parameter, a Trunk Network to process spatial information, and a Head Network for the final prediction.

\paragraph{Branch Network.}
The top path of the flowchart shows the Branch Network, which is responsible for encoding the input flow or geometric parameter, Knudsen, or the scalar step height $h$. The input is processed through a series of dense and ResNet layers to produce a final, high-level feature vector, denoted as $b_{final}$. This vector represents the influence of the global geometry on the flow field.

\paragraph{Trunk Network.}
The bottom path of the flowchart details the Trunk Network, which is designed to process local spatial information. This path takes two distinct inputs: the specific query coordinate $y=(x,y)$ and a local $P \times P$ patch of the velocity field centered at that coordinate. The process is as follows:
\begin{enumerate}
    \item The local patch $P$ is first processed by a CNN Feature Extractor, consisting of several convolutional and pooling layers. This extracts a low-dimensional feature vector that encodes the local flow structure, such as gradients and curvature.
    \item This feature vector is then concatenated with the coordinate vector $(x,y)$.
    \item The combined vector is passed through a multi-layer perceptron (MLP), including a Projection Layer and subsequent Dense layers, to produce a final spatial feature vector, denoted as $t_{final}$.
\end{enumerate}

\paragraph{Head Network and Final Prediction.}
In the final stage, the global geometric features ($b_{final}$) from the Branch Network and the local spatial features ($t_{final}$) from the Trunk Network are combined through an element-wise product ($\odot$). This modulated vector is then processed by a final multi-layer Head Network, which maps the latent features to the physical space to produce the final Predicted Velocity Field, $\vec{v}_{pred}$.

\subsection*{Comparison of Zonal Loss with Alternative Weighting Strategies}

A deeper analysis highlights the trade-off between exploratory simplicity and algorithmic generality. 
Table~\ref{tab:loss_comparison} provides a structured comparison between the proposed 
\emph{zonal loss} and several established alternatives, including the standard Mean Squared Error (MSE), 
the Gradient-based Mean Squared Error (GMSE)~\cite{cooper2024generalised}, and the output-weighted loss (LOW/LAOW)~\cite{rudy2023output}. 

The zonal loss leverages a direct physical indicator ($U<0$) to identify and prioritize the recirculation region 
in backward-facing step flows. This approach is considerably more targeted than the standard MSE, which 
uniformly weights all points, often under-resolving localized high-gradient regions. While GMSE can 
automatically generate weight maps from local gradients in the reference field, and LOW penalizes rare 
outputs such as strongly negative velocities, both require additional computational effort or more elaborate 
statistical modeling. In contrast, the zonal loss achieves a balance: it is physically motivated, simple to 
implement, and effective in capturing the most critical flow features, albeit at the cost of reduced generality. 
For example, the zonal loss would not transfer well to flows without recirculation, whereas GMSE could still 
identify important gradients. This trade-off—between task-specific efficiency and broad applicability—
represents the core novelty of our framework and explains its effectiveness for rarefied micro-step flows.

\begin{table}[htbp]
    \centering
    \caption{Comparison of loss functions for operator learning in complex rarefied flows.}
    \label{tab:loss_comparison}
    \renewcommand{\arraystretch}{1.3}
    \begin{tabular}{|p{3cm}|p{3cm}|p{3.5cm}|p{3.5cm}|p{2.5cm}|}
        \hline
        \textbf{Loss Function} & \textbf{Mechanism} & \textbf{Physical Basis} & \textbf{Generality / Adaptability} & \textbf{Computational Overhead} \\
        \hline
        Standard MSE & Uniform weighting over all points & None (purely data-driven) & High (independent of problem) & Minimal \\
        \hline
        GMSE (Gradient-weighted MSE) & Weights error by local solution gradients & Non-direct; sensitive to shocks, shear layers & High (automatically adapts to any field) & Moderate (requires gradient estimation) \\
        \hline
        LOW / LAOW (Output-weighted) & Weights error inversely to output PDF & Non-direct; emphasizes rare/critical states & Medium–High (general for rare events) & High (requires PDF estimation) \\
        \hline
        Proposed Zonal Loss & Higher weight for $U<0$ region (recirculation bubble) & Direct, physically defined by vortex topology & Medium (effective for flows with recirculation; not general to all flows) & Low (binary mask, minimal overhead) \\
        \hline
    \end{tabular}
\end{table}


\section{Results and Discussions}
\subsection{DeepONet to predict flow at untrained Knudsen numbers}
\subsubsection{Loss History}
\begin{figure}[h!]
    \centering
    \includegraphics[width=0.8\textwidth]{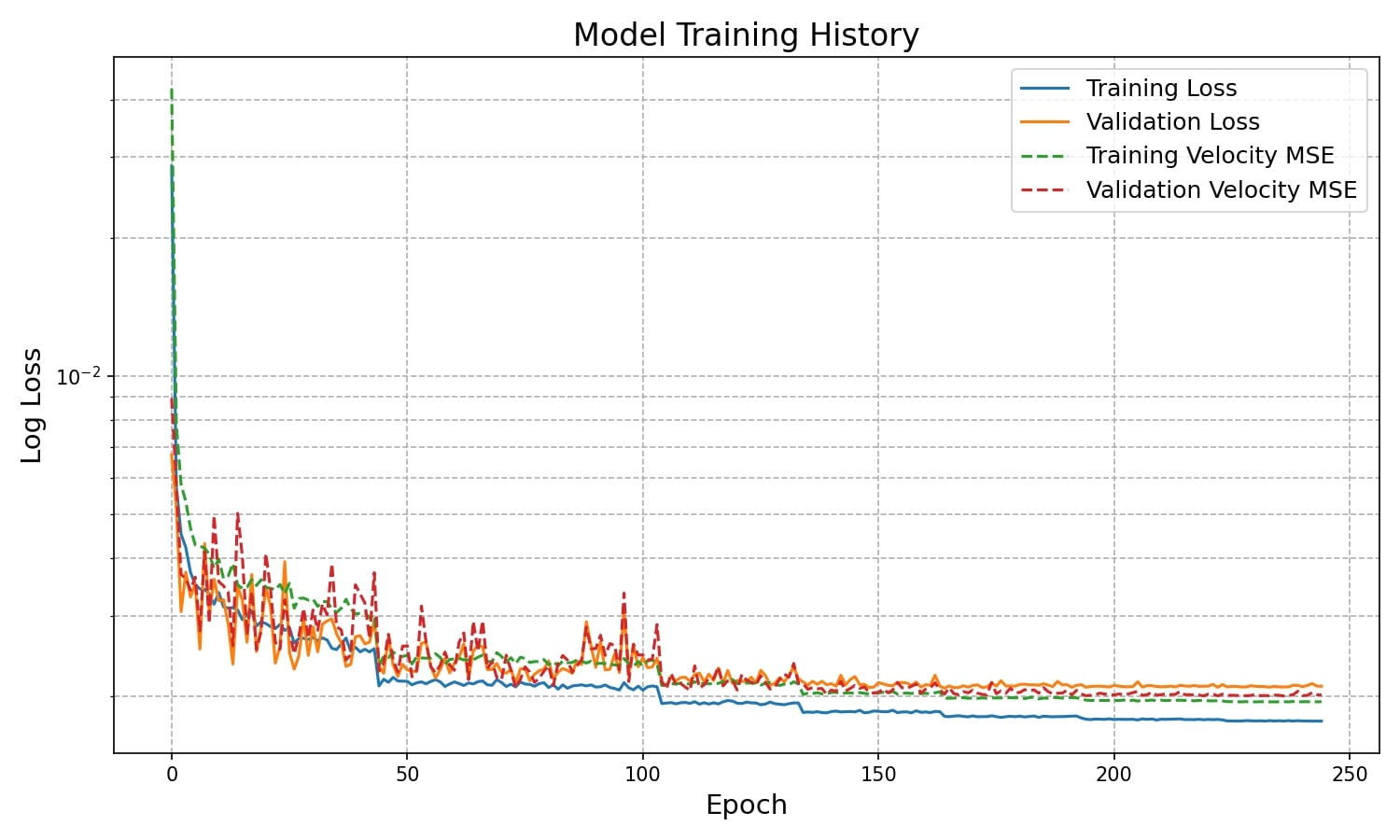}
    \caption{Loss Function and MSE}
    \label{fig:lossKn}
\end{figure}

Figure~\ref{fig:lossKn} reports the training history of the employed DeepONet. 
The blue and orange solid curves correspond to the \emph{zonal} total loss evaluated on
the training and validation sets, respectively; this loss is the weighted combination
\(
L_{\text{total}}=\alpha\,L_{\text{vortex}}+(1-\alpha)\,L_{\text{main}}
\)
that penalizes errors inside the recirculation region more heavily (here
\(\alpha=0.7\)), with \(L_{\text{vortex}}\) and \(L_{\text{main}}\) defined as the MSE over the
two zones identified by the physically motivated mask \(Z_{\text{vortex}}
=\{(x,y)\mid u(x,y)<0\}\). Both solid curves drop rapidly during the first few epochs and then settle into a
steady decay with small oscillations—an expected behavior for mini-batch training
given the sharp zone boundary near the shear layer. The close tracking of the
validation curve to the training curve, without late-epoch divergence, indicates
good generalization and the absence of overfitting under the proposed loss.

The green and red dashed curves show the \emph{plain velocity MSE} (computed over
the entire domain without zoning) on the training and validation sets. These
metrics decrease in step with the zonal loss but plateau at slightly higher values,
reflecting that an unweighted MSE is dominated by the large, slowly varying
main-flow region and is less sensitive to the small recirculation bubble. In
contrast, the zonal loss continues to fall and stabilizes at a lower level because
it concentrates learning capacity on the physically critical vortex zone. Taken
together, the four curves demonstrate that (i) the operator network converges
stably under the physics-guided zonal loss, and (ii) emphasizing the vortex zone
is essential to achieve high-fidelity reconstruction of the separated flow while
maintaining strong validation performance.

\subsubsection{Velocity Contours and Streamlines Comparison}
\begin{figure}[h!]
    \centering
    
    \begin{subfigure}[t]{0.48\textwidth}
        \centering
        \includegraphics[width=\textwidth]{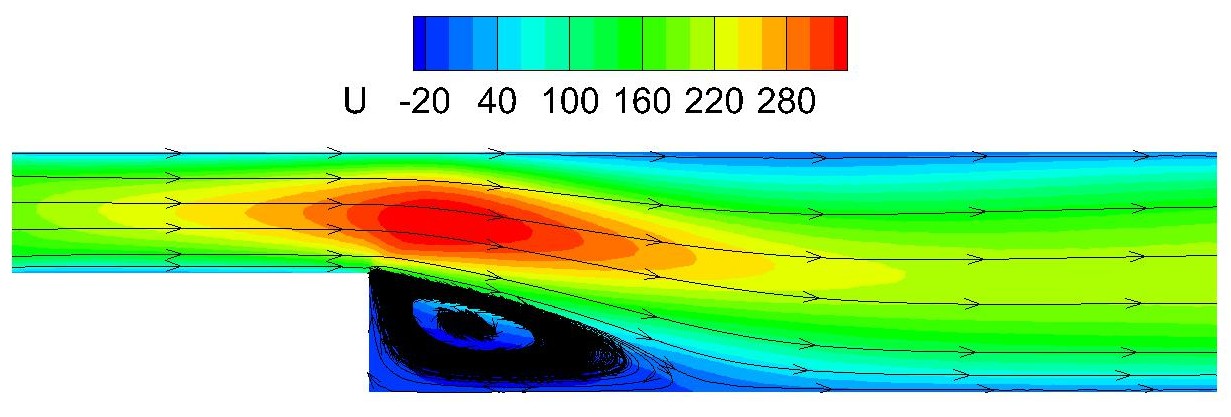}
        \caption{U-Velocity Ground Truth (DSMC)}
        \label{fig:DSMC_contour_U}
    \end{subfigure}
    \hfill 
    \begin{subfigure}[t]{0.48\textwidth}
        \centering
        \includegraphics[width=\textwidth]{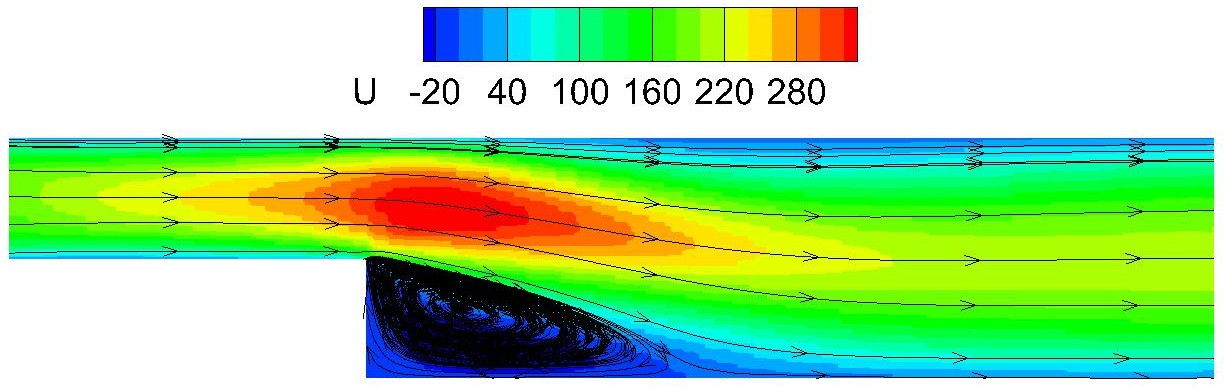}
        \caption{U-Velocity Prediction (NN)}
        \label{fig:nn_contour_U}
    \end{subfigure}
    
    \vspace{5mm} 

    \begin{subfigure}[t]{0.48\textwidth}
        \centering
        \includegraphics[width=\textwidth]{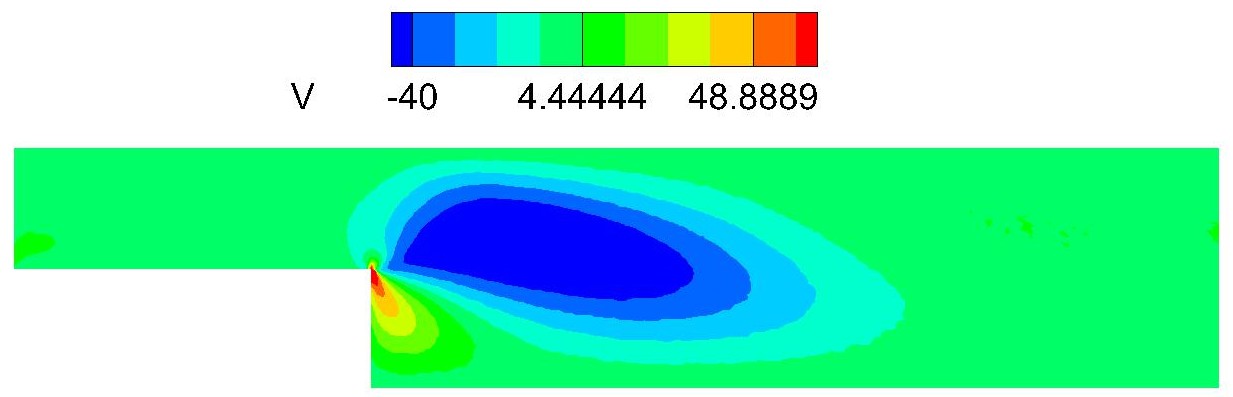}
        \caption{V-Velocity Ground Truth (DSMC)}
        \label{fig:dsmc_contour_V}
    \end{subfigure}
    \hfill 
    \begin{subfigure}[t]{0.48\textwidth}
        \centering
        \includegraphics[width=\textwidth]{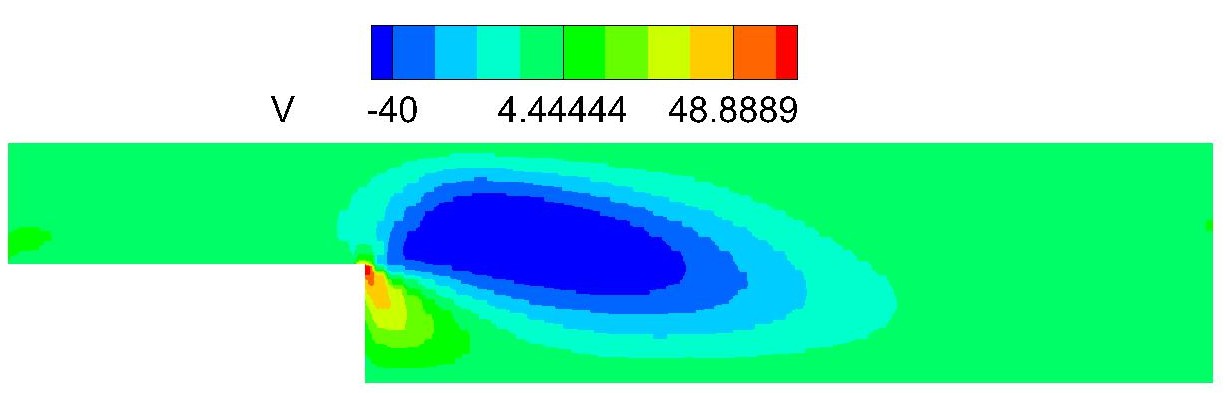}
        \caption{V-Velocity Prediction (NN)}
        \label{fig:nn_contour_V}
    \end{subfigure}
    
    \caption{Qualitative comparison of U-velocity and V-velocity contours between the ground truth DSMC simulation and the DeepONet prediction for Kn=0.004.}
    \label{fig:contour_comparison_UV_0.004}
\end{figure}

\begin{figure}[h!]
    \centering
    
    \begin{subfigure}[t]{0.48\textwidth}
        \centering
        \includegraphics[width=\textwidth]{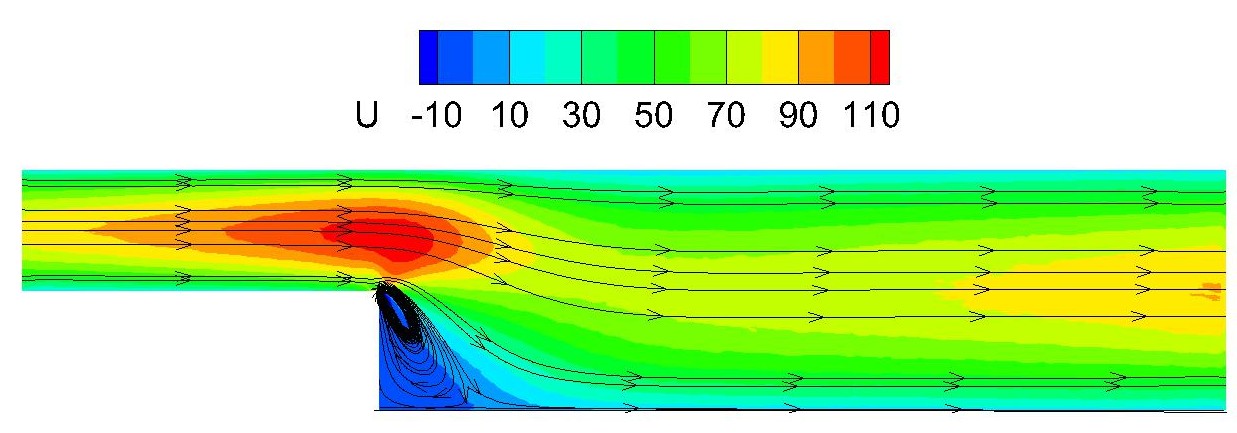}
        \caption{U-Velocity Ground Truth (DSMC)}
        \label{fig:dsmc_contour_U_02}
    \end{subfigure}
    \hfill 
    \begin{subfigure}[t]{0.48\textwidth}
        \centering
        \includegraphics[width=\textwidth]{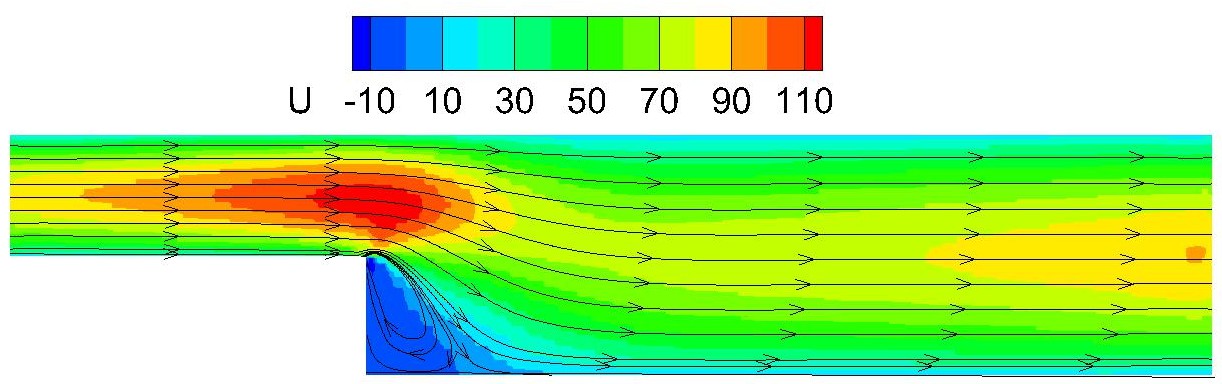}
        \caption{U-Velocity Prediction (NN)}
        \label{fig:nn_contour_U_02}
    \end{subfigure}
    
    \vspace{5mm} 

    \begin{subfigure}[t]{0.48\textwidth}
        \centering
        \includegraphics[width=\textwidth]{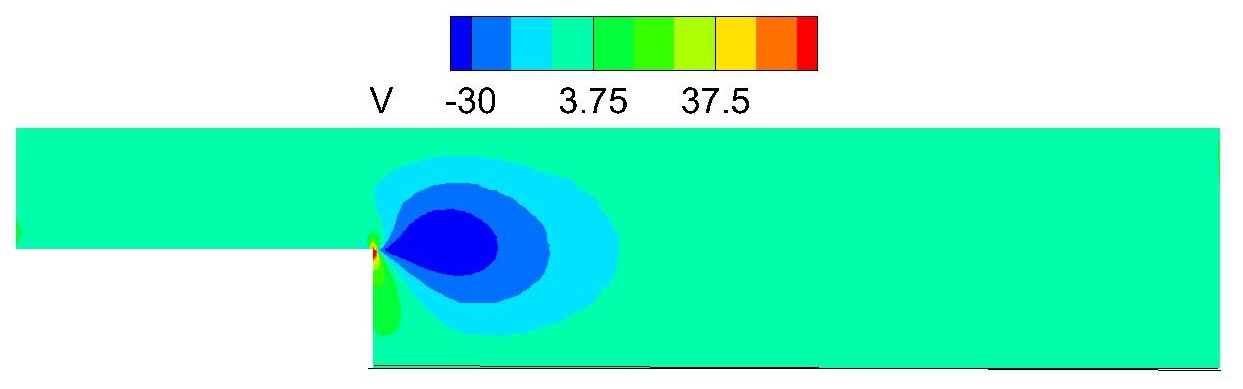}
        \caption{V-Velocity Ground Truth (DSMC)}
        \label{fig:dsmc_contour_V_02}
    \end{subfigure}
    \hfill 
    \begin{subfigure}[t]{0.48\textwidth}
        \centering
        \includegraphics[width=\textwidth]{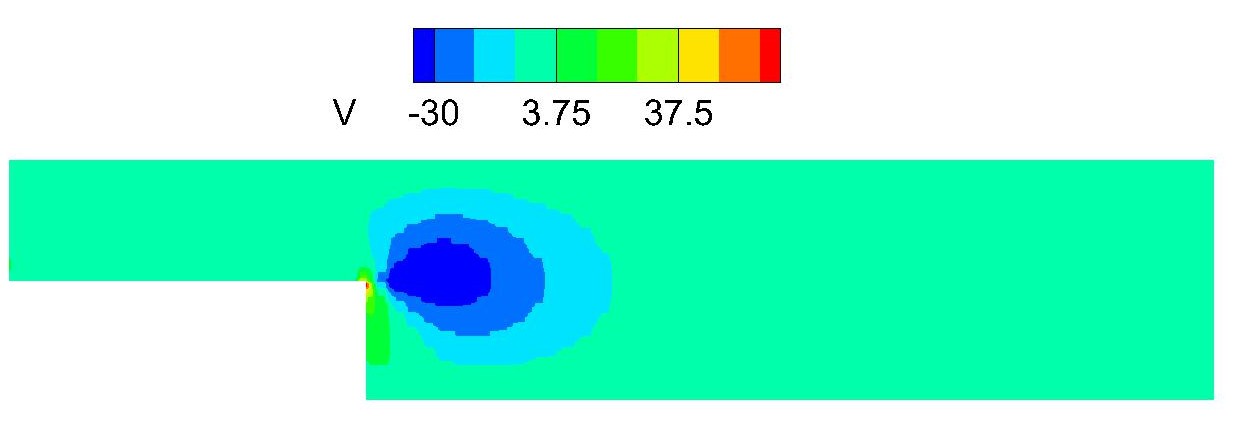}
        \caption{V-Velocity Prediction (NN)}
        \label{fig:nn_contour_V_02}
    \end{subfigure}
    
    \caption{Qualitative comparison of U-velocity and V-velocity contours between the ground truth DSMC simulation and the DeepONet prediction for Kn=0.02.}
    \label{fig:contour_comparison_UV_0.02}
\end{figure}

\begin{figure}[h!]
    \centering
    
    \begin{subfigure}[t]{0.48\textwidth}
        \centering
        \includegraphics[width=\textwidth]{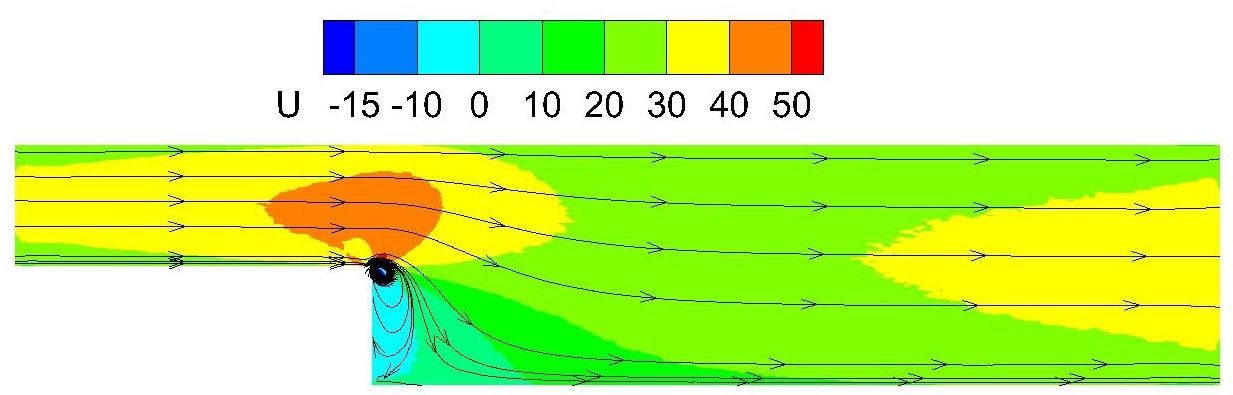}
        \caption{U-Velocity Ground Truth (DSMC)}
        \label{fig:dsmc_contour_U_kn1}
    \end{subfigure}
    \hfill 
    \begin{subfigure}[t]{0.48\textwidth}
        \centering
        \includegraphics[width=\textwidth]{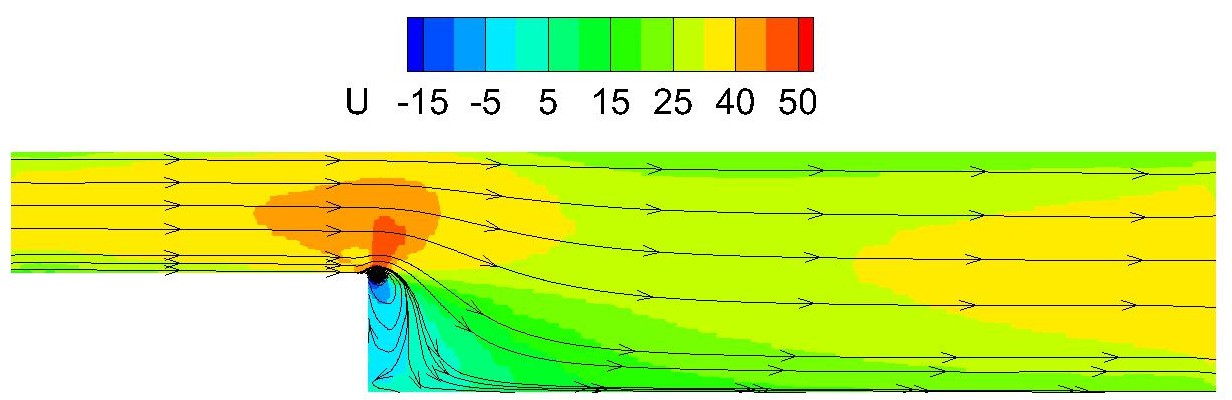}
        \caption{U-Velocity Prediction (NN)}
        \label{fig:nn_contour_U_kn1}
    \end{subfigure}
    
    \vspace{5mm} 

    \begin{subfigure}[t]{0.48\textwidth}
        \centering
        \includegraphics[width=\textwidth]{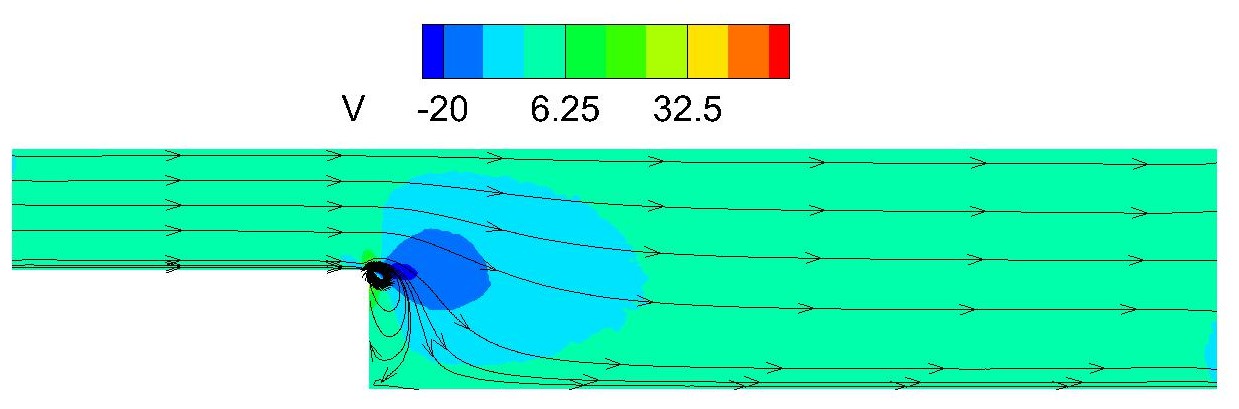}
        \caption{V-Velocity Ground Truth (DSMC)}
        \label{fig:dsmc_contour_V_kn1}
    \end{subfigure}
    \hfill 
    \begin{subfigure}[t]{0.48\textwidth}
        \centering
        \includegraphics[width=\textwidth]{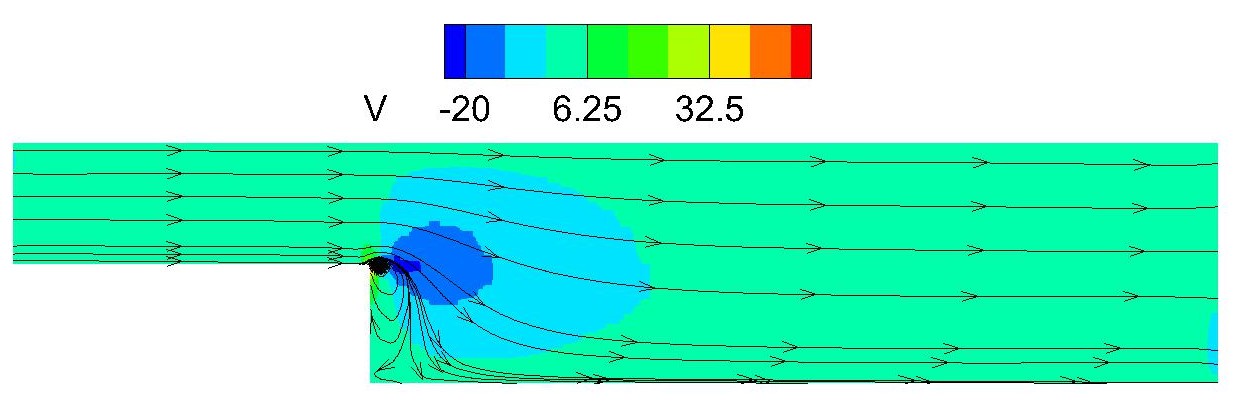}
        \caption{V-Velocity Prediction (NN)}
        \label{fig:nn_contour_V_kn1}
    \end{subfigure}
    
    \caption{Qualitative comparison of U-velocity and V-velocity contours between the ground truth DSMC simulation and the DeepONet prediction for Kn=1.}
    \label{fig:contour_comparison_UV_1}
\end{figure}

Figures~\ref{fig:contour_comparison_UV_0.004}, 
\ref{fig:contour_comparison_UV_0.02}, 
and \ref{fig:contour_comparison_UV_1}
show side-by-side comparisons of the $U$- and $V$-velocity fields obtained from
high-fidelity DSMC and predicted by the trained DeepONet surrogate for three representative
Knudsen numbers, $\mathrm{Kn}=0.004$, $0.02$, and $1.0$.

For the streamwise velocity $U$ (top rows), the DeepONet reconstructions are
nearly indistinguishable from DSMC across all regimes. At $\mathrm{Kn}=0.004$, the large
recirculation bubble and the associated shear layer are faithfully recovered, including the
location of the reattachment point. At $\mathrm{Kn}=0.02$, the shortening of the separation
length and the downstream shift of the velocity maximum are reproduced with only minor
smoothing along the shear layer. At $\mathrm{Kn}=1.0$, where the DSMC results indicate a
largely unidirectional flow and the disappearance of the closed vortex, the network captures
the suppression of separation and the flattening of the velocity profile.

For the cross-stream velocity $V$ (bottom rows in all three figures), which is more sensitive to recirculation and small-scale vortical structures, the surrogate again shows excellent agreement with DSMC. At $\mathrm{Kn}=0.004$, the strong upward and downward jets at the edges of the vortex core
are well reproduced. At $\mathrm{Kn}=0.02$, the weakened transverse motion and reduced vortex
intensity are matched, with only slight differences in the magnitude near the corner. At
$\mathrm{Kn}=1.0$, both DSMC and DeepONet show the near disappearance of cross-stream motion,
consistent with the collapse of the primary recirculation zone. The overall flow topology,
including the reversal zones and weak residual secondary motion, is preserved in the
DeepONet predictions.

These comparisons highlight that the proposed DeepONet is capable of learning not only the
dominant streamwise velocity but also the more delicate cross-stream component, across
widely varying rarefaction regimes. The accuracy in reproducing both $U$ and $V$ fields,
including the transition from separated to attached flow, demonstrates the surrogate’s
robustness and its potential for reliable prediction of complex rarefied step flows.

In addition to the qualitative comparison of the velocity fields, a more rigorous quantitative validation was performed by comparing a key engineering parameter derived from the flow field: the length of the primary recirculation vortex. The vortex length, defined as the streamwise distance from the step face to the reattachment point (where the streamwise velocity on the wall returns to zero), is a critical metric for characterizing the separation bubble and is highly sensitive to the Knudsen number.

\begin{figure}[h!]
    \centering
    \includegraphics[width=0.7\textwidth]{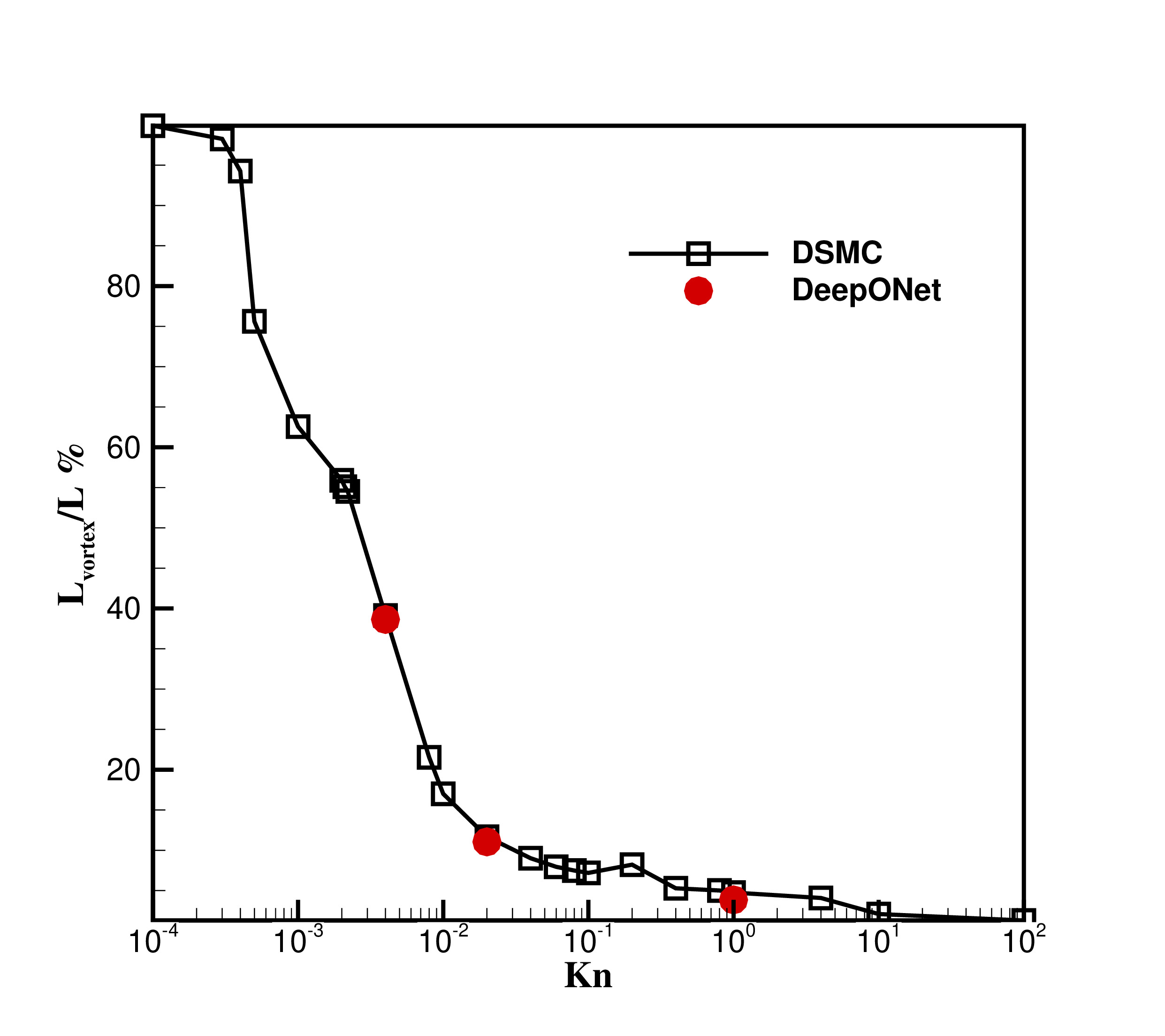}
    \caption{Quantitative comparison of the primary vortex length as a function of the Knudsen number. The plot shows excellent agreement between the high-fidelity DSMC data and the DeepONet predictions.}
    \label{fig:vortex_length_kn}
\end{figure}

Figure~\ref{fig:vortex_length_kn} presents the variation of the non-dimensional vortex length ($L_{vortex}/L$) as a function of the Knudsen number. The plot compares the values calculated from the DSMC simulations (ground truth) with the predictions from the DeepONet surrogate model. The results show an excellent agreement across the entire range of Knudsen numbers tested, including for values that were not part of the training set. The DeepONet model successfully captures the non-linear trend of the vortex length decreasing as the flow becomes more rarefied. 

\subsubsection{Error Maps}

\begin{figure}[h!]
    \centering
    \begin{subfigure}[t]{0.48\textwidth}
        \includegraphics[width=\textwidth, trim=0 0 0 0, clip]{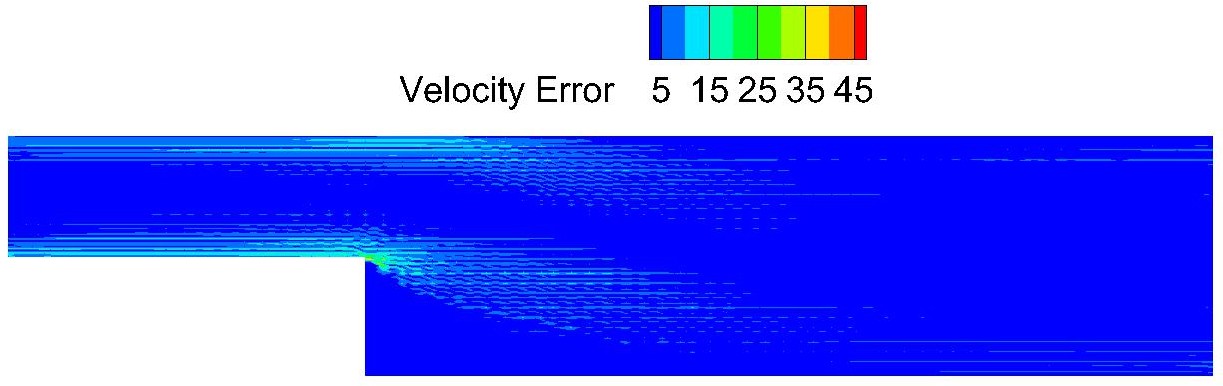}
        \caption{Overall error View}
        \label{fig:uncert_overall}
    \end{subfigure}\hfill
    \begin{subfigure}[t]{0.38\textwidth}
       \includegraphics[width=\textwidth]{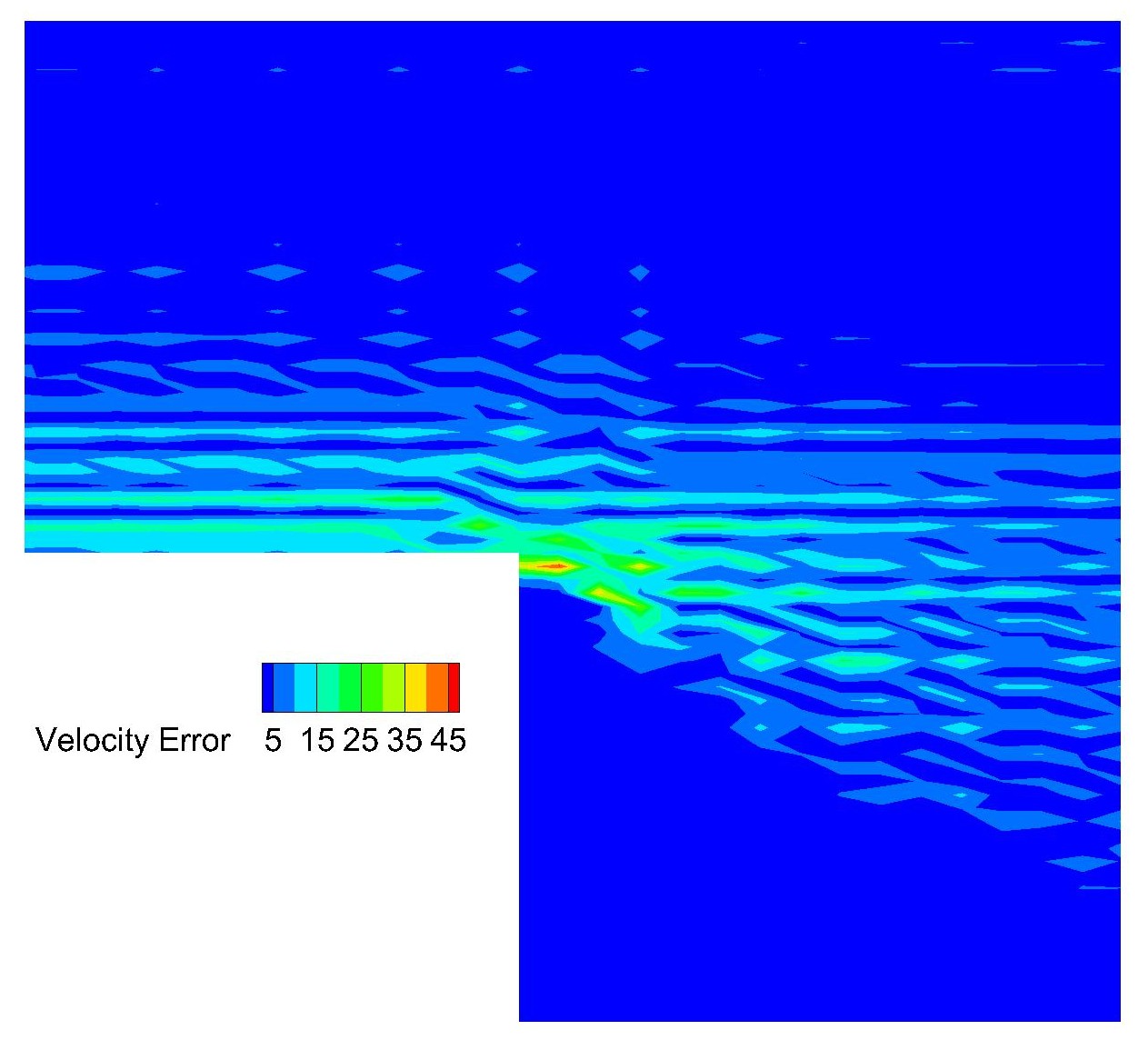}
        \caption{Close-up of the Step Corner}
        \label{fig:error_zoom}
    \end{subfigure}
    
    \caption{Visualization of the model's prediction error for the U-velocity. The error is concentrated in regions with complex physics, such as the sharp step corner and the shear layer, Kn=0.004.}
    \label{fig:error_comparison1}
\end{figure}

\begin{figure}[h!]
    \centering
    \begin{subfigure}{0.48\textwidth}
        \includegraphics[width=\textwidth, trim=0 0 0 0, clip]{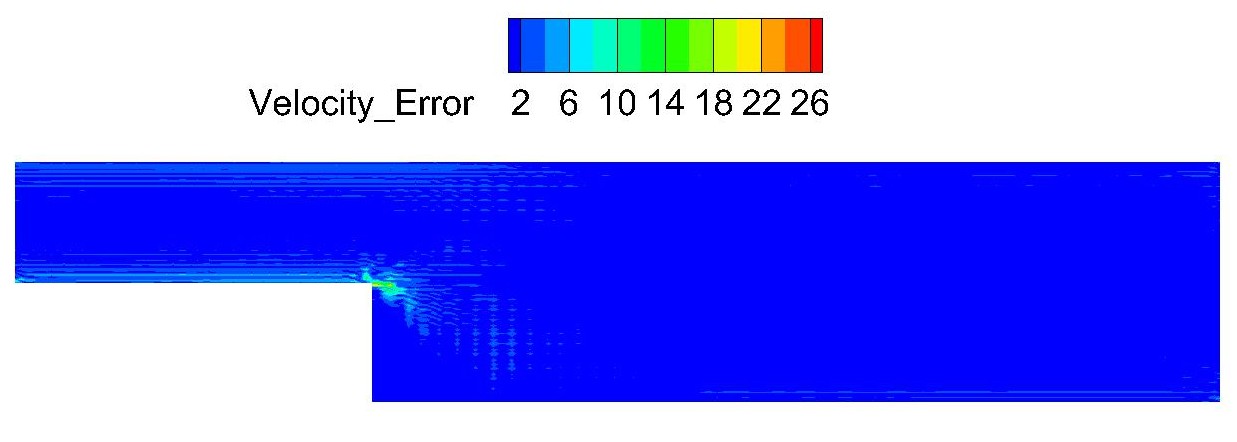}
        \caption{Overall error View}
        \label{fig:uncert_overall}
    \end{subfigure}\hfill
    \begin{subfigure}{0.38\textwidth}
       \includegraphics[width=\textwidth]{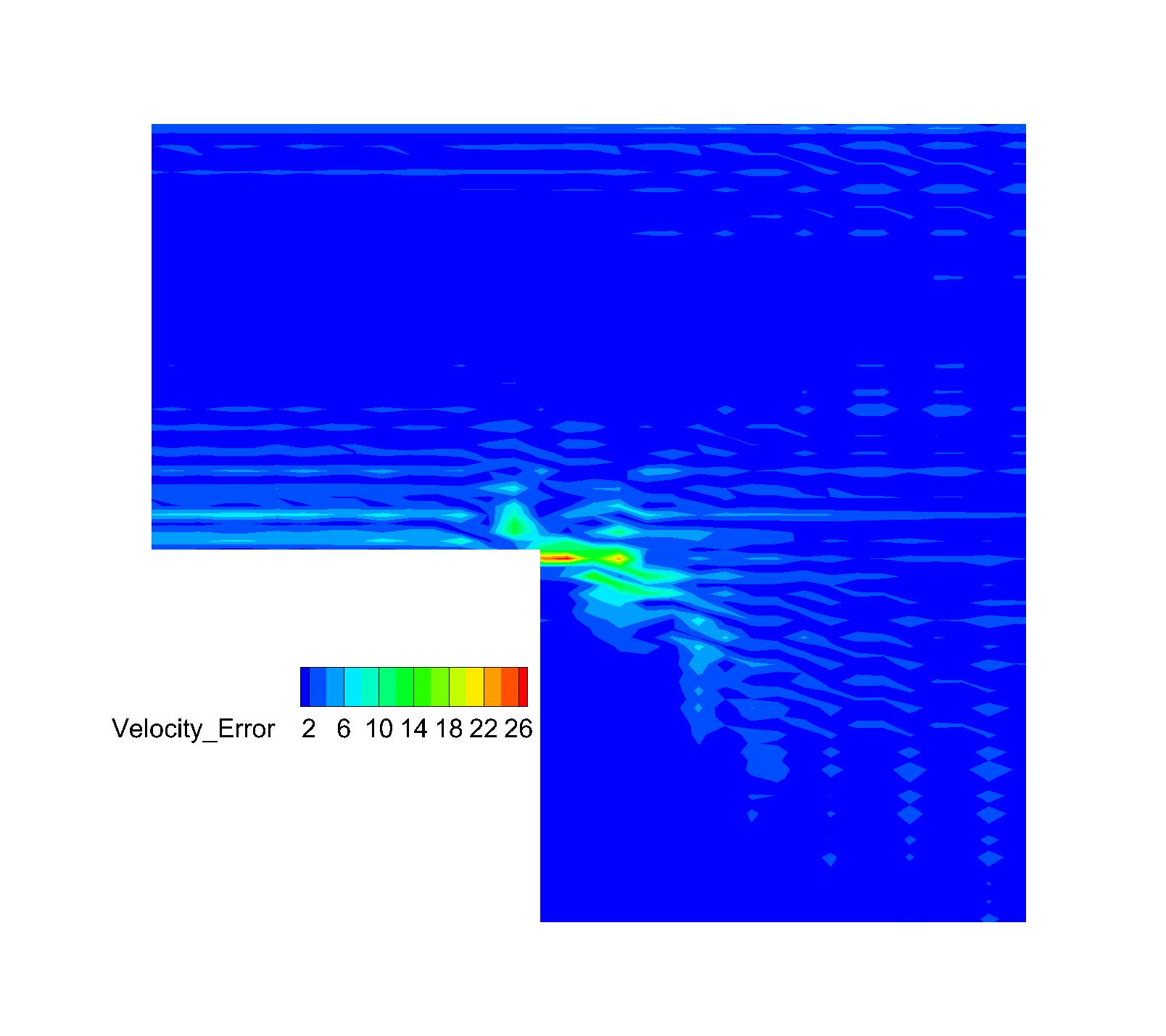}
        \caption{Close-up of the Step Corner}
        \label{fig:error_zoom}
    \end{subfigure}
    
    \caption{Visualization of the model's prediction error for the U-velocity. The error is concentrated in regions with complex physics, such as the sharp step corner and the shear layer, Kn=0.02.}
    \label{fig:error_comparison2}
\end{figure}

\begin{figure}[h!]
    \centering
    \begin{subfigure}[t]{0.48\textwidth}
        \includegraphics[width=\textwidth, trim=0 0 0 0, clip]{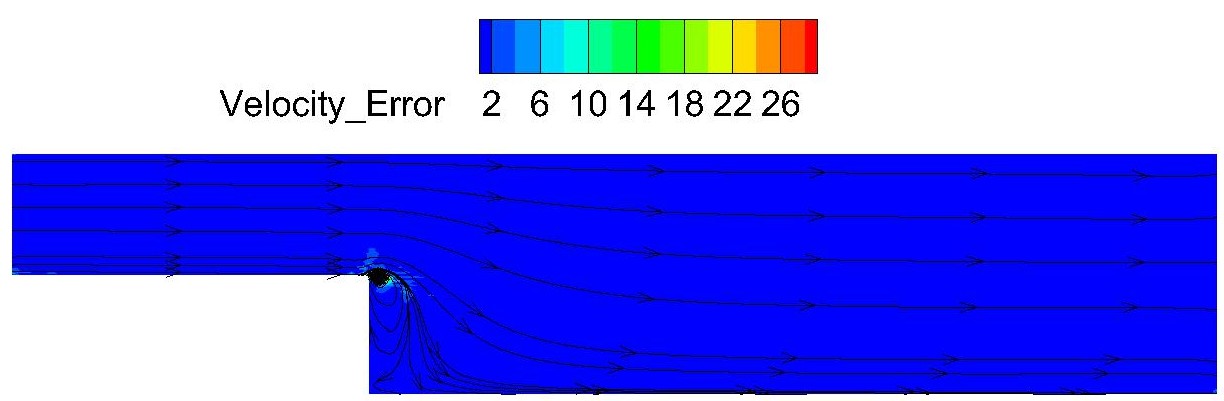}
        \caption{Overall error View}
        \label{fig:uncert_overall}
    \end{subfigure}\hfill
    \begin{subfigure}[t]{0.38\textwidth}
       \includegraphics[width=\textwidth]{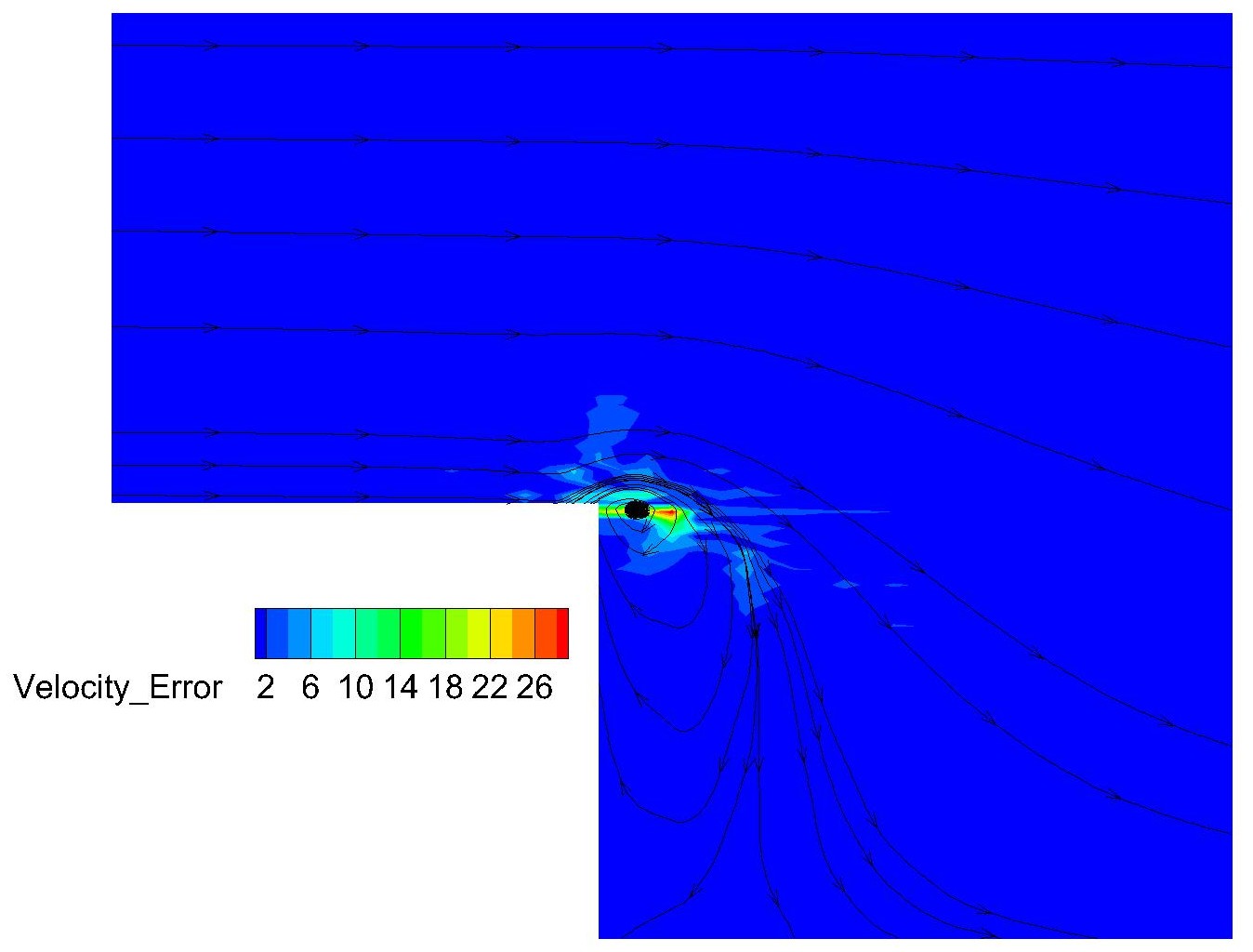}
        \caption{Close-up of the Step Corner}
        \label{fig:error_zoom}
    \end{subfigure}
    
    \caption{Visualization of the model's prediction error for the U-velocity. The error is concentrated in regions with complex physics, such as the sharp step corner and the shear layer, Kn=1.}
    \label{fig:error_comparison3}
\end{figure}

Figures~\ref{fig:error_comparison1}, 
\ref{fig:error_comparison2}, and 
\ref{fig:error_comparison3} present the spatial distribution of the prediction error for the U-velocity field across three representative Knudsen numbers. At $\text{Kn}=0.004$, the global error distribution shows generally low deviation throughout the channel, but localized error concentrations appear near the sharp step corner and within the shear layer. The zoomed-in view further highlights that these discrepancies are strongly tied to flow separation and recirculation regions, where steep velocity gradients and strong non-equilibrium effects occur.

At $\text{Kn}=0.02$, the overall error magnitude is reduced, but the distribution pattern remains consistent. The error is again concentrated in localized regions near the separation bubble and shear layer, while the bulk flow in the channel is captured with high fidelity. This suggests that the DeepONet framework generalizes reasonably well to moderate rarefaction regimes, although challenges remain in predicting localized vortical structures.

At $\text{Kn}=1$, corresponding to a transitional rarefaction regime, the error field becomes more structured, reflecting the increasing complexity of the flow physics. The global error map shows slightly larger deviations, with the step corner and the primary recirculation bubble dominating the error distribution. Overall, as Kn increases to 1.0, the magnitude of error grows but remains confined to the shear layer region; importantly, no large-scale errors appear elsewhere in the domain. These results indicate that prediction errors are not uniformly distributed but are tightly linked to the emergence of complex recirculation structures and strong gradients, which represent the most challenging features for data-driven surrogates. 

\subsubsection{Uncertainty Maps}

\begin{figure}[h!]
    \centering
    \begin{subfigure}{0.48\textwidth}
        \includegraphics[width=\textwidth, trim=0 0 0 0, clip]{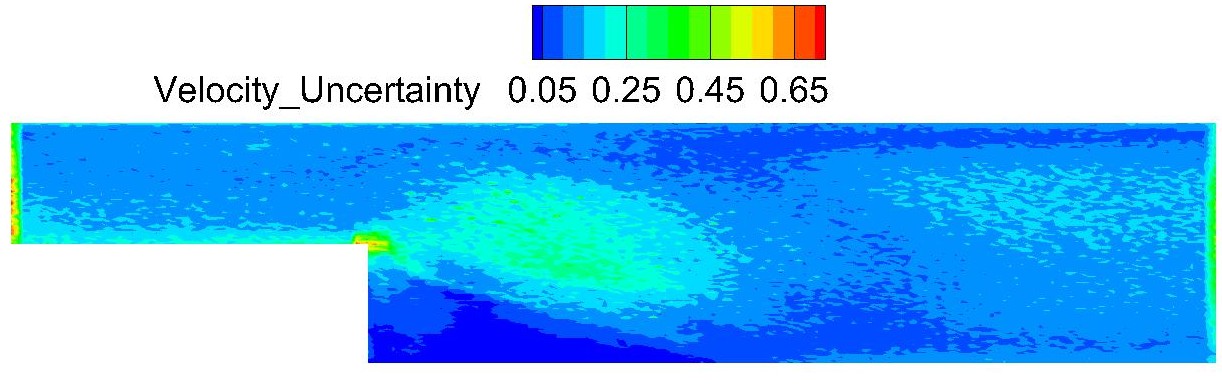}
        \caption{Overall error View}
        \label{fig:uncert_overall}
    \end{subfigure}\hfill
    \begin{subfigure}{0.38\textwidth}
       \includegraphics[width=\textwidth]{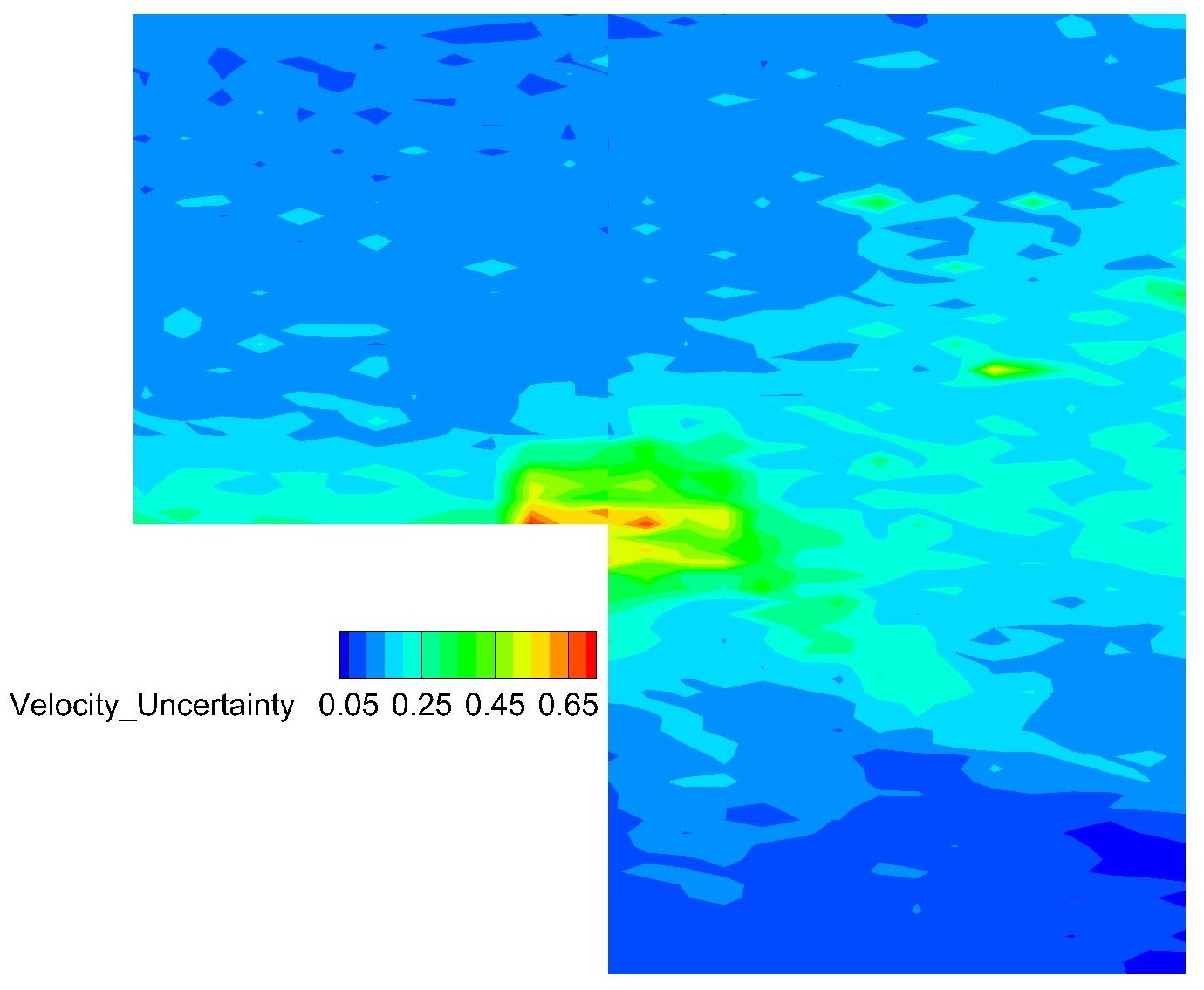}
        \caption{Close-up of the Step Corner}
        \label{fig:error_zoom}
    \end{subfigure}
    
    \caption{Visualization of the model's prediction uncertainty for the U-velocity. The uncertainty is concentrated in regions with complex physics, such as the sharp step corner and the shear layer, Kn=0.004.}
    \label{fig:uncertainty_comparison1}
\end{figure}

\begin{figure}[h!]
    \centering
    \begin{subfigure}{0.48\textwidth}
        \includegraphics[width=\textwidth, trim=0 0 0 0, clip]{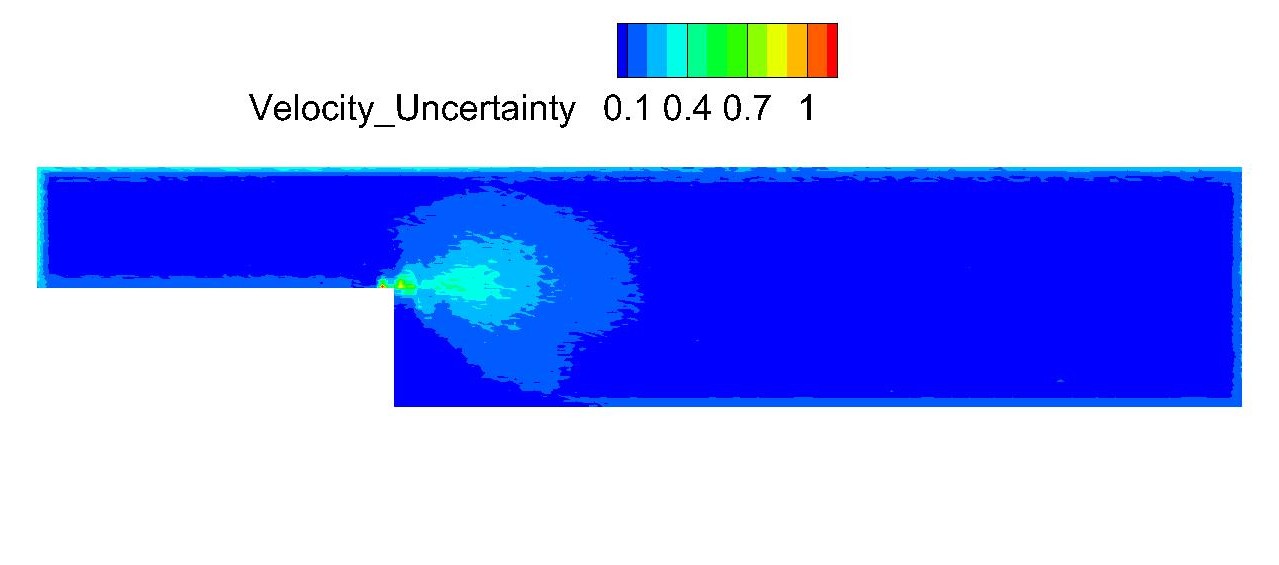}
        \caption{Overall Uncertainty View}
        \label{fig:uncert_overall}
    \end{subfigure}\hfill
    \begin{subfigure}{0.38\textwidth}
       \includegraphics[width=\textwidth]{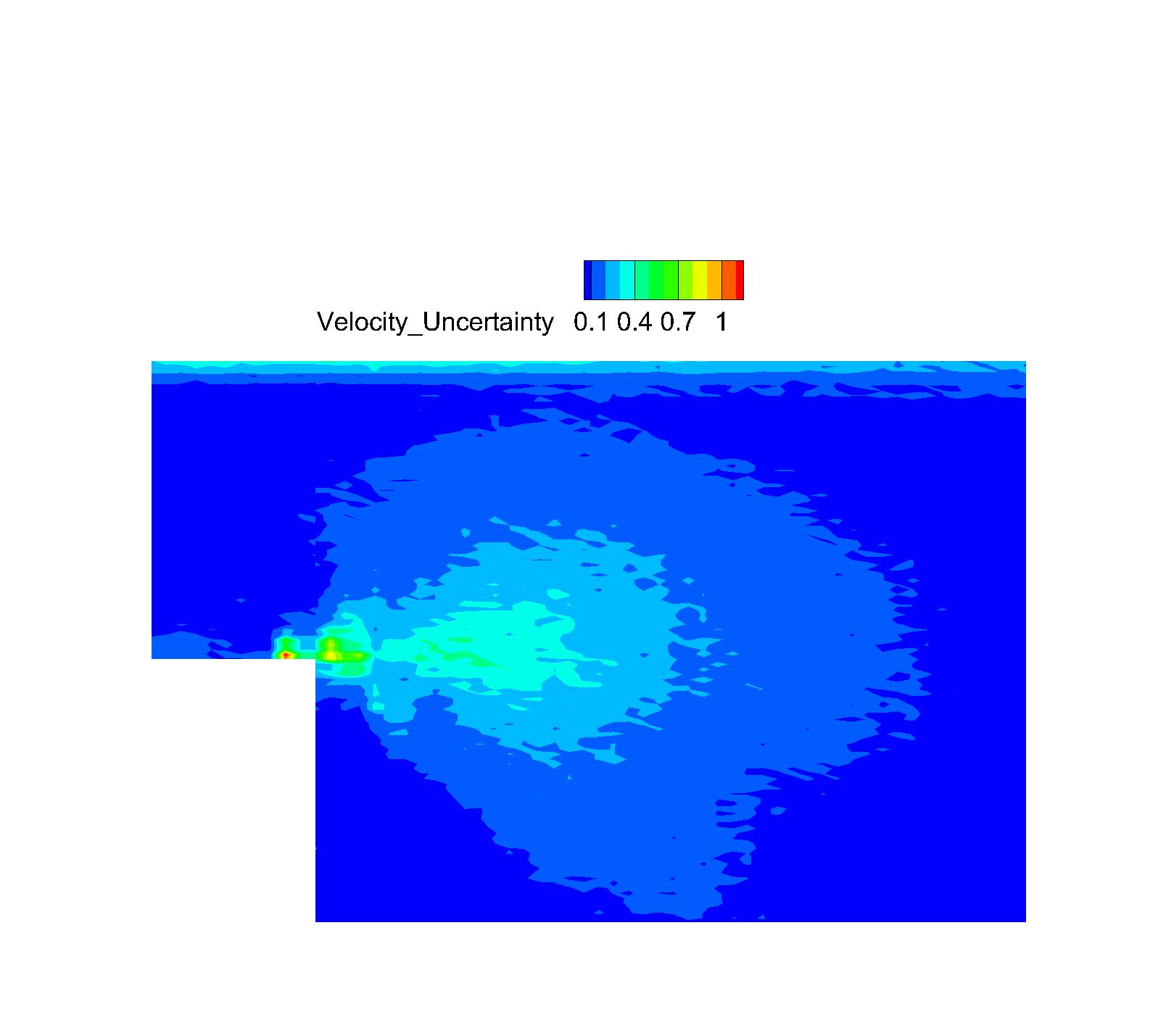}
        \caption{Close-up of the Step Corner}
        \label{fig:uncert_zoom}
    \end{subfigure}
    
    \caption{Visualization of the model's prediction uncertainty for the U-velocity. The uncertainty is concentrated in regions with complex physics, such as the sharp step corner and the shear layer, Kn=0.02.}
    \label{fig:uncertainty_comparison2}
\end{figure}

\begin{figure}[h!]
    \centering
    \begin{subfigure}{0.48\textwidth}
        \includegraphics[width=\textwidth, trim=0 0 0 0, clip]{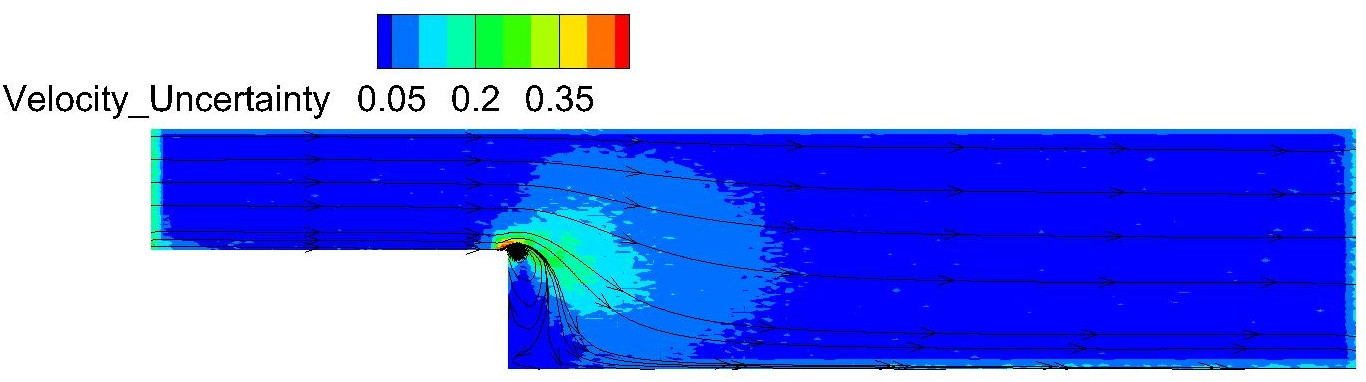}
        \caption{Overall Uncertainty View}
        \label{fig:uncert_overallkn=1}
    \end{subfigure}\hfill
    \begin{subfigure}{0.38\textwidth}
       \includegraphics[width=\textwidth]{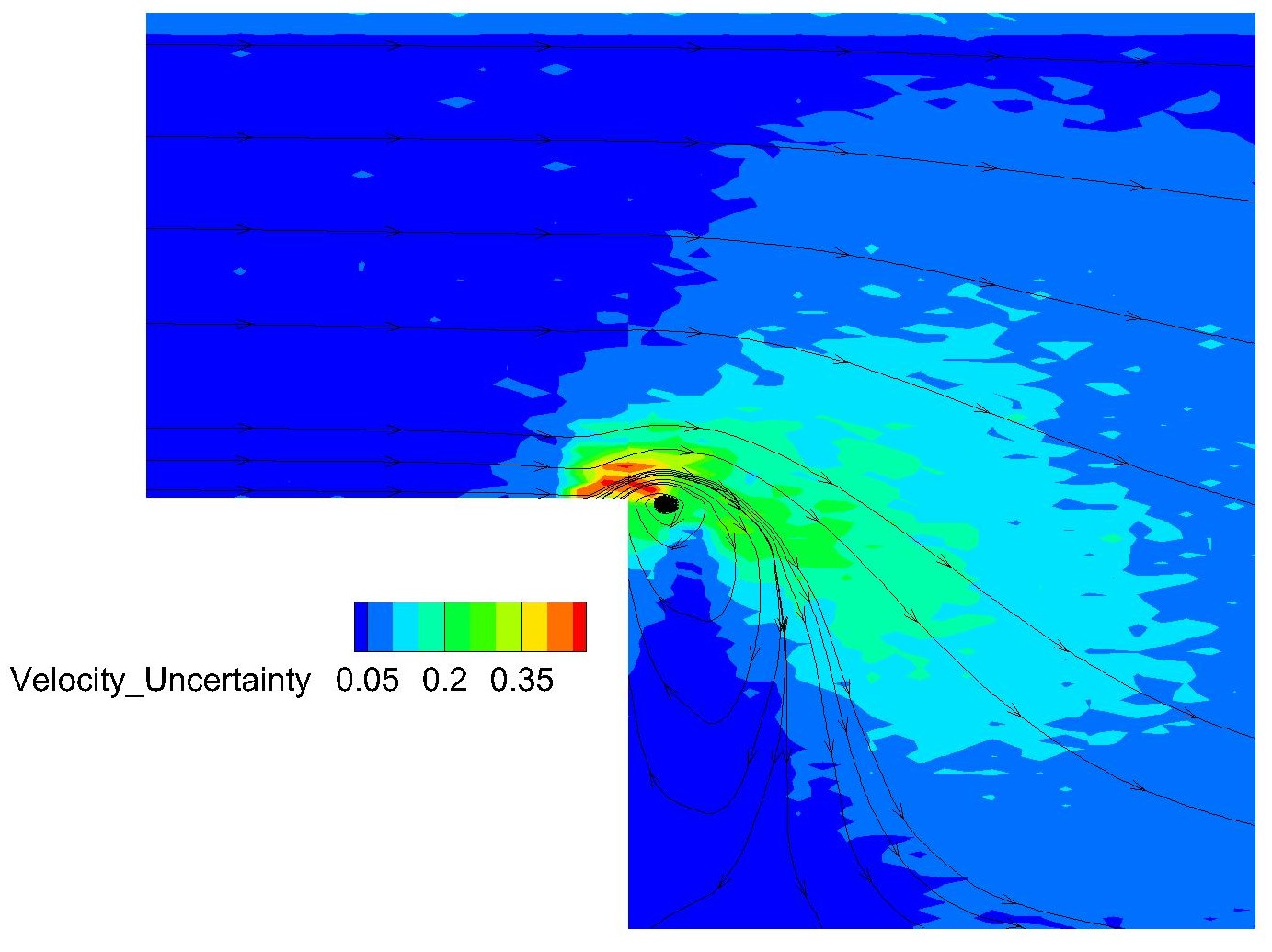}
        \caption{Close-up of the Step Corner}
        \label{fig:uncert_zoomkn=1}
    \end{subfigure}
    
    \caption{Visualization of the model's prediction uncertainty for the U-velocity. The uncertainty is concentrated in regions with complex physics, such as the sharp step corner and the shear layer, Kn=1}
    \label{fig:uncertainty_comparison3}
\end{figure}

Complementary to the error maps, Figures~\ref{fig:uncertainty_comparison1}, 
\ref{fig:uncertainty_comparison2}, and 
\ref{fig:uncertainty_comparison3} visualize the model’s prediction uncertainty for the U-velocity. At $\text{Kn}=0.004$, the uncertainty field mirrors the error distribution, with elevated values near the sharp step corner and the shear layer. This indicates that the model is capable of identifying regions where predictions are less reliable, a critical feature for practical applications of surrogate modeling.

At $\text{Kn}=0.02$, the uncertainty levels remain concentrated around the recirculation bubble and shear layer, but the affected region becomes slightly broader than in the error maps. This suggests that while the model can still produce accurate predictions, it anticipates difficulty in capturing rapid velocity variations and possible secondary vortical structures, signaling caution in those regions.

At $\text{Kn}=1$, the uncertainty is strongly localized, with maximum values clustered around the step corner and the vortex reattachment zone. The close-up views show a clear correspondence between high-gradient regions induced by rarefaction effects and elevated uncertainty. Importantly, the localization of uncertainty confirms that the DeepONet framework not only delivers accurate predictions but also quantifies the reliability of those predictions, providing a robust and interpretable surrogate for rarefied gas flow simulations.

\subsubsection{Velocity Profiles Comparison}

\begin{figure}[h!]
    \centering
    \begin{subfigure}[t]{0.48\textwidth}
        \centering
        \includegraphics[width=\textwidth]{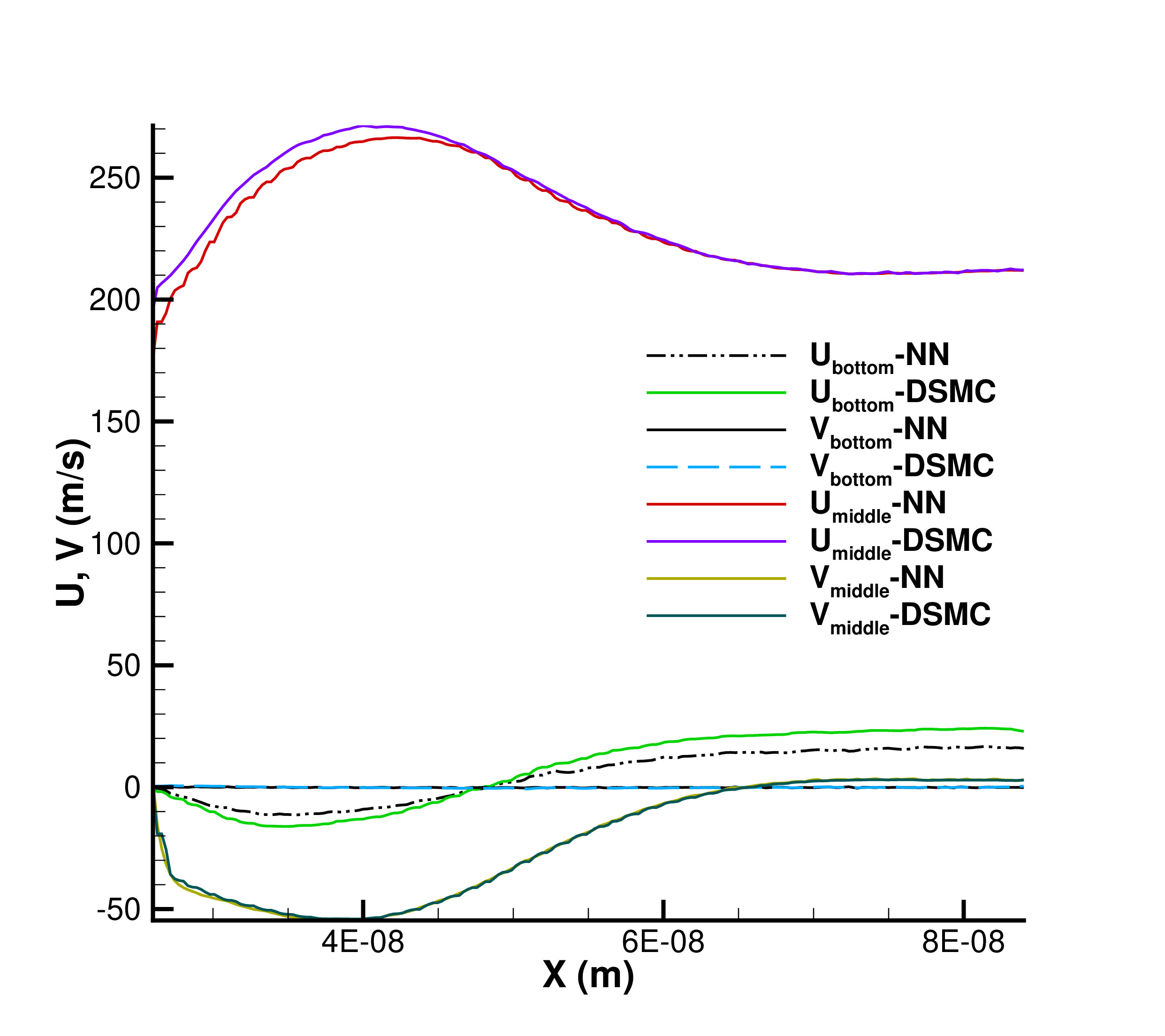}
        \label{fig:line_comparisonkn=0.004}
    \end{subfigure}
    \hfill 
    \begin{subfigure}[t]{0.48\textwidth}
        \centering
        \includegraphics[width=\textwidth]{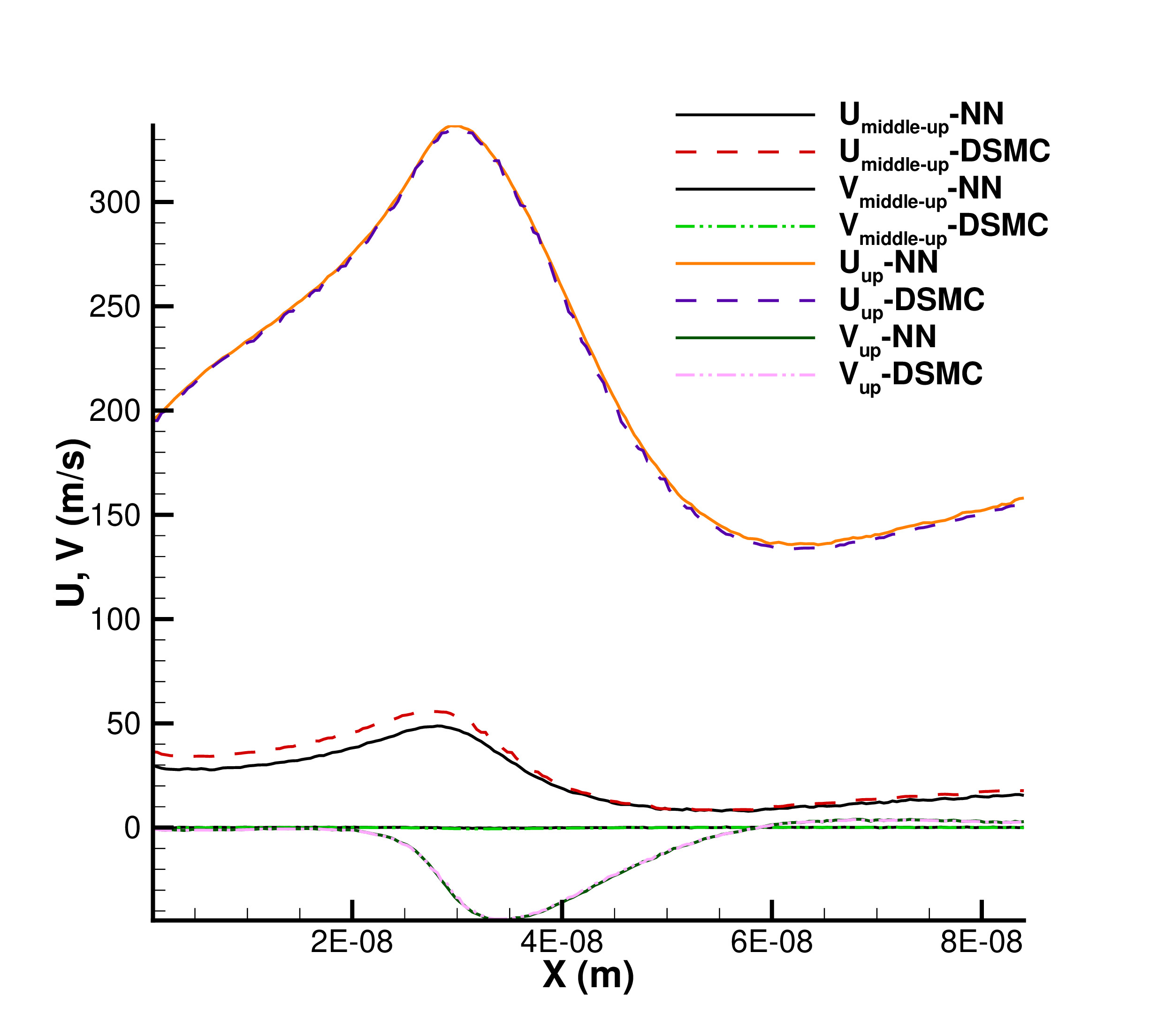}
        \label{fig:error_contourkn=0.004}
    \end{subfigure}
    
    \caption{Quantitative comparison of velocity profiles along four horizontal lines. Solid lines are DSMC data, dashed lines are DeepONet predictions., Kn=0.004.}
    \label{fig:overall_performance1}
\end{figure}

\begin{figure}[h!]
    \centering
    \begin{subfigure}[t]{0.48\textwidth}
        \centering
        \includegraphics[width=\textwidth]{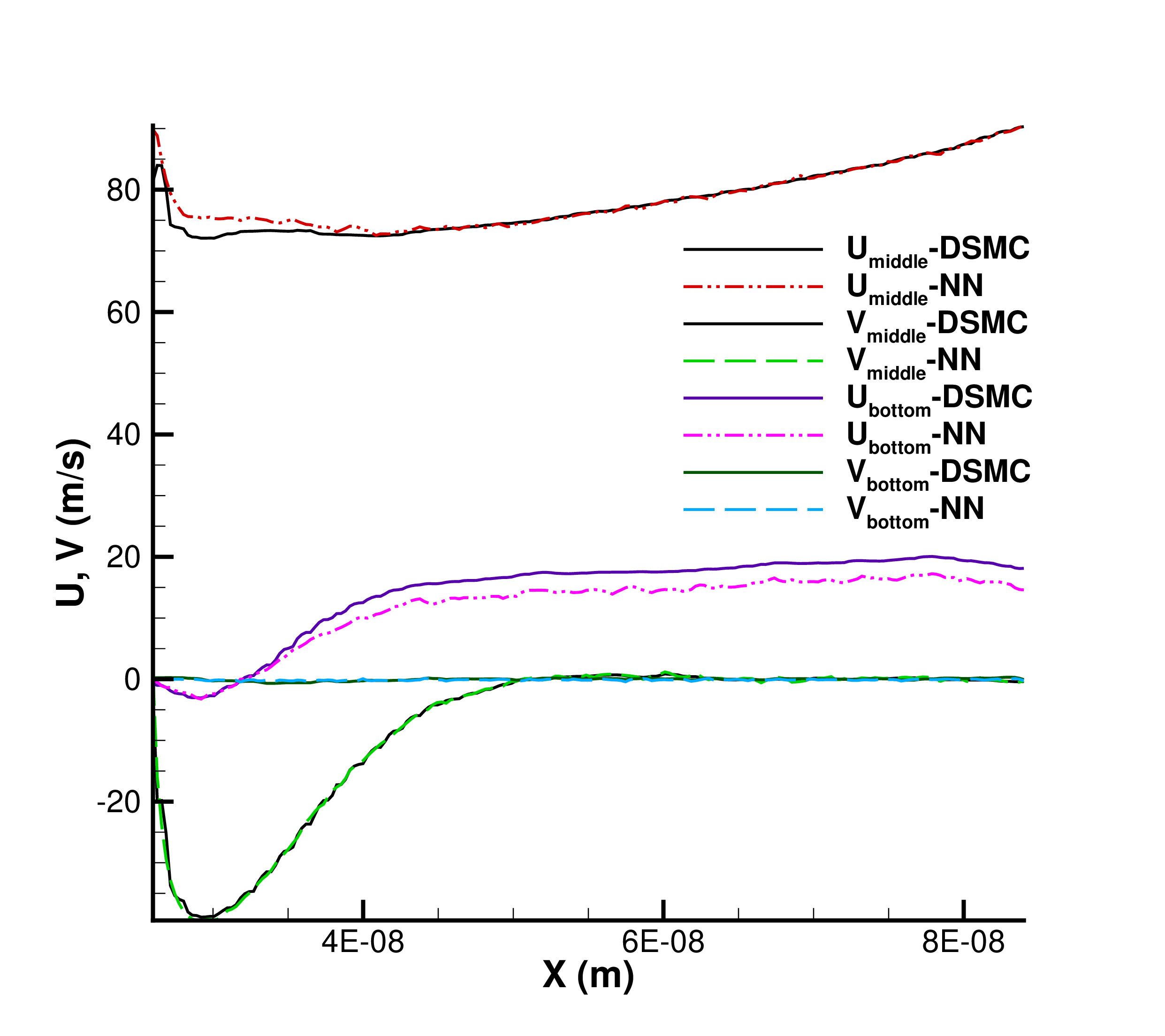}
        \label{fig:line_comparison}
    \end{subfigure}
    \hfill 
    \begin{subfigure}[t]{0.48\textwidth}
        \centering
        \includegraphics[width=\textwidth]{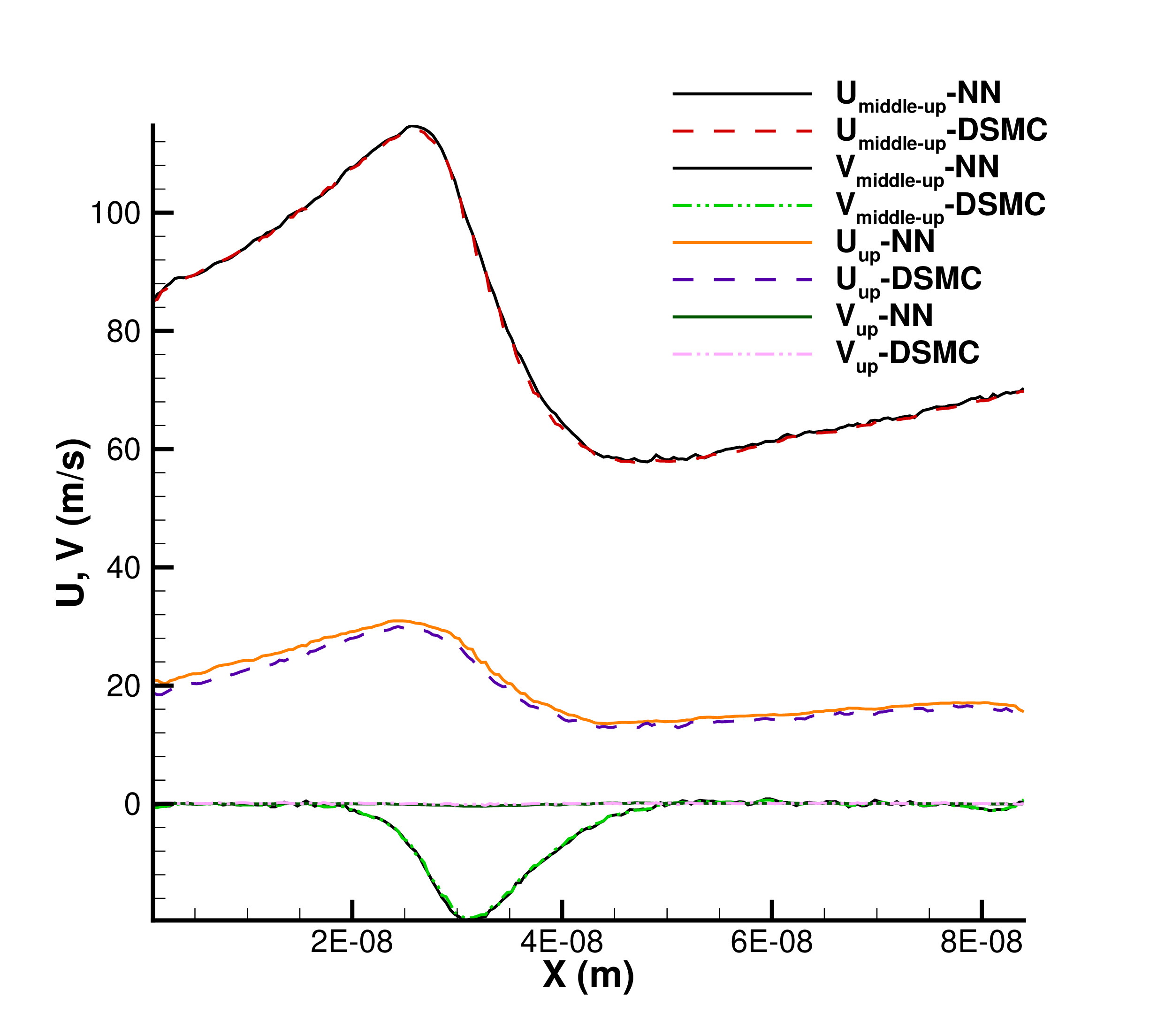}
        \label{fig:error_contour}
    \end{subfigure}
    
    \caption{Quantitative comparison of velocity profiles along four horizontal lines. Solid lines are DSMC data, dashed lines are DeepONet predictions, Kn=0.02.}
    \label{fig:overall_performance2}
\end{figure}

\begin{figure}[h!]
    \centering
    \begin{subfigure}[t]{0.48\textwidth}
        \centering
        \includegraphics[width=\textwidth]{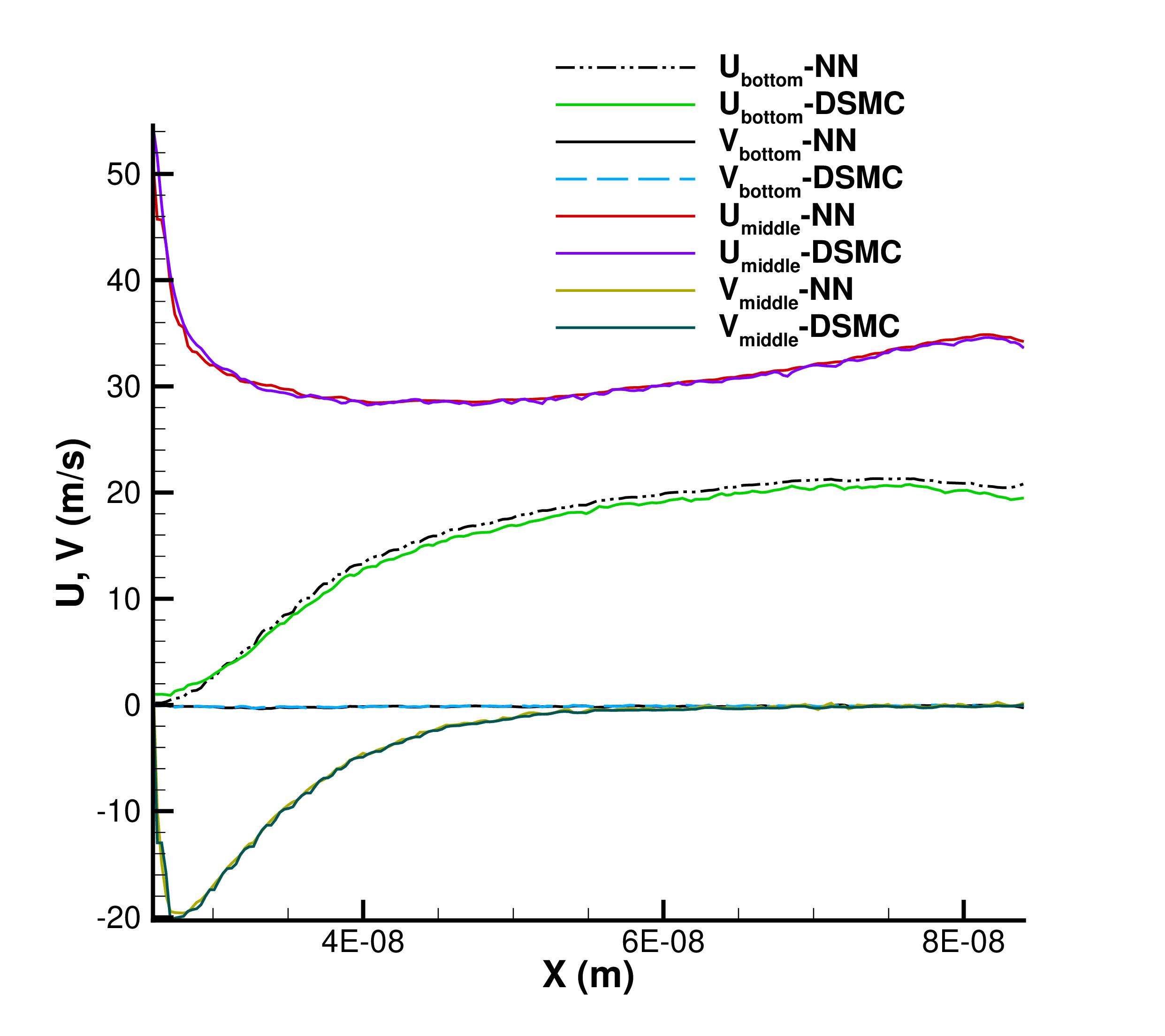}
        \label{fig:line_comparison}
    \end{subfigure}
    \hfill 
    \begin{subfigure}[t]{0.48\textwidth}
        \centering
        \includegraphics[width=\textwidth]{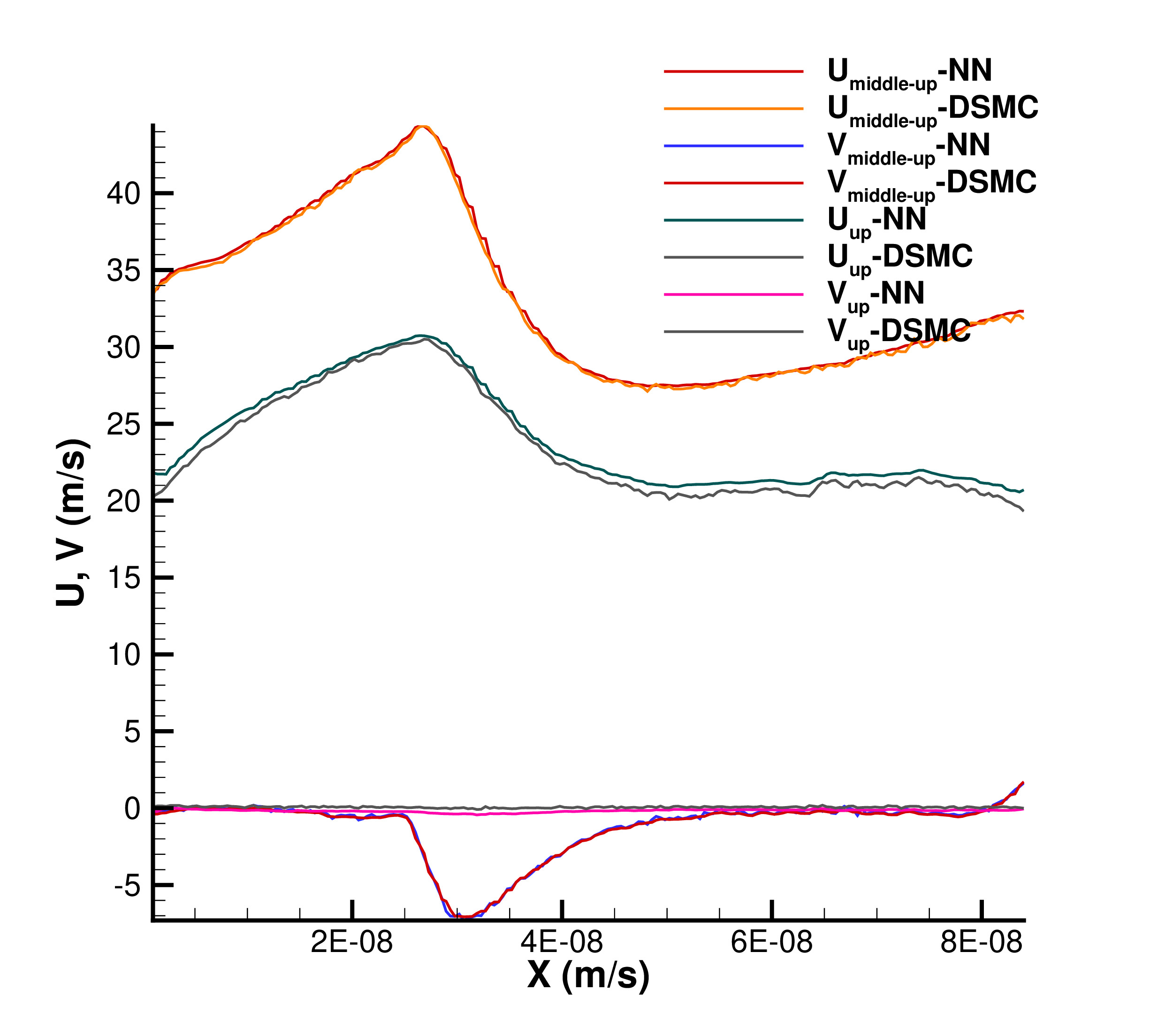}
        \label{fig:error_contour}
    \end{subfigure}
    
    \caption{Quantitative comparison of velocity profiles along four horizontal lines. Solid lines are DSMC data, dashed lines are DeepONet predictions, Kn=1.}
    \label{fig:overall_performance3}
\end{figure}

To further assess the robustness of the proposed framework, Figures~\ref{fig:overall_performance1}--\ref{fig:overall_performance3} present a quantitative validation against high-fidelity DSMC results along several representative horizontal lines. Four lines, 
1) one on the bottom wall (wall 3 in Fig. 1, called bottom in Fig. ~\ref{fig:overall_performance1}), 
2) one on the upper wall (wall 4 in Fig. 1, called up in Fig. ~\ref{fig:overall_performance1}), 
3) a line over wall 1 extending towards the exit of the channel (called "middle" in Fig. ~\ref{fig:overall_performance1}), 
4) and a line in the step at the middle between line "middle" and "up", called middle-up, are considered. Both components of the velocity are compared on these lines. Solid curves denote DSMC reference data, while dashed curves correspond to the DeepONet predictions. 

In Figure~\ref{fig:overall_performance1} ($Kn=0.004$), the agreement is strikingly close for both the $U$- and $V$-velocity components. In particular, the DeepONet successfully captures the negative $U$-velocity region inside the vortex, which is the most challenging portion of the flow field due to strong non-equilibrium effects. Minor deviations are visible in the flow recovery zone downstream of the reattachment point, where the model slightly underpredicts the velocity magnitude. This discrepancy is an anticipated trade-off, as the zonal loss function was tailored to emphasize accuracy in the vortex-dominated region.

Figure~\ref{fig:overall_performance2} ($Kn=0.02$) confirms that the model generalizes well as the Knudsen number increases, maintaining a high level of fidelity across all velocity components. The predicted $V$-velocity profiles show excellent agreement with DSMC, even in regions with sharp gradients near the separation and reattachment points. Slight mismatches occur in the far-field streamwise recovery region, but the overall error remains small and localized.

At $Kn=1$ (Figure~\ref{fig:overall_performance3}), corresponding to the transition between slip and transitional regimes, the model continues to reproduce the main features of the DSMC data. The DeepONet predictions capture both the overall structure of the $U$-velocity profiles and the subtle variations of the $V$-velocity component. Some discrepancies appear in the weaker secondary recirculation zone close to the step corner, but the predicted profiles remain within acceptable error margins. The preservation of accuracy across three orders of magnitude in Knudsen number demonstrates the strong generalization capability of the network.

Overall, these comparisons establish that the proposed DeepONet not only replicates DSMC results in regions explicitly emphasized during training but also provides reliable predictions across a broad range of flow regimes. The ability to resolve both primary vortex structures and recovery dynamics underscores the suitability of the framework for applications involving rarefied gas flows.

\subsection{Application to Geometric Parameter Variation: Step Height}

To showcase the versatility of the proposed framework, the DeepONet surrogate model was adapted to predict the velocity field as a function of a geometric parameter---the ratio of the step height to the channel height ($h/H$). This demonstrates the model's capability to learn the operator mapping from the domain geometry to the flow solution, a critical task for shape optimization problems. 

\begin{figure}[h!]
    \centering
    \includegraphics[width=0.8\textwidth]{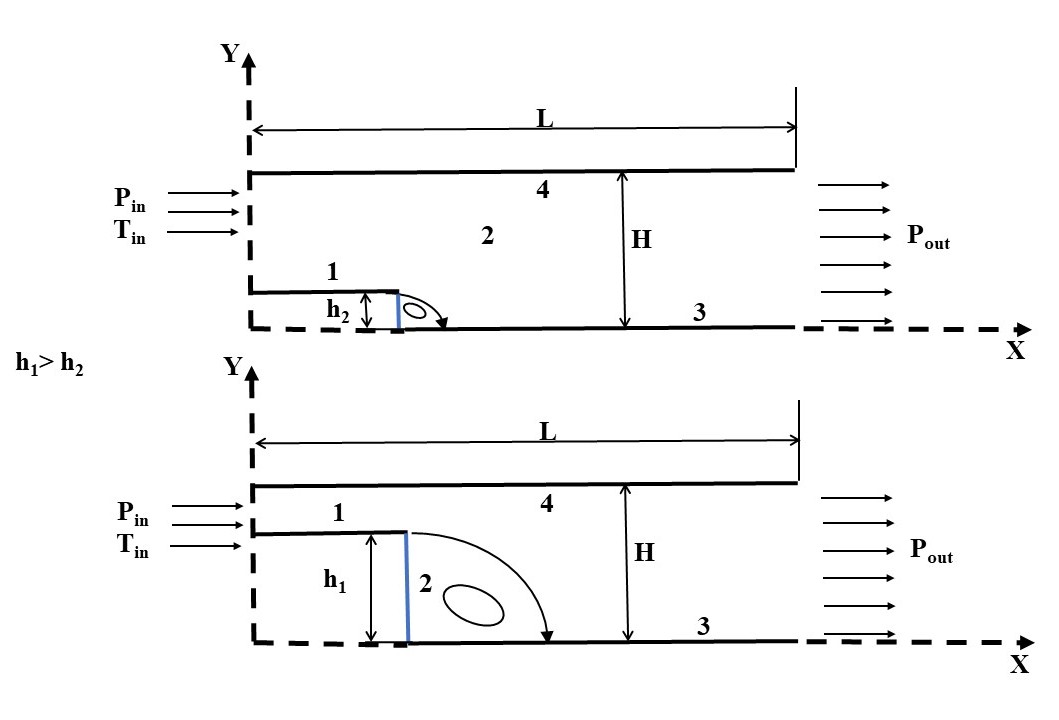}
    \caption{Schematic of the simulated geometry for height study}
    \label{fig:SchematicH}
\end{figure}

Figure~\ref{fig:SchematicH} schematizes the geometry used to quantify the influence of the
step–height ratio on separation. The channel has total height $H$ and length $L$; upstream of
the step, the local height is $h$ (with $h<H$), and the backward-facing step produces an
expansion into the lower channel. Walls are labeled 1–4 consistent with the problem
statement; the flow is driven by an inlet/outlet pressure ratio while all walls are kept
isothermal. Knudsen number is set as Kn=0.01. The control parameter for this study is the non-dimensional step height $h/H$,
which we vary systematically to modify the strength and extent of the recirculation region
behind the step. This schematic clarifies how increasing $h/H$ enlarges the expansion and,
consequently, tends to intensify separation and shift the reattachment location downstream. At the investigated Knudsen number, there is no vorticity on the top wall of the step geometry.

\subsubsection{DSMC snapshots across height ratios.}
\begin{figure}[h!]
    \centering
    
    \begin{subfigure}[t]{0.48\textwidth}
        \centering
        \includegraphics[width=\textwidth]{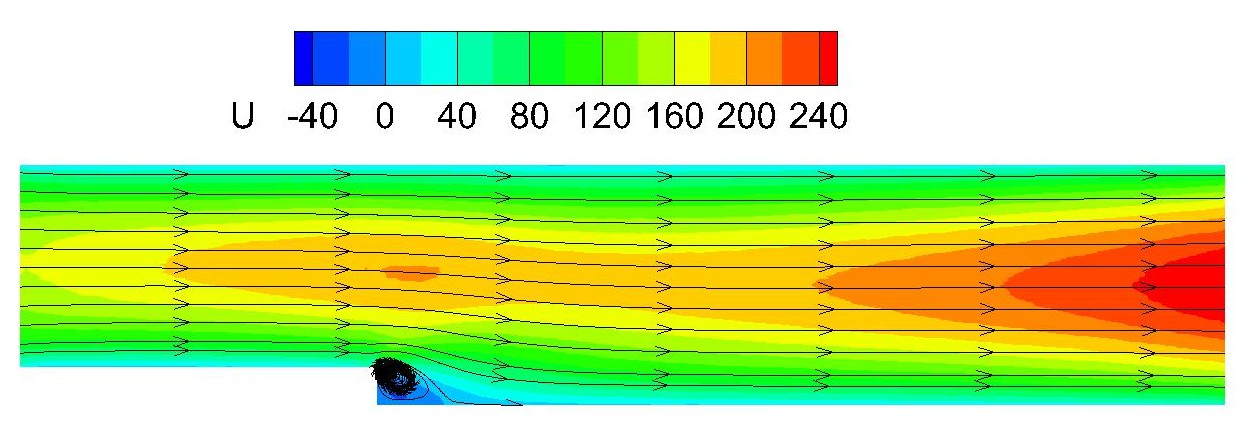}
        \caption{h/H=\SI{16}{\percent}}
        \label{fig:H16}
    \end{subfigure}\hfill 
    \begin{subfigure}[t]{0.48\textwidth}
        \centering
        \includegraphics[width=\textwidth]{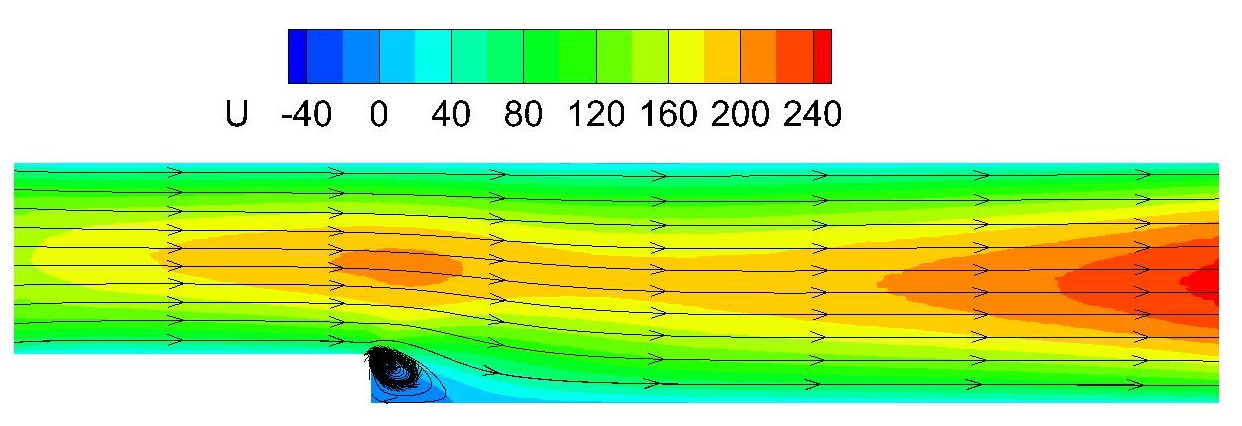}
        \caption{h/H=\SI{21}{\percent}}
        \label{fig:H21}
    \end{subfigure}
    
    \vspace{5mm} 

    \begin{subfigure}[t]{0.48\textwidth}
        \centering
        \includegraphics[width=\textwidth]{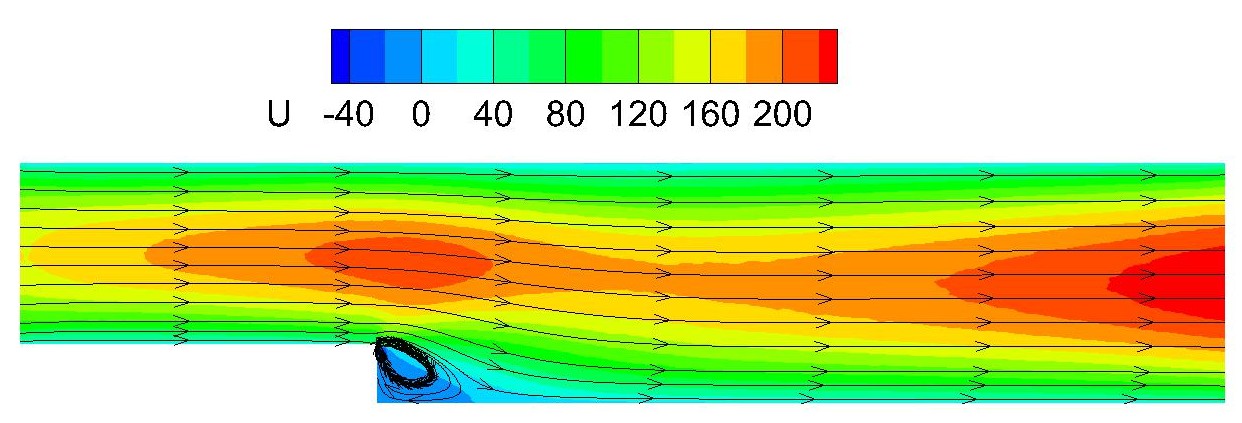}
        \caption{h/H=\SI{25}{\percent}}
        \label{fig:H25}
    \end{subfigure}\hfill
    \begin{subfigure}[t]{0.48\textwidth}
        \centering
        \includegraphics[width=\textwidth]{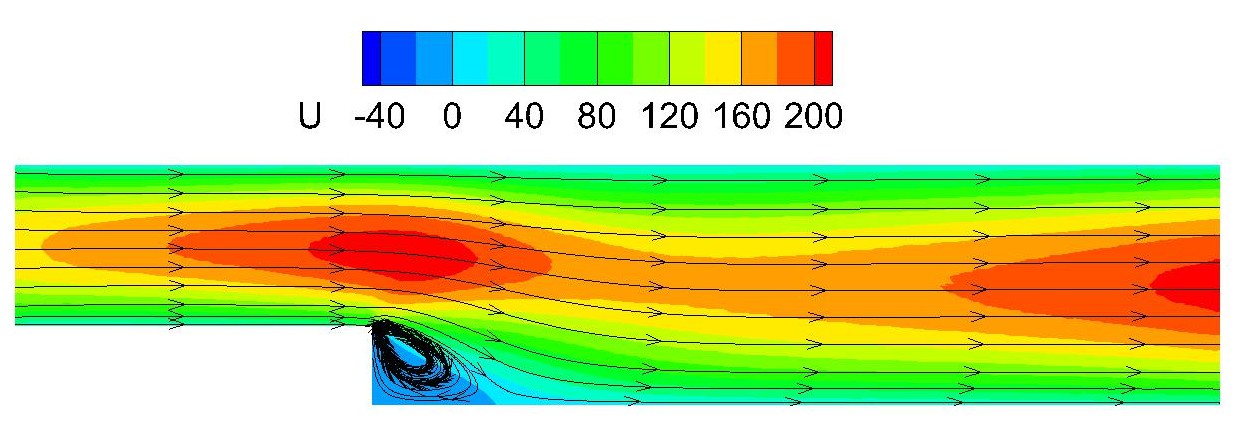}
        \caption{h/H=\SI{33}{\percent}}
        \label{fig:H33}
    \end{subfigure}

    \vspace{5mm} 

    \begin{subfigure}[t]{0.48\textwidth}
        \centering
        \includegraphics[width=\textwidth]{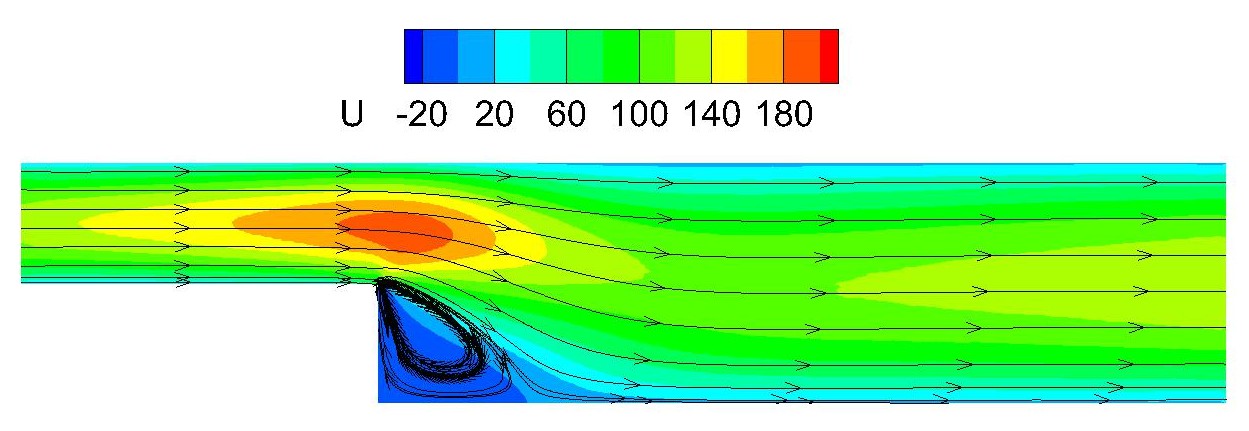}
        \caption{h/H=\SI{50}{\percent}}
        \label{fig:H50}
    \end{subfigure}\hfill
    \begin{subfigure}[t]{0.48\textwidth}
        \centering
        \includegraphics[width=\textwidth]{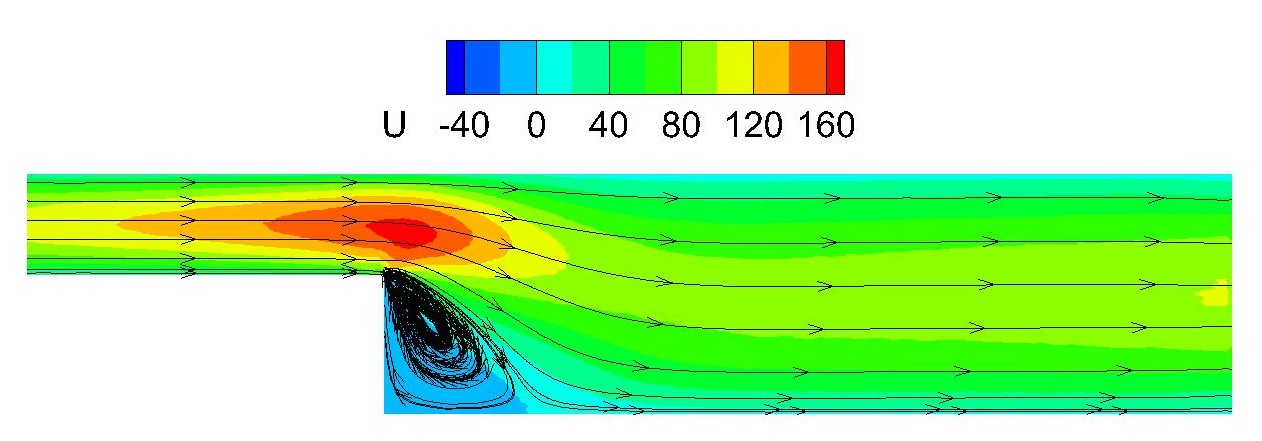}
        \caption{h/H=\SI{58}{\percent}}
        \label{fig:H58}
    \end{subfigure}
    
    \vspace{5mm} 

    \begin{subfigure}[t]{0.48\textwidth}
        \centering
        \includegraphics[width=\textwidth]{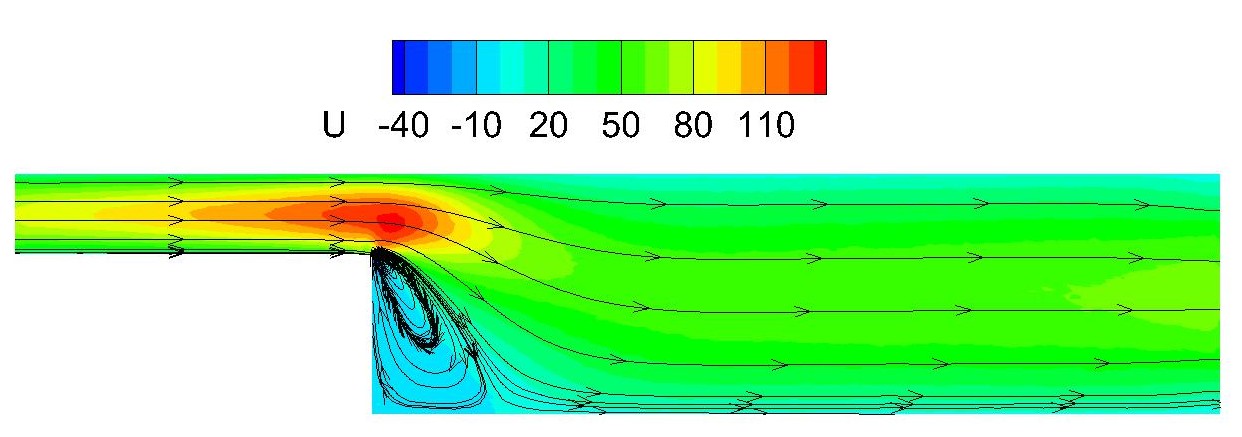}
        \caption{h/H=\SI{67}{\percent}}
        \label{fig:H67}
    \end{subfigure}\hfill
    \begin{subfigure}[t]{0.48\textwidth}
        \centering
        \includegraphics[width=\textwidth]{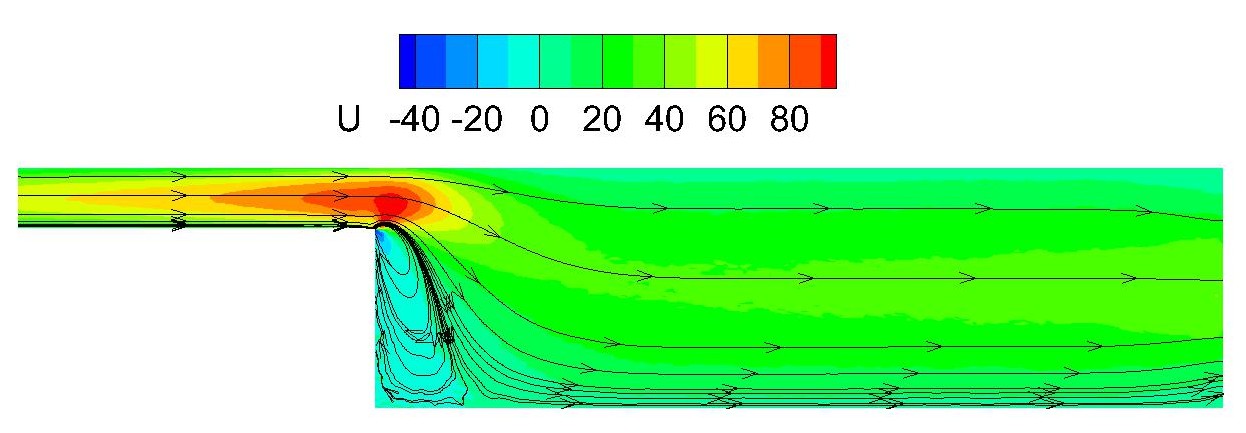}
        \caption{h/H = \SI{75}{\percent}}
        \label{fig:H75}
    \end{subfigure}
    
    \caption{Qualitative comparison of the U-velocity contour and streamlines for representative step to channel heights (h/H) ratios in the dataset.}
    \label{fig:height_range_comparison}
\end{figure}

Figure~\ref{fig:height_range_comparison} presents DSMC results for eight representative values of
the step–height ratio, showing streamlines overlaid on the $U$-velocity contours.
A clear, monotonic trend emerges: as $h/H$ increases from $0.16$ to $0.75$, the primary
recirculation bubble expands and its center moves downstream; the shear layer becomes thicker
and more curved; and the peak streamwise velocity in the core decreases due to the stronger
expansion. For small $h/H$ the separation is compact and the reattachment occurs relatively
close to the corner, whereas for large $h/H$ the separation length grows appreciably and the
low-speed region occupies a larger portion of the lower channel. These qualitative changes in
both streamlines and $U$-contours provide a rich set of flow topologies for learning.

The core methodology, including the convolutional trunk and the physics-guided zonal loss function, remains consistent with the Knudsen number study.

The specific hyperparameters for the DeepONet architecture used in this study are detailed as follows. The model is composed of a Branch network, a convolutional Trunk network, and a Head network that combines their outputs.

\begin{itemize}
    \item \textbf{Branch Network:} The branch network, which processes the scalar step height value, consists of 6 ResNet blocks with a width of 384 neurons.
    
    \item \textbf{Trunk Network:} The trunk network first processes local field information using two convolutional layers with 32 and 64 filters, respectively, operating on input patches of size 5x5. The flattened output is then combined with the point coordinates and passed through an MLP of 4 dense layers with a width of 384 neurons.
    
    \item \textbf{Head Network:} The outputs of the branch and trunk networks are merged and processed by a final head network composed of 4 dense layers with a width of 768 neurons.
\end{itemize}

Crucially, the physics-guided zonal loss function was employed to ensure high fidelity in the recirculation zone. A weighting factor of $\alpha = 0.6$ was used, prioritizing the accuracy of the negative velocity region behind the step. A dropout rate of 0.3 was applied for regularization and for enabling uncertainty quantification via Monte Carlo Dropout.

The DSMC fields shown in Figure~\ref{fig:height_range_comparison} constitute the geometry-dependent
dataset for training the neural operator. In the Branch network, the scalar input is the
height ratio $h/H$, while the Trunk network takes spatial coordinates $(x,y)$; the network’s
output is the velocity field at those coordinates. We generated DSMC solutions at nine values
of $h/H$ in the interval $[0.16,\,0.75]$; eight of these, shown in Figure~\ref{fig:height_range_comparison} are used for training and one
\emph{unseen} case, $h/H=0.44$, is withheld for validation. The held-out prediction assesses
the model’s ability to generalize to a geometry it did not observe during training, i.e.,
to interpolate accurately within the parameter space solely from the operator mapping it
has learned.

\subsubsection{Training dynamics for height variation.}
\begin{figure}[h!]
    \centering
    \includegraphics[width=0.8\textwidth]{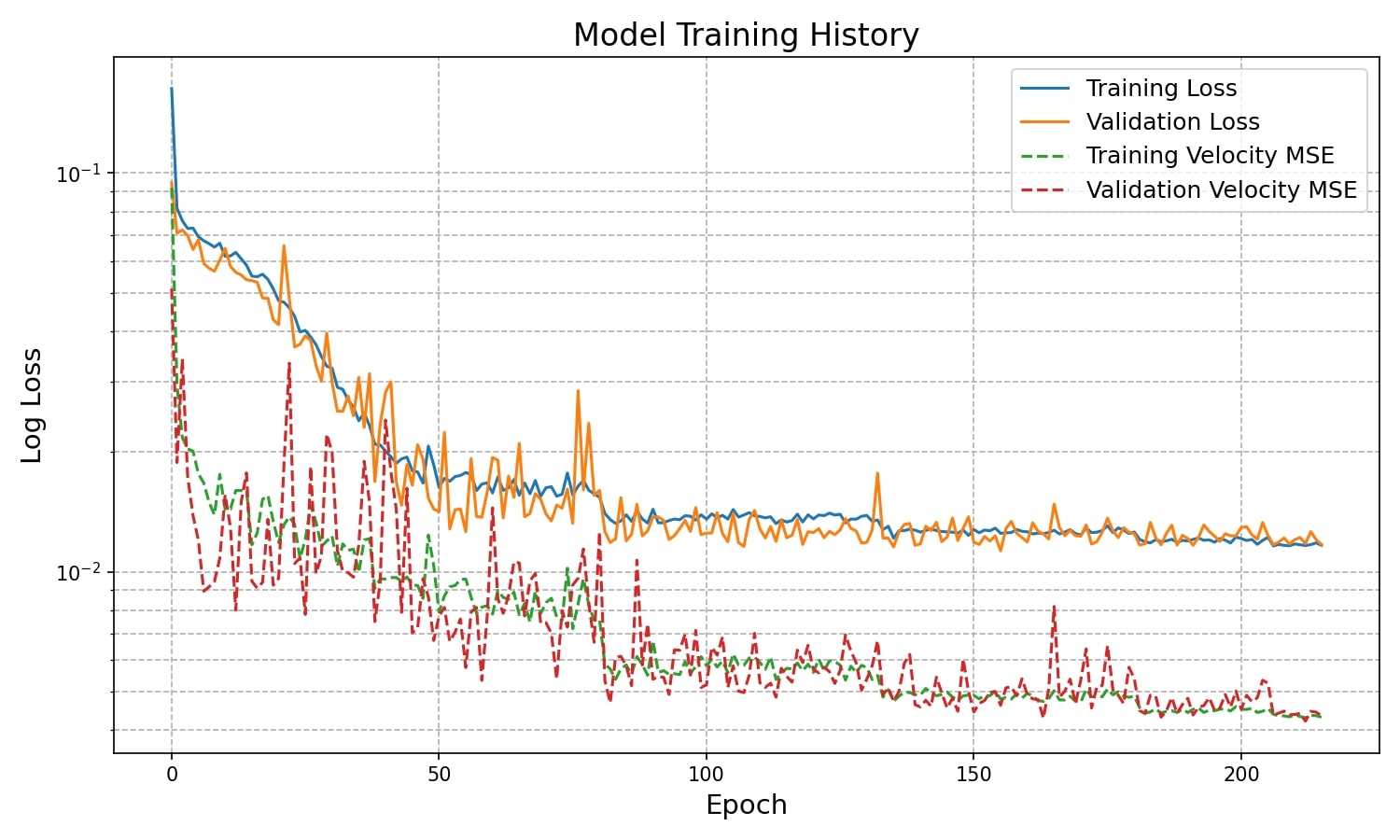}
    \caption{Loss Function and MSE for the step with height variation}
    \label{fig:loss_Height}
\end{figure}

Figure~\ref{fig:loss_Height} illustrates the loss history for the DeepONet trained on the
height–ratio dataset. Compared to the Knudsen–number study, where the model was exposed to
20 training cases spanning a wide range of flow regimes, here only 8 training geometries are
available. This reduced dataset naturally limits the richness of the operator mapping and
slows down convergence. The training and validation losses (solid blue and orange curves)
both decrease steadily, but they saturate at higher values than in the Knudsen case, where
denser sampling allowed the network to interpolate more accurately. Similarly, the mean
squared error (MSE) of the velocity components (dashed green and red curves) remains an
order of magnitude higher than in the Knudsen-number experiment. Nevertheless, the
oscillations damp out after about 100 epochs, and both training and validation curves track
each other closely, confirming that the model does not overfit despite the smaller dataset.
This comparison highlights the strong dependence of neural operator accuracy on the density
and diversity of training samples: with only eight height ratios, the generalization error is
larger than for the Knudsen-number mapping, yet still sufficiently low to capture the
underlying flow physics and predict the unseen case at $h/H=0.44$.

Overall, while the prediction error is higher than in the Knudsen-number study due to the 
reduced number of training samples, the results demonstrate that DeepONet can still capture 
the essential vortex dynamics and velocity distribution even with sparse data, underscoring 
its robustness as a surrogate model for rarefied gas flows.

\begin{figure}[h!]
    \centering
    
    \begin{subfigure}[t]{0.48\textwidth}
        \centering
        \includegraphics[width=\textwidth]{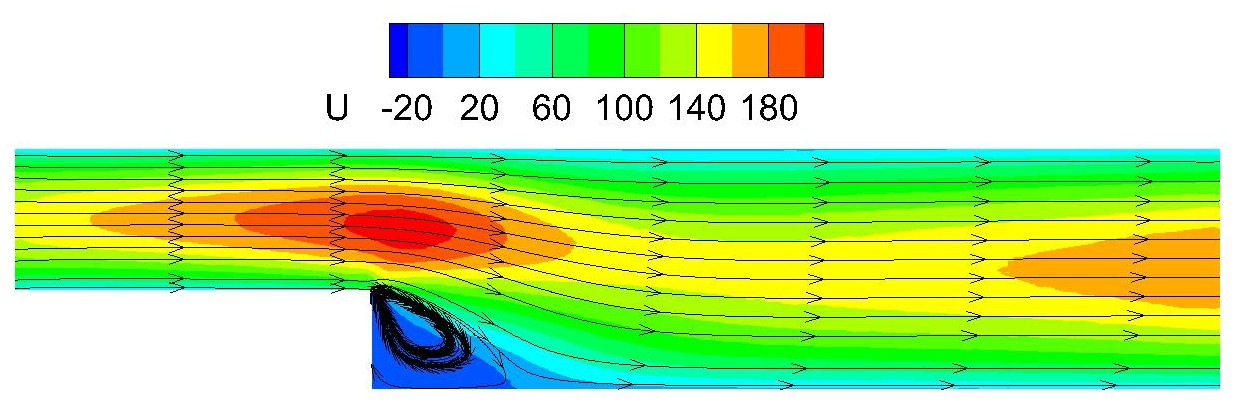}
        \caption{Ground Truth (DSMC), U-Velocity}
        \label{fig:dsmc_u_h44}
    \end{subfigure}\hfill 
    \begin{subfigure}[t]{0.48\textwidth}
        \centering
        \includegraphics[width=\textwidth]{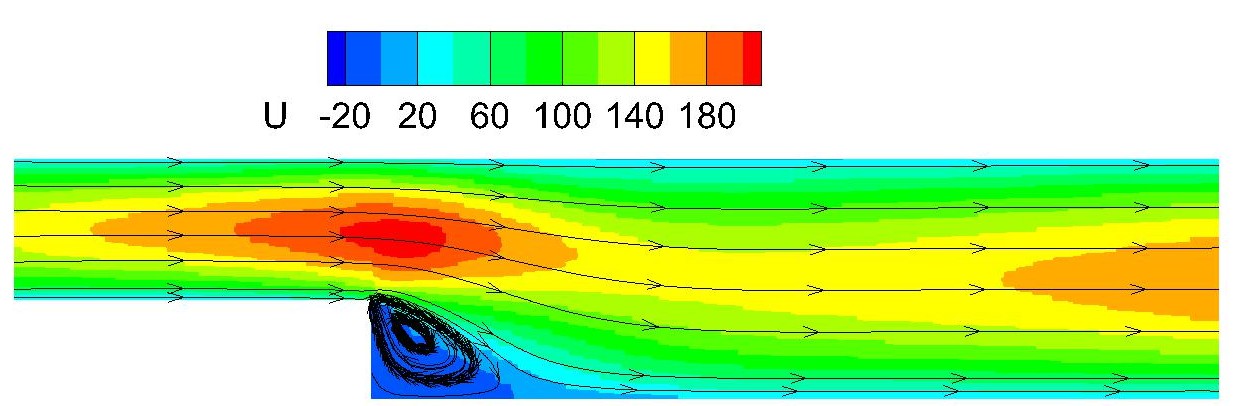}
        \caption{DeepONet Prediction (NN), U-Velocity}
        \label{fig:nn_u_h44}
    \end{subfigure}
    
    \vspace{5mm} 

    \begin{subfigure}[t]{0.48\textwidth}
        \centering
        \includegraphics[width=\textwidth]{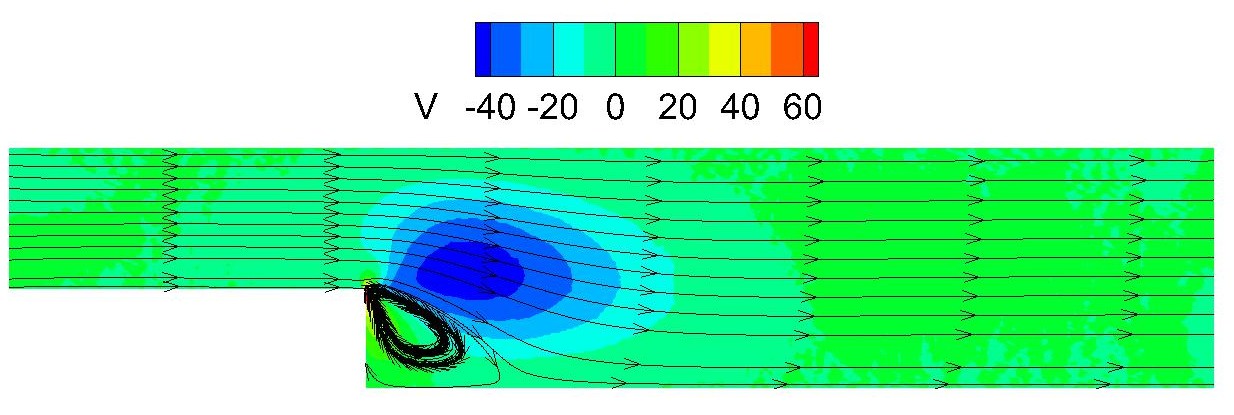}
        \caption{Ground Truth (DSMC), V-Velocity}
        \label{fig:dsmc_v_h44}
    \end{subfigure}\hfill
    \begin{subfigure}[t]{0.48\textwidth}
        \centering
        \includegraphics[width=\textwidth]{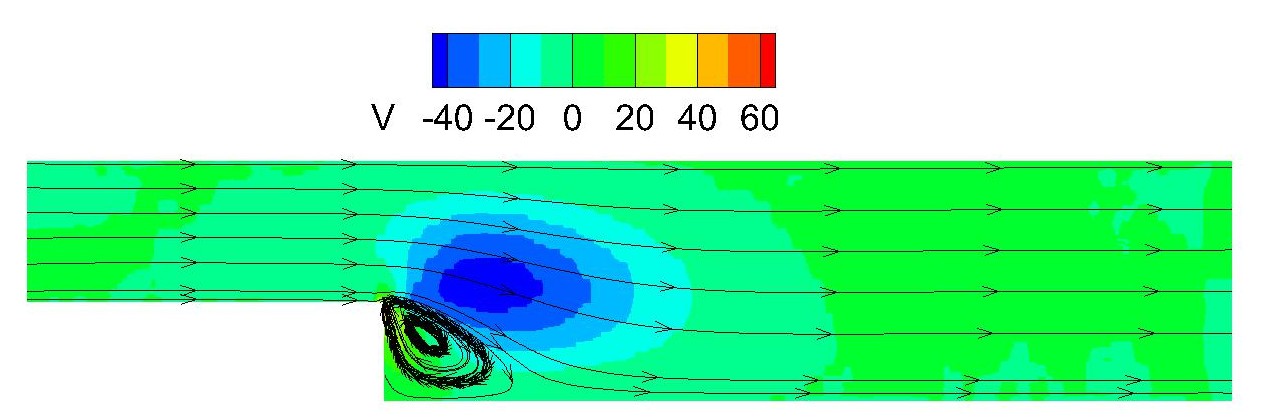}
        \caption{DeepONet Prediction (NN), V-Velocity}
        \label{fig:nn_v_h44}
    \end{subfigure}
    
    \caption{Qualitative comparison of the U-velocity and V-velocity contours between the ground truth DSMC simulation and the DeepONet prediction for the unseen step height ratio of h/H=\SI{44}{\percent}.}
    \label{fig:height_comparison_h44}
\end{figure}

\begin{figure}[h!]
    \centering 

    \begin{subfigure}{0.48\textwidth}
        \includegraphics[width=\linewidth]{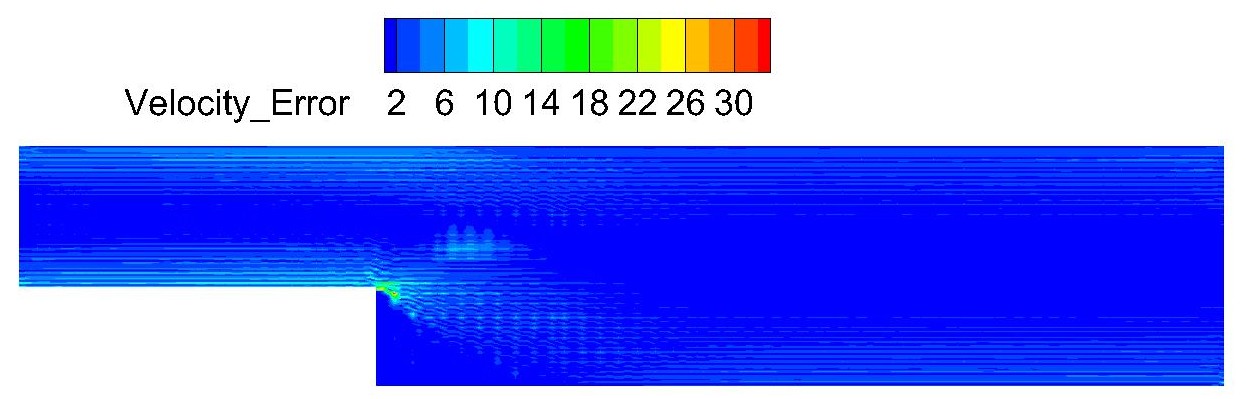}
        \caption{Velocity Error}
        \label{fig:velocity-error2}
    \end{subfigure}
    \hfill 
    \begin{subfigure}{0.48\textwidth}
        \includegraphics[width=\linewidth]{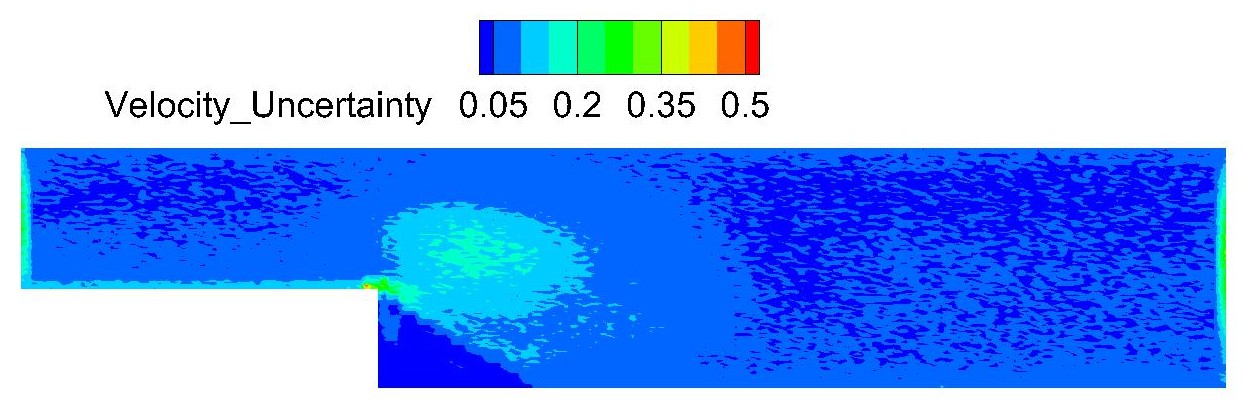}
        \caption{Velocity Uncertainty}
        \label{fig:velocity-uncertainty2}
    \end{subfigure}

    \caption{Velocity error and uncertainty for the case of h/H=\SI{44}{\percent}.}
    \label{fig:error-uncertainty-comparison}
\end{figure}

Figures~\ref{fig:height_comparison_h44} and \ref{fig:error-uncertainty-comparison} 
demonstrate the extrapolative capability of the DeepONet framework when applied to an 
unseen geometric configuration with a step height ratio of $h/H=44\%$. 
In Figure~\ref{fig:height_comparison_h44}, the predicted U- and V-velocity fields are shown 
to be in excellent agreement with the DSMC ground truth, successfully capturing the onset and 
extent of the primary recirculation zone, the detachment and reattachment points, and the 
subtle shear-layer dynamics in the main channel. This highlights the ability of the network to 
generalize beyond the training set and approximate the nonlinear operator mapping geometric 
parameters $(h/H)$ to the full velocity field.  

The corresponding error and uncertainty distributions in 
Figure~\ref{fig:error-uncertainty-comparison} reveal that discrepancies are primarily localized 
to regions of strong gradients and non-equilibrium effects, such as the sharp step corner and 
shear layer. Importantly, the model’s uncertainty quantification exhibits strong spatial 
correlation with the error distribution, indicating that the DeepONet not only learns the flow 
operator but also provides a meaningful estimate of prediction reliability.  
Although the test case lies outside the training manifold, the magnitude of the error remains 
low across the majority of the flow domain, underscoring the robustness of the operator-learning 
paradigm. This result confirms that DeepONet is capable of faithfully representing the complex 
multi-scale dynamics of rarefied gas flows under geometric variations, a task that conventional 
neural architectures often fail to achieve.

\subsubsection{Validation with Multiple Hold-Out Cases and Increased Data Sparsity}

To further test the model's limits, a more challenging scenario was devised. The model was trained on a reduced dataset of only seven simulation cases, while two cases at step height ratios of $h/H = 44\%$ and $h/H = 67\%$ were held out for testing. This scenario simulates a situation with extreme data scarcity.

\begin{figure}[h!]
    \centering
    \includegraphics[width=0.7\textwidth]{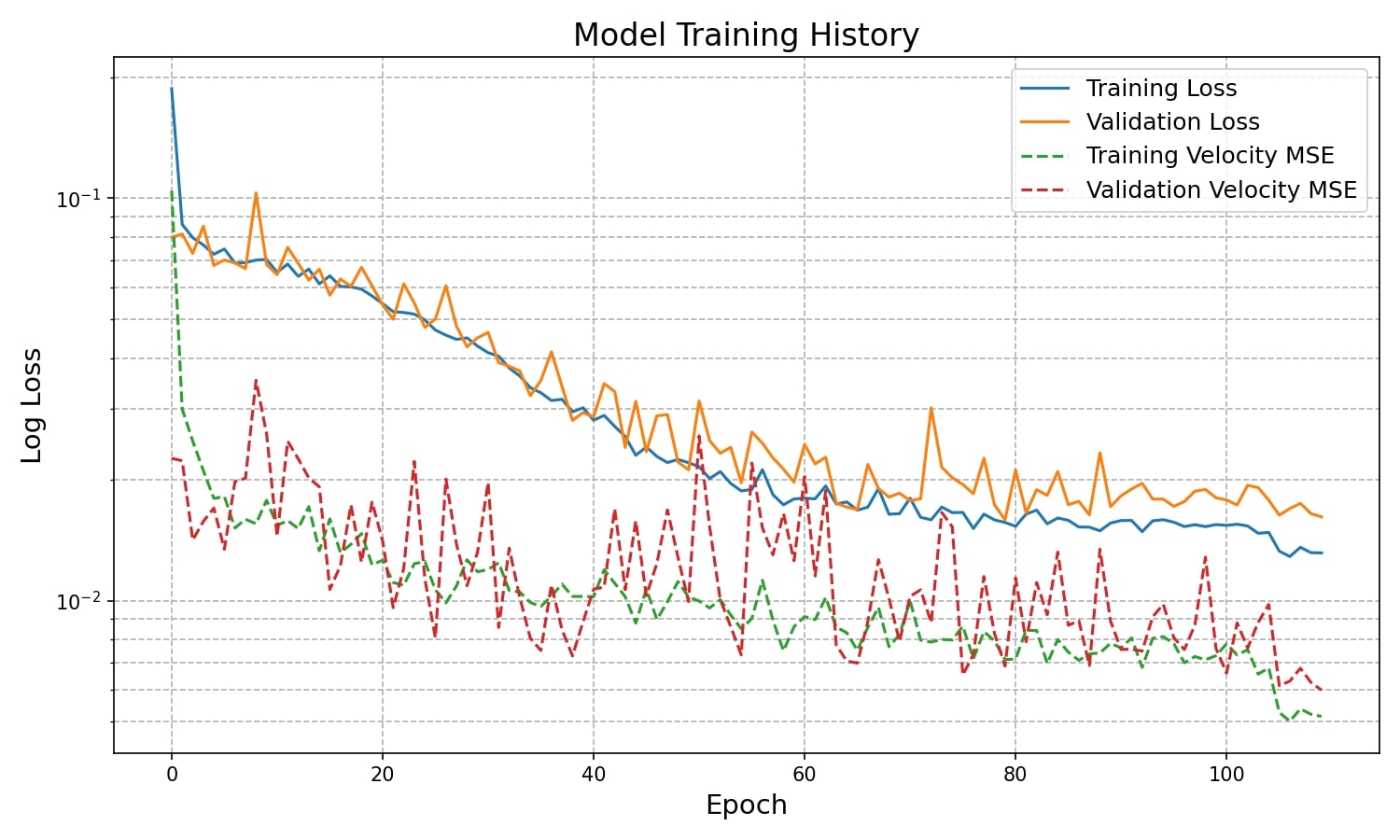}
    \caption{Training and validation loss history for the simplified DeepONet model trained on cases with varying step heights.}
    \label{fig:loss_history_height}
\end{figure}

The training history for this more demanding setup is shown in Figure~\ref{fig:loss_history_height}. In contrast to the previous case, a small but noticeable gap emerges between the training loss and the validation loss. This gap is a classic sign of mild overfitting. With fewer training examples, the high-capacity DeepONet model begins to memorize the specific characteristics of the training set, slightly hindering its ability to generalize to the two unseen test cases.

\begin{figure}[h!]
    \centering
    \begin{subfigure}{0.48\textwidth}
        \includegraphics[width=\linewidth]{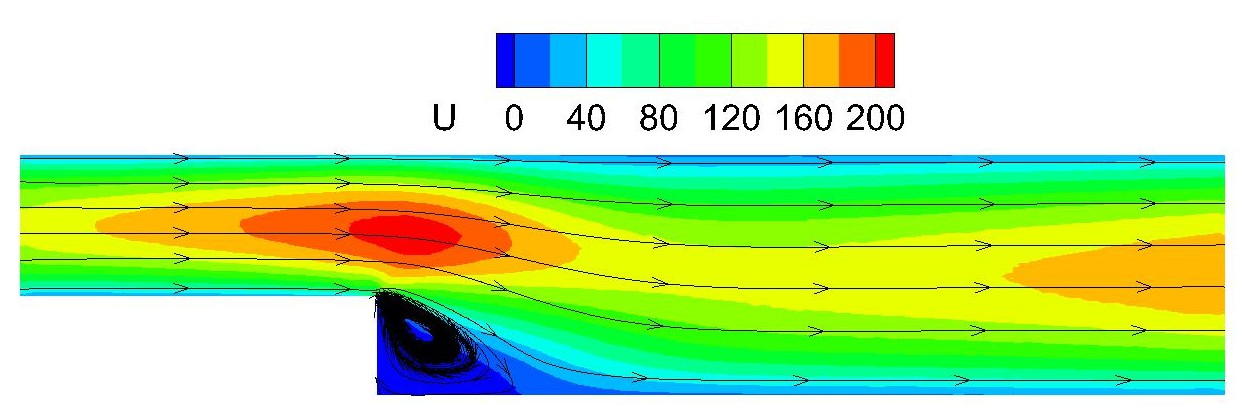}
        \caption{DSMC Ground Truth, U-velocity}
        \label{fig:h44_u_dsmc}
    \end{subfigure}
    \hfill 
    \begin{subfigure}{0.48\textwidth}
        \includegraphics[width=\linewidth]{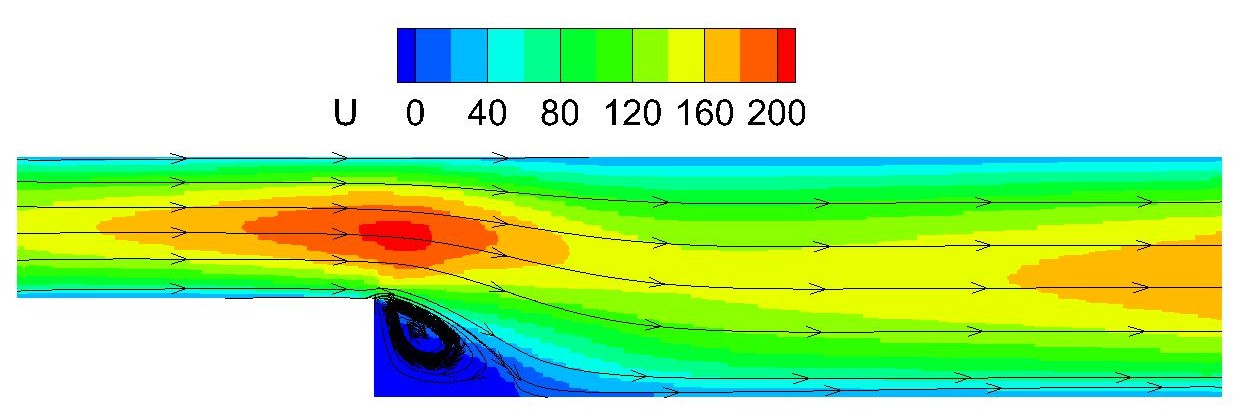}
        \caption{DeepONet Prediction, U-velocity}
        \label{fig:h44_u_nn}
    \end{subfigure}
    
    \vspace{0.5cm} 
    
    \begin{subfigure}{0.48\textwidth}
        \includegraphics[width=\linewidth]{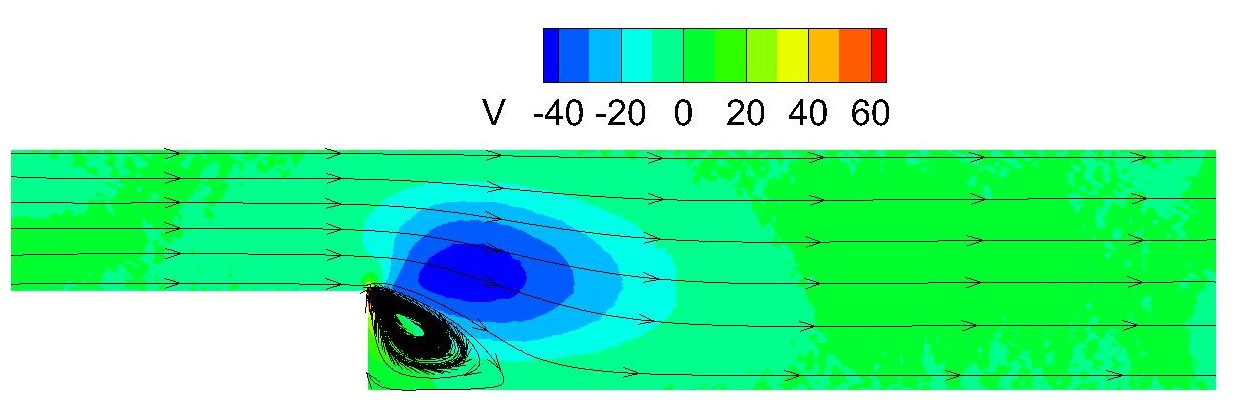}
        \caption{DSMC Ground Truth, V-velocity}
        \label{fig:h44_v_dsmc}
    \end{subfigure}
    \hfill 
    \begin{subfigure}{0.48\textwidth}
        \includegraphics[width=\linewidth]{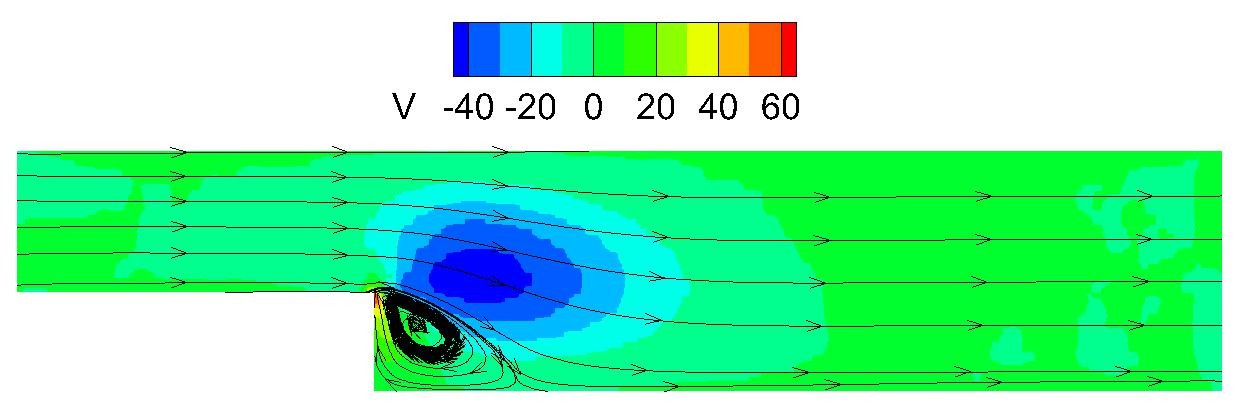}
        \caption{DeepONet Prediction, V-velocity}
        \label{fig:h44_v_nn}
    \end{subfigure}
    
    \caption{Qualitative comparison of U-velocity (top row) and V-velocity (bottom row) contours between the ground truth DSMC simulation and the DeepONet prediction for the unseen step height ratio of $h/H = 44\%$.}
    \label{fig:comparison_h44}
\end{figure}

\begin{figure}[h!]
    \centering
    \begin{subfigure}{0.48\textwidth}
        \includegraphics[width=\linewidth]{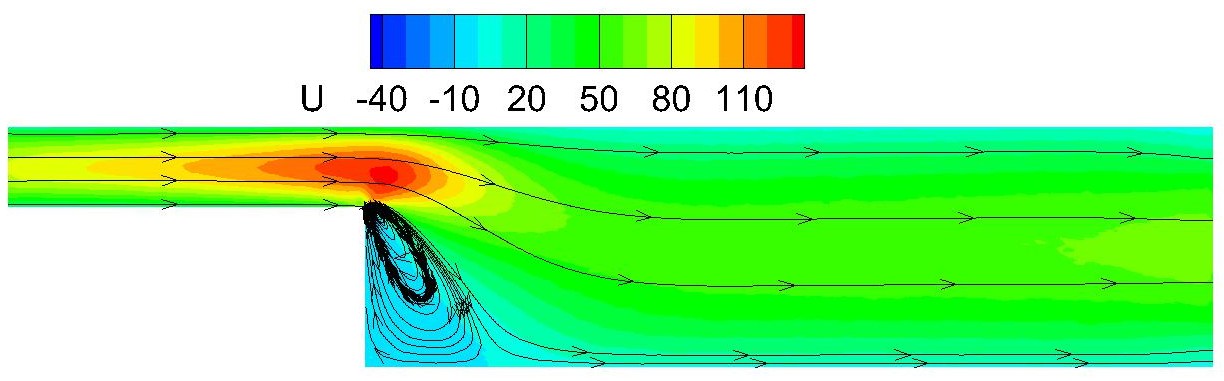}
        \caption{DSMC Ground Truth, U-velocity}
        \label{fig:h67_u_dsmc}
    \end{subfigure}
    \hfill 
    \begin{subfigure}{0.48\textwidth}
        \includegraphics[width=\linewidth]{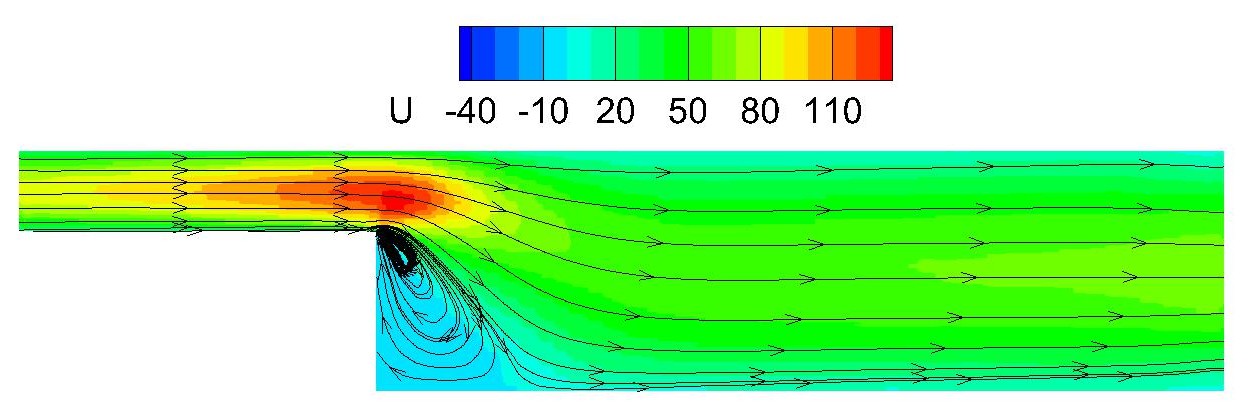}
        \caption{DeepONet Prediction, U-velocity}
        \label{fig:h67_u_nn}
    \end{subfigure}
    
    \vspace{0.5cm} 
    
    \begin{subfigure}{0.48\textwidth}
        \includegraphics[width=\linewidth]{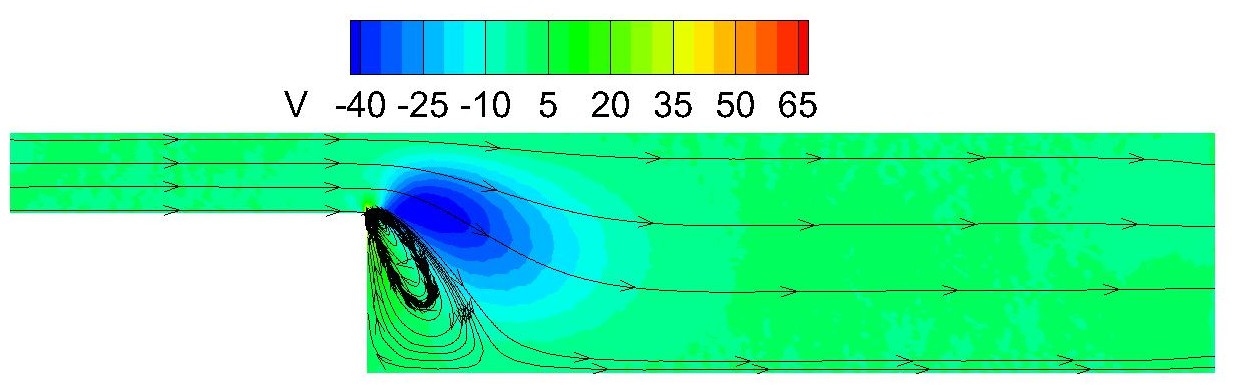}
        \caption{DSMC Ground Truth, V-velocity}
        \label{fig:h67_v_dsmc}
    \end{subfigure}
    \hfill 
    \begin{subfigure}{0.48\textwidth}
        \includegraphics[width=\linewidth]{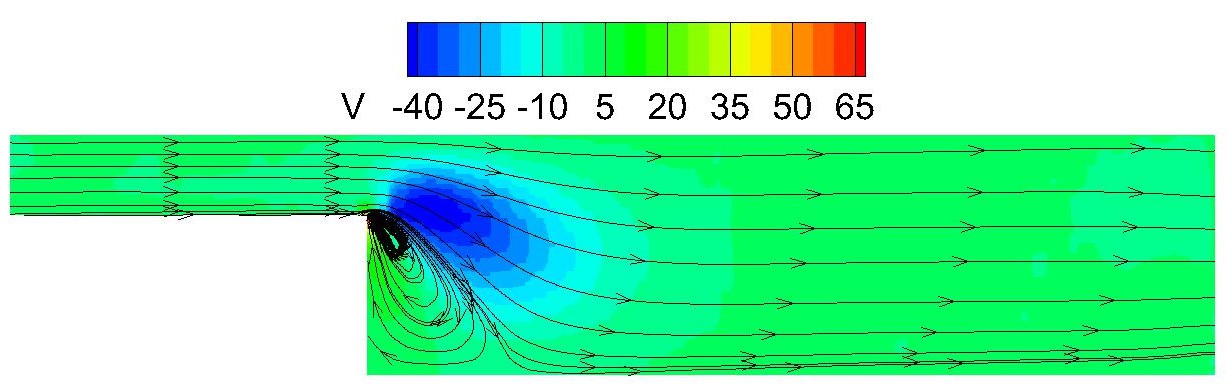}
        \caption{DeepONet Prediction, V-velocity}
        \label{fig:h67_v_nn}
    \end{subfigure}
    
    \caption{Qualitative comparison of U-velocity (top row) and V-velocity (bottom row) contours between the ground truth DSMC simulation and the DeepONet prediction for the unseen step height ratio of $h/H = 67\%$.}
    \label{fig:comparison_h67}
\end{figure}

Despite this mild overfitting, the model's predictive capabilities remain strong across both held-out cases. Figures~\ref{fig:comparison_h44} and \ref{fig:comparison_h67} present the qualitative velocity comparisons for the test cases at $h/H = 44\%$ and $h/H = 67\%$, respectively. In both scenarios, the DeepONet's predictions for both U- and V-velocity show a strong qualitative agreement with the DSMC ground truth. The model correctly identifies the critical flow structures, such as the recirculation zone and the main channel flow, for both an interpolating case ($h/H=44\%$) and a more extrapolating case ($h/H=67\%$). This result is significant, as it demonstrates the robustness of the surrogate modeling framework. Even when pushed to the limits of data availability where mild overfitting occurs, the model does not fail catastrophically and still produces physically plausible and largely accurate predictions across the parameter space.

\subsection{Comparison of Various Loss Function Formulations}

In this section, we compare three distinct loss functions to train the DeepONet model. The choice of loss function directly influences the model's learning process and its ability to handle localized physical phenomena. We compare our Zonal Loss Function with the Mean Squared Error (MSE) Loss and Gradient-Weighted Mean Squared Error (GMSE) Loss suggested in~\cite{cooper2024generalised}.

\subsubsection{Mean Squared Error (MSE) Loss}
The Mean Squared Error is a standard, widely used loss function that treats all data points with equal importance. It computes the average squared difference between the predicted values ($\hat{y}_i$) and the true values ($y_i$) over all $N$ points in the training batch. The formulation is given by:
\begin{equation}
    L_{\text{MSE}} = \frac{1}{N} \sum_{i=1}^{N} \| y_i - \hat{y}_i \|^2
\end{equation}
While robust and straightforward, the MSE loss function does not account for the underlying physical structure of the problem. For flows with localized, high-gradient regions, its global averaging nature can lead to the "smearing" of errors from critical zones across the entire computational domain.

\subsubsection{Gradient-Weighted Mean Squared Error (GMSE) Loss}
The Gradient-Weighted Mean Squared Error (GMSE) is an advanced loss function proposed by Cooper-Baldock et al. designed to enhance model training for datasets like those in Computational Fluid Dynamics (CFD), where small, high-variance regions contain the most critical information~\cite{cooper2024generalised}. It modifies the standard Mean Squared Error by assigning a spatially-varying weight, $w(\mathbf{x}_i)$, to the error at each point $\mathbf{x}_i$. This compels the model to focus more on regions with complex features and strong gradients~\cite{cooper2024generalised}.

The general form of the GMSE loss is given by:
\begin{equation}
    L_{\text{GMSE}} = \frac{1}{N} \sum_{i=1}^{N} w(\mathbf{x}_i) \| y_i - \hat{y}_i \|^2
\end{equation}

The dynamic weight matrix $w(\mathbf{x}_i)$ (denoted as $W_i$ in the paper) is calculated for each ground truth field through a multi-step process:

\paragraph{1. Disparity Calculation:}
First, the gradient field is computed to identify regions of high pixel disparity. This is achieved by calculating the element-wise difference along the x and y axes and then determining the magnitude of the resulting vector[cite: 199, 205]. This non-linear operation effectively captures high-frequency details[cite: 214].
\begin{align}
    W_{d,x} &= W_{x} - W_{x-1} \\
    W_{d,y} &= W_{y} - W_{y-1} \\
    W_{d} &= \sqrt{W_{d,x}^{2} + W_{d,y}^{2}}
\end{align}

\paragraph{2. Gaussian Blur:}
A Gaussian blur is applied to the disparity array $W_{d}$ to smooth and slightly enlarge the identified areas of importance[cite: 217]. The Gaussian function is defined as:
\begin{equation}
    W_{\text{blur}}(x,y) = \frac{1}{2\pi\sigma^{2}} e^{ - \frac{x^{2} + y^{2}}{2\sigma^{2}} }
\end{equation}
where $\sigma$ controls the standard deviation of the blur.

\paragraph{3. Gamma Correction:}
The blurred array, $W_{\text{blur}}$, is then raised to the power of a fixed value $\gamma$ to enhance the contrast, which strengthens or weakens the gradient weights[cite: 222].
\begin{equation}
    W_{\gamma} = W_{\text{blur}}^{\gamma}
\end{equation}

\paragraph{4. Normalization and Offset:}
The resulting array $W_{\gamma}$ is normalized to a range of [0, 1]. To ensure that low-gradient regions (e.g., the freestream) are not entirely ignored, an offset $C_o$ is applied to establish a non-zero lower bound for the weights[cite: 252]. The final weight matrix $W_i$ is calculated as:
\begin{align}
    W_{\text{norm}} &= \frac{W_{\gamma} - \min(W_{\gamma})}{\max(W_{\gamma}) - \min(W_{\gamma})} \\
    W_{i} &= (W_{\text{norm}} \cdot [1 - C_{o}]) + C_{o}
\end{align}
This comprehensive weighting strategy provides a "soft" but highly effective focus on physically significant areas like shear layers and wakes, leading to faster convergence and more accurate reconstructions[cite: 38, 595].

\subsection{Results of Ablation Study: Comparison of Prediction Errors}

\begin{figure}[h!]
    \centering
    \begin{subfigure}[b]{0.5\textwidth}
        \centering
        \includegraphics[width=\textwidth]{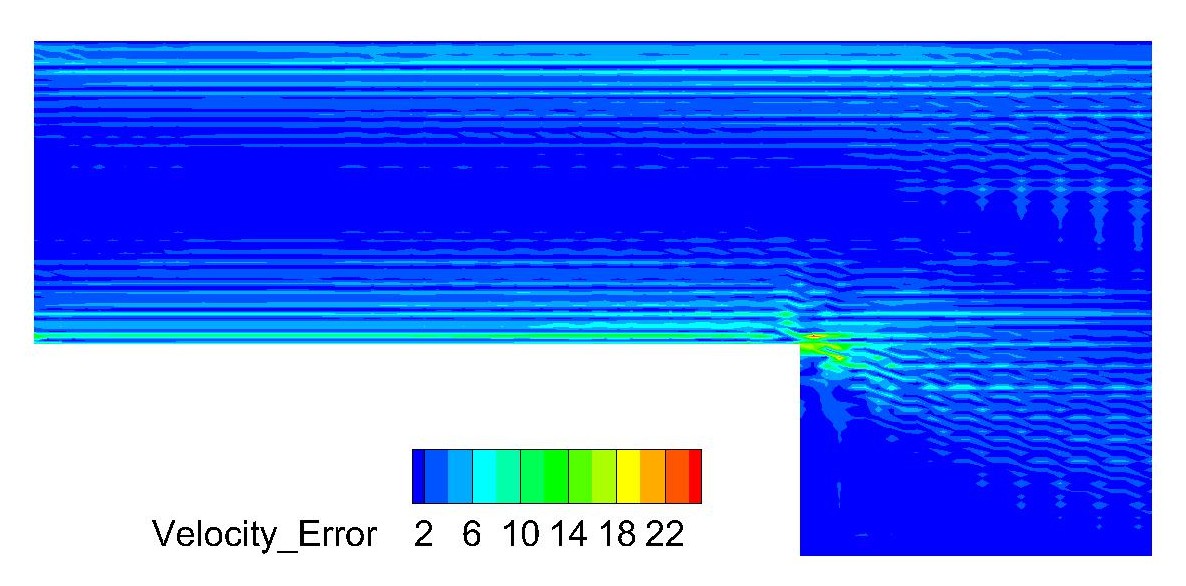}
        \caption{Error distribution using the MSE loss function.}
        \label{fig:mse_error}
    \end{subfigure}
    \hfill
    \begin{subfigure}[b]{0.5\textwidth}
        \centering
        \includegraphics[width=\textwidth]{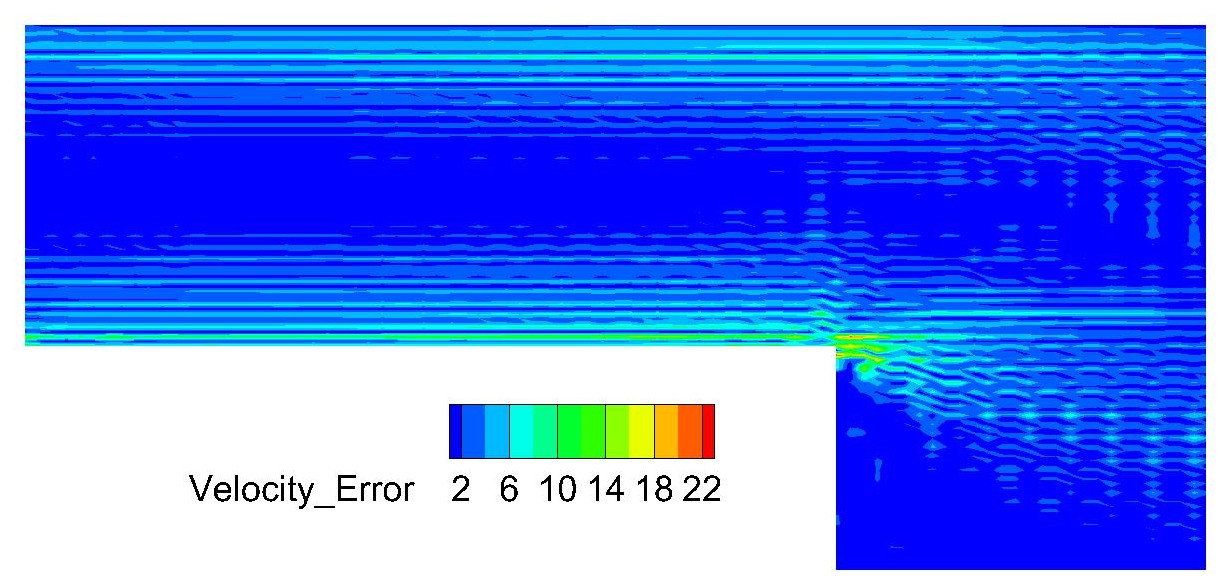}
        \caption{Error distribution using the GMSE loss function.}
        \label{fig:gmse_error}
    \end{subfigure}
    \hfill
    \begin{subfigure}[b]{0.5\textwidth}
        \centering
        \includegraphics[width=\textwidth]{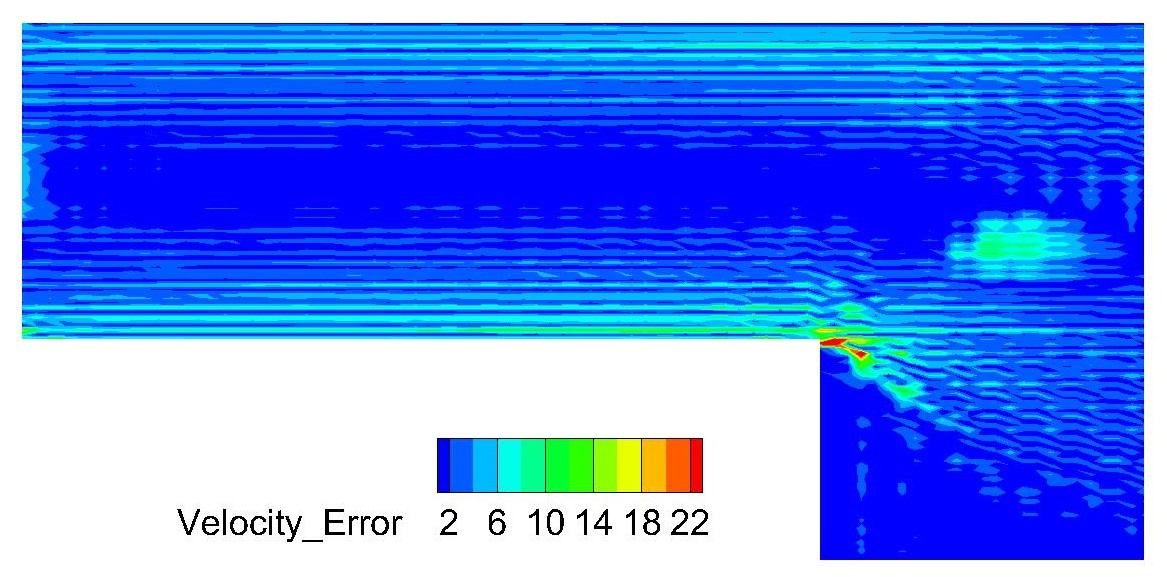}
        \caption{Error distribution using the Zonal loss function.}
        \label{fig:zonal_error}
    \end{subfigure}
    \caption{A comparative visualization of the prediction error for the U-velocity field near the step, resulting from models trained with three different loss functions.}
    \label{fig:error_comparison}
\end{figure}

The efficacy of each loss function is visually assessed by examining the spatial distribution of the prediction error for the U-velocity component. Figure \ref{fig:error_comparison} presents a zoomed-in view of the error fields in the region immediately downstream of the backward-facing step.

A direct comparison of the error distributions (Fig.~\ref{fig:mse_error}--\ref{fig:zonal_error}) shows distinct characteristics for each loss function. The \textbf{MSE loss} (Fig.~\ref{fig:mse_error}) produces relatively moderate error levels, but these errors are dispersed widely across the flow domain, extending far downstream in oscillatory bands. The \textbf{GMSE loss} (Fig.~\ref{fig:gmse_error}) leads to a more localized error distribution compared to MSE, with stronger concentration near the shear layer and recirculation zone, though some downstream contamination remains visible. In contrast, the \textbf{Zonal loss} (Fig.~\ref{fig:zonal_error}) yields the highest peak error magnitude in the immediate vicinity of the step corner, but this error is strongly confined to the vortex region. The rest of the domain remains relatively clean, with substantially reduced background error compared to the other two methods.

These observations indicate that MSE favors smoother but more widespread errors, GMSE provides partial improvement by capturing gradients yet still allows error leakage, while the Zonal strategy produces sharper but localized errors concentrated in the physically complex region of the flow.

\subsubsection{Quantitative Error Analysis}

To quantitatively assess the performance of each loss function, the L2 relative error norm was computed for the hold-out test case. 
The L2 relative error norm is defined by the following equation:
\begin{equation}
    \text{L2 Error} = \frac{\| y - \hat{y} \|_{2}}{\| y \|_{2}} = \frac{\sqrt{\sum_{i=1}^{N} (y_i - \hat{y}_i)^2}}{\sqrt{\sum_{i=1}^{N} y_i^2}}
\end{equation}
where $N$ is the total number of data points in the domain being evaluated, and $\| \cdot \|_{2}$ denotes the Euclidean norm (L2 norm) of a vector. The resulting value is a dimensionless quantity, often expressed as a percentage, which indicates the overall predictive discrepancy.

In this work, the L2 relative error norm is calculated based on the velocity vector field, which includes both the horizontal (U) and vertical (V) velocity components. It quantifies the difference between the true velocity vector, $\mathbf{u}$, and the predicted velocity vector, $\hat{\mathbf{u}}$, across all $N$ points in the domain.

The formula is defined as:
\begin{equation}
    \text{L2 Error} = \frac{\| \mathbf{u} - \hat{\mathbf{u}} \|_{2}}{\| \mathbf{u} \|_{2}} = \frac{\sqrt{\sum_{i=1}^{N} \left( (U_i - \hat{U}_i)^2 + (V_i - \hat{V}_i)^2 \right)}}{\sqrt{\sum_{i=1}^{N} (U_i^2 + V_i^2)}}
\end{equation}
where $(U_i, V_i)$ are the components of the true velocity vector at point $i$, and $(\hat{U}_i, \hat{V}_i)$ are the components of the predicted velocity vector.

The analysis was performed on both the full computational domain and specifically within the vortex recirculation region (where the true velocity U < 0). The results are summarized in Table \ref{tab:l2_error_summary}.

\begin{table}[h!]
    \centering
    \caption{L2 Relative Error Comparison for the Test Case.}
    \label{tab:l2_error_summary}
    \begin{tabular}{|l|c|c|} 
        \hline
        \textbf{Method} & \textbf{L2 Error (Full Domain)} & \textbf{L2 Error (Vortex Region)} \\
        \hline
        MSE & \textbf{2.1739\%} & 14.6135\% \\
        \hline
        GMSE & 2.2070\% & 17.4516\% \\
        \hline
        Zonal & 2.2254\% & \textbf{11.9413\%} \\
        \hline
    \end{tabular}
\end{table}

The quantitative results reveal a crucial performance trade-off. While the standard \textbf{MSE} loss achieves the lowest error on the full domain, this is likely because the model's optimization is dominated by the large, low-gradient freestream areas, which are easier to predict. This global optimization, however, comes at the cost of reduced accuracy in the more complex and physically significant vortex region.

Conversely, the Zonal loss method demonstrates the lowest error within the vortex region by a significant margin. This superior performance is a direct result of its design. The Zonal loss function explicitly segregates the domain based on a physical criterion (U < 0) and applies a higher weight to the loss calculated within the vortex. This forces the model to prioritize learning the complex flow dynamics in that specific zone, resulting in a more physically faithful reconstruction where it matters most.

This analysis suggests that for engineering applications where the fidelity of a specific, complex flow feature (e.g., recirculation, separation bubbles, shockwaves) is more critical than the overall global error metric, a targeted approach like the Zonal loss function is the more effective strategy.


\subsection{A Convolutional Fusion-DeepONet for Rarefied Flows}

To create a surrogate model capable of accurately predicting geometry-dependent rarefied flow fields from sparse data, we developed a novel hybrid neural operator architecture, which we term the **Convolutional Fusion-DeepONet (CF-DeepONet)**. This architecture is designed to learn the operator $\mathcal{G}: h \mapsto \vec{v}(y)$ that maps a scalar geometric parameter, the step height $h$, to the corresponding 2D velocity field $\vec{v}$ at any query coordinate $y=(x,y)$ in the domain. The CF-DeepONet uniquely combines the patch-based spatial feature extraction of a convolutional neural network (CNN) with the multi-scale conditioning mechanism of the Fusion-DeepONet, an architecture proposed by Peyvan et al. for continuum flows. Our model is composed of three primary sub-networks: a Branch Network to process the geometric parameter, a Trunk Network to process spatial information, and a Head Network for the final prediction, as detailed below.

\paragraph{Branch Network.}
The Branch Network is responsible for encoding the input geometric parameter, the scalar step height $h \in \mathbb{R}$. The input $h$ is first passed through a dense layer to expand its dimensionality to the model's hidden width, $d_{model}$. This feature vector is then processed by a stack of $L_b$ ResNet blocks, which apply a series of nonlinear transformations. Crucially, the outputs of the first $L_t$ hidden layers are stored as a set of conditioning vectors $\{z^{(1)}, z^{(2)}, \dots, z^{(L_t)}\}$, where each $z^{(l)} \in \mathbb{R}^{d_{model}}$. These vectors serve as the geometry-aware signals that are fused into the Trunk Network at multiple scales. The final output of the branch network after all $L_b$ blocks is a single feature vector denoted by $b_{final}$.

\paragraph{Trunk Network.}
The Trunk Network is designed to process spatial information from two distinct sources: the specific query coordinate $y$ and a local snapshot of the velocity field around that coordinate. It consists of a convolutional front-end followed by a fusion-informed MLP.
\begin{itemize}
    \item \textbf{Convolutional Front-End:} To provide the model with local physical context, we supply a $P_s \times P_s$ patch of the velocity field, $P \in \mathbb{R}^{P_s \times P_s \times 2}$, centered at the query coordinate $y$. This patch is processed by a series of 2D convolutional and average pooling layers to extract a low-dimensional feature vector, $y_{feat} = \text{CNN}(P)$. This vector, which encodes the local flow structure (e.g., gradients, curvature), is then concatenated with the coordinate vector $y=(x,y)$.
    
    \item \textbf{Fusion-Informed MLP:} The concatenated feature vector $[y, y_{feat}]$ is first projected by an initial dense layer into a high-dimensional space, yielding the initial trunk vector $t^{(0)} \in \mathbb{R}^{d_{model}}$. This vector then enters the core of the architecture: a fusion block consisting of $L_t$ layers. At each layer $l$, the trunk vector $t^{(l-1)}$ is fused with the corresponding conditioning signal $z^{(l)}$ from the Branch Network via an element-wise product ($\odot$). The result is then passed through a dense layer with a tanh activation function to produce the next-level trunk vector. This fusion process is described by the recurrence relation:
    $$
    t^{(l)} = \text{Dense}_{\tanh}\left(t^{(l-1)} \odot z^{(l)}\right), \quad \text{for } l=1, \dots, L_t
    $$
    This multi-scale conditioning allows the geometric information from the branch to influence the trunk's processing of spatial features at every level, creating a set of powerful, geometry-informed basis functions. The final output of this process is the trunk vector $t_{final} = t^{(L_t)}$.
\end{itemize}

\paragraph{Head Network and Final Prediction.}
In the final stage, the geometry-aware spatial features ($t_{final}$) are modulated by the global parameter features ($b_{final}$) through another element-wise product: $m = b_{final} \odot t_{final}$. This combined and richly-informed feature vector $m$ is then processed by a final multi-layer perceptron, termed the Head Network, which consists of $L_h$ dense layers. The Head Network maps the latent features to the physical space, producing the predicted 2D velocity vector $\vec{v}_{pred} \in \mathbb{R}^2$ at the query coordinate $y$.

\begin{figure}[h!]
    \centering
    
    \resizebox{0.9\textwidth}{!}{%
    
\begin{tikzpicture}[node distance=1cm and 1cm]
        \tikzset{
            block/.style={rectangle, draw, fill=blue!10, text width=7em, text centered, rounded corners, minimum height=3em},
            smallblock/.style={rectangle, draw, fill=blue!10, text width=6em, text centered, rounded corners, minimum height=2.5em},
            op/.style={circle, draw, fill=orange!20, minimum size=2em},
            io/.style={rectangle, rounded corners, draw, fill=green!10, text width=5em, text centered, minimum height=3em},
            line/.style={draw, -{Stealth[length=2mm]}},
            fusion/.style={draw, dashed, -{Stealth[length=2mm]}, color=red!80}
        }
        
        \node[io] (input_h) {Step Height Input \\ $h$};
        \node[io, below=3.5cm of input_h] (input_patch) {Local Patch Input \\ $P$};
        \node[io, below=1.5cm of input_patch] (input_coords) {Coordinate Input \\ $(x, y)$};

        \node[block, right=of input_h] (branch_net) {Branch Network};
        
        \node[smallblock, right=of input_patch] (cnn) {CNN Feature Extractor};
        \node[op, right=1.2cm of cnn] (concat) {Concatenate};
        \node[smallblock, right=1.2cm of concat] (trunk_proj) {Projection Layer};
        \node[op, right=of trunk_proj] (fusion_mult1) {$\odot$};
        \node[smallblock, right=of fusion_mult1] (fusion_dense1) {Dense};
        
        \node[op, right=2cm of branch_net] (final_mult) {$\odot$};
        \node[block, right=of final_mult] (head_net) {Head Network};
        \node[io, right=of head_net] (output) {Predicted Velocity Field \\ $\vec{v}_{pred}$};
        
        \path[line] (input_h) -- (branch_net);
        \path[line] (input_patch) -- (cnn);
        \path[line] (input_coords) -- (concat);
        \path[line] (cnn) -- (concat);
        \path[line] (concat) -- (trunk_proj);
        \path[line] (trunk_proj) -- (fusion_mult1);
        \path[line] (fusion_mult1) -- (fusion_dense1);
        
        \path[line] (branch_net.east) -- node[above] {$b_{final}$} (final_mult.west);
        \path[line] (fusion_dense1.north) to[out=90,in=-20]
            node[above, pos=0.35] {$t_{final}$} (final_mult);
        \path[line] (final_mult.east) -- (head_net.west);
        \path[line] (head_net.east) -- (output.west);
        
        \path[fusion] (branch_net.south) -- ++(0,-0.5cm) -| (fusion_mult1.north)
            node[pos=0.25, right] {$z^{(1)}$};
        \path[fusion] ([xshift=1.2cm]branch_net.south) -- ++(0,-1.5cm) -| (fusion_dense1.north)
            node[pos=0.25, right] {$z^{(2)}$};
        
    \end{tikzpicture}
    } 
    \caption{Flowchart of the Fusion-DeepONet Architecture}
    \label{fig:fusion-flowchart}
\end{figure}
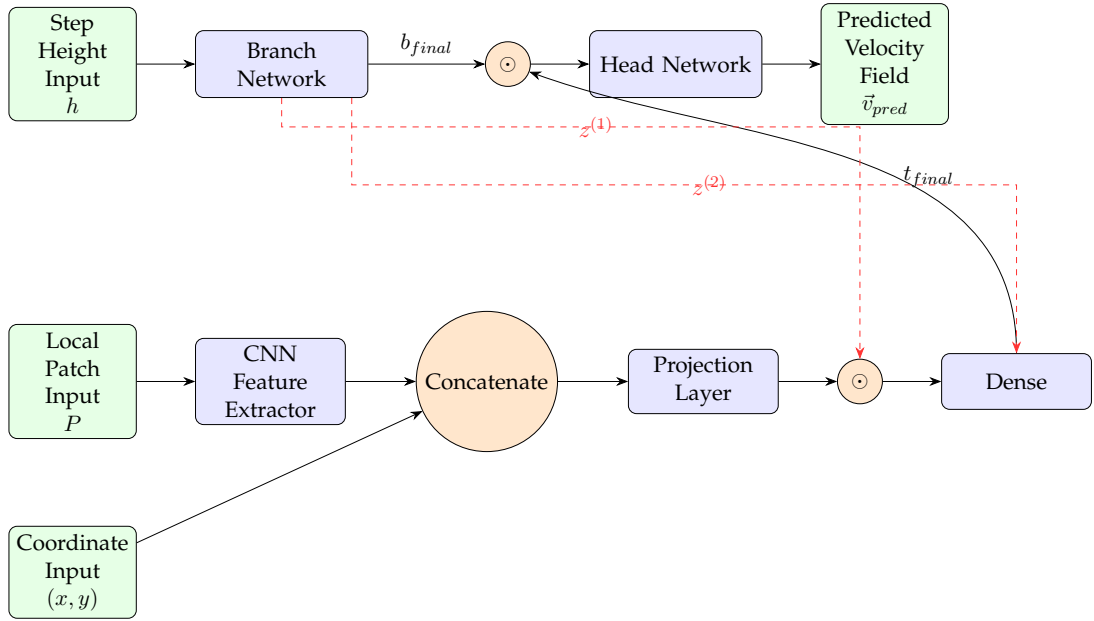

\subsubsection{Discussions on the Fusion Model Performance}

The performance of both the standard Convolutional DeepONet and the more advanced Convolutional Fusion-DeepONet was evaluated on the held-out test case ($h/H = 44\%$). A qualitative comparison of the predicted U-velocity and V-velocity fields against the ground truth DSMC data is presented in Figure~\ref{fig:comparison_3Methods}. Visually, both models successfully capture the primary features of the flow, including the location and general shape of the recirculation zone and the velocity distribution in the main channel.

For a quantitative assessment, the Root Mean Squared Error (RMSE) was calculated across the entire velocity field for both models. The results are summarized in Table~\ref{tab:rmse_comparison}. Surprisingly, the analysis reveals that the standard DeepONet architecture achieves a lower overall error compared to the more complex Fusion-DeepONet.

\begin{table}[h!]
    \centering
    \caption{Quantitative performance comparison of the surrogate models on the test case.}
    \label{tab:rmse_comparison}
    \begin{tabular}{l|c}
        \hline
        \textbf{Model Architecture} & \textbf{Root Mean Squared Error (RMSE)} \\
        \hline
        Standard DeepONet & 2.689 \\
        Fusion-DeepONet   & 3.589 \\
        \hline
    \end{tabular}
\end{table}

\subsubsection{Analysis of Model Performance and Overfitting}

The superior performance of the simpler architecture in this study highlights a critical concept in machine learning: the trade-off between model complexity and data availability. The Fusion-DeepONet, with approximately 1.5 million trainable parameters, is a significantly more expressive model than the standard DeepONet, which has approximately 1.1 million parameters. While this increased capacity is theoretically advantageous, it also makes the model more susceptible to overfitting, especially when trained on a sparse dataset.

\begin{figure}[h!]
    \centering
    \includegraphics[width=0.8\textwidth]{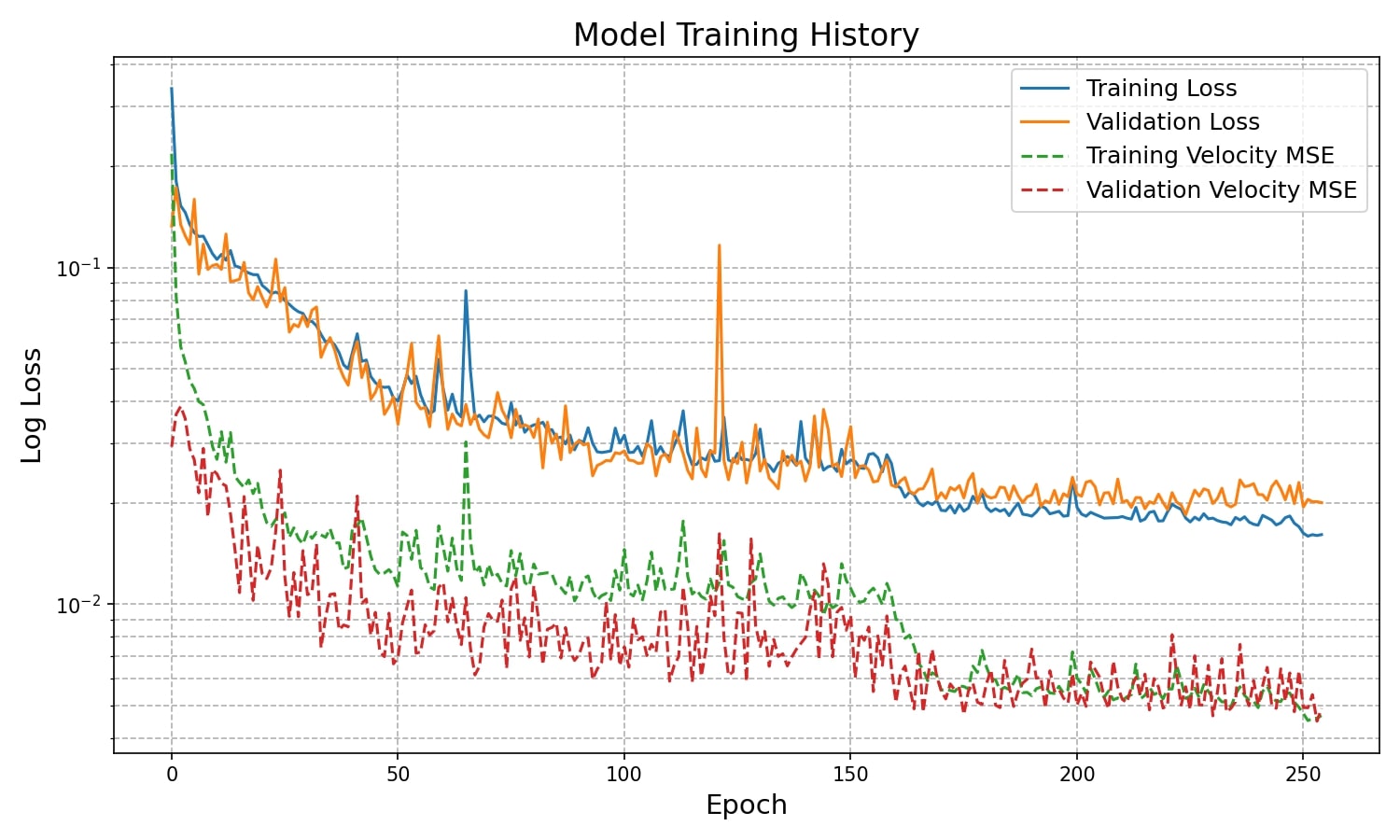}
    \caption{Loss Function and MSE for the Fusion Algorithm}
    \label{fig:lossfusion}
\end{figure}

Direct evidence of this overfitting can be observed by comparing the training histories of the two models. The loss plot for the standard DeepONet (Figure~\ref{fig:loss_Height}) shows that the training and validation losses track each other closely, indicating good generalization. In contrast, the loss plot for the Fusion-DeepONet (Figure~\ref{fig:lossfusion}) reveals a significant and persistent gap between the training and validation loss curves. This divergence is the classic signature of overfitting: the model has begun to memorize the specific noise and artifacts of the seven training cases rather than learning the underlying physical operator. Consequently, its ability to generalize to the unseen test case is compromised, leading to a higher overall error.

This result is not a failure of the Fusion-DeepONet architecture itself, but rather a crucial finding on its data requirements for this class of problems. It suggests that for rarefied flow simulations where generating high-fidelity data is computationally expensive, a well-regularized, simpler architecture may provide more robust and accurate predictions than a more complex model that cannot be adequately constrained by the limited data.

As shown in Figure~\ref{fig:loss_Height}, the DeepONet trained on step-height variations exhibits a relatively smooth convergence, with both the training and validation losses decreasing steadily and eventually reaching low values of the order of $10^{-2}$. Although the error level is higher than in the Knudsen-number case (with 20 input samples), the DeepONet still achieves a robust approximation given that only 8 input variations were provided. 

In contrast, the Fusion algorithm (Figure~\ref{fig:lossfusion}) does not reach the same level of accuracy. While the overall trend of the loss curves indicates convergence, the final training and validation errors plateau at significantly higher levels compared to the DeepONet case. Furthermore, the velocity MSE curves remain noisy and do not consistently decrease to the same extent. This performance gap highlights a fundamental limitation: with only 8 training samples for the height variation case, the Fusion model does not appear to fully leverage its more complex architecture. The Fusion strategy likely requires a larger and more diverse dataset to realize its potential advantages over the standard DeepONet. These results suggest that while DeepONet can generalize well even with relatively scarce data, Fusion-based models may demand substantially more training data to achieve competitive accuracy.

\begin{figure}[htbp]
    \centering
    \begin{subfigure}{0.45\textwidth}
        \centering
        \includegraphics[width=\linewidth]{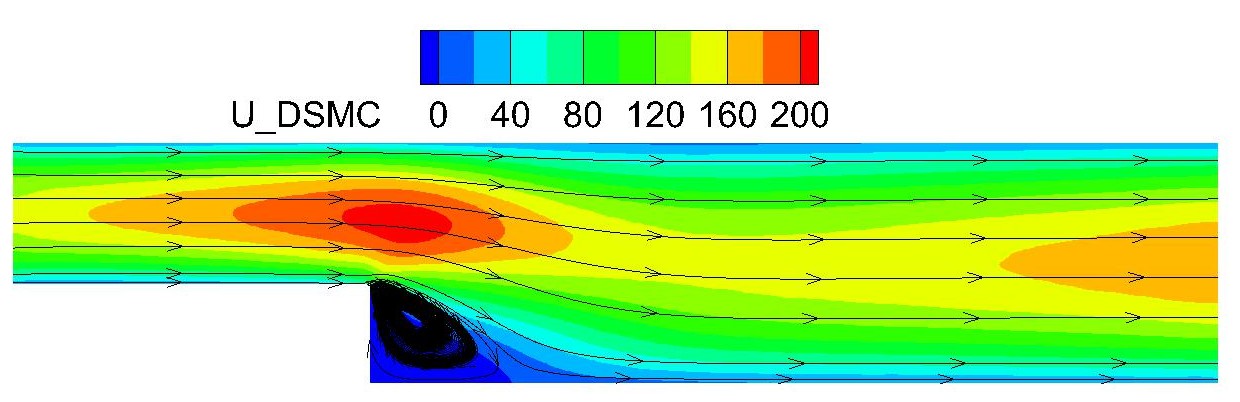}
        \caption{DSMC, U-velocity}
    \end{subfigure}
    \begin{subfigure}{0.45\textwidth}
        \centering
        \includegraphics[width=\linewidth]{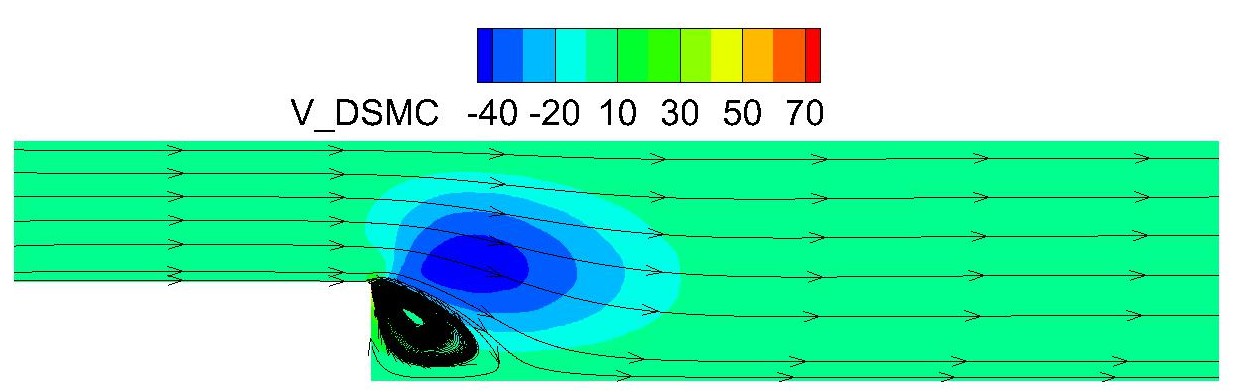}
        \caption{DSMC, V-velocity}
    \end{subfigure}

    \vspace{0.3cm}

    \begin{subfigure}{0.45\textwidth}
        \centering
        \includegraphics[width=\linewidth]{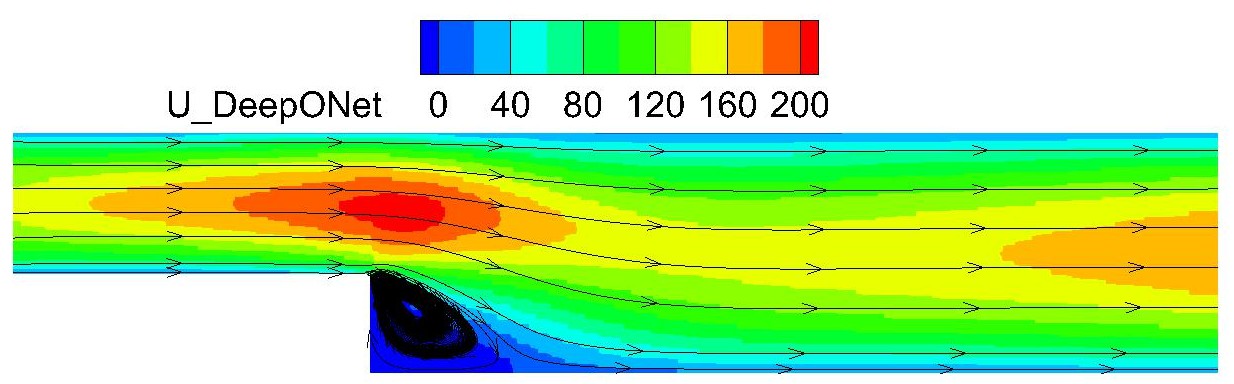}
        \caption{DeepONet, U-velocity}
    \end{subfigure}
    \begin{subfigure}{0.45\textwidth}
        \centering
        \includegraphics[width=\linewidth]{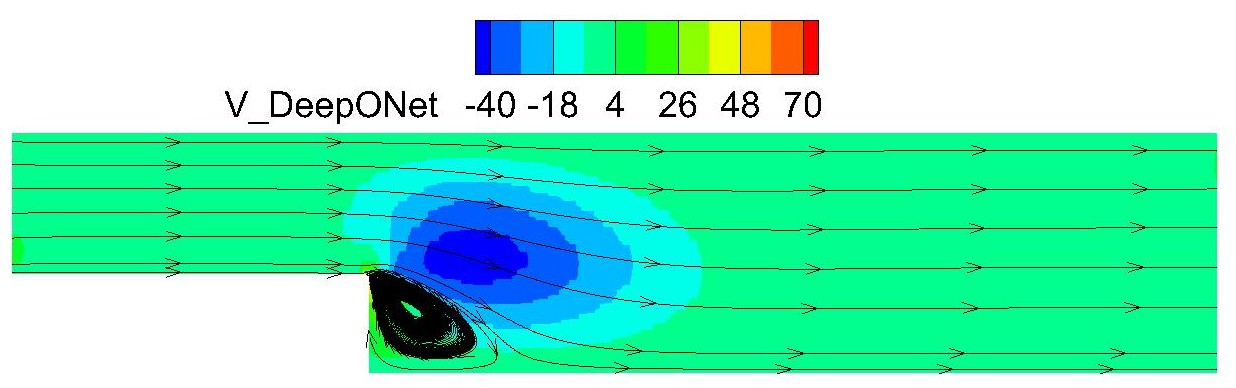}
        \caption{DeepONet, V-velocity}
    \end{subfigure}

    \vspace{0.3cm}

    \begin{subfigure}{0.45\textwidth}
        \centering
        \includegraphics[width=\linewidth]{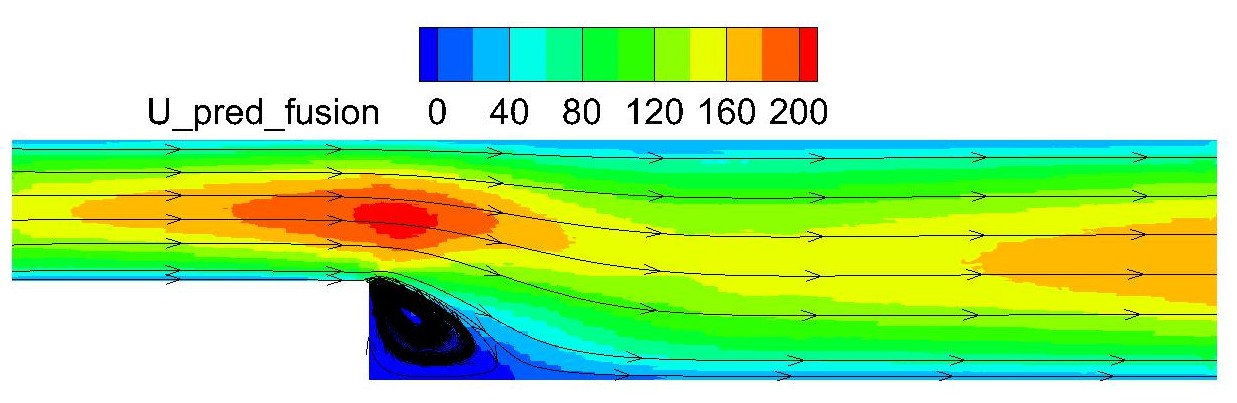}
        \caption{Fusion-DeepONet, U-velocity}
    \end{subfigure}
    \begin{subfigure}{0.45\textwidth}
        \centering
        \includegraphics[width=\linewidth]{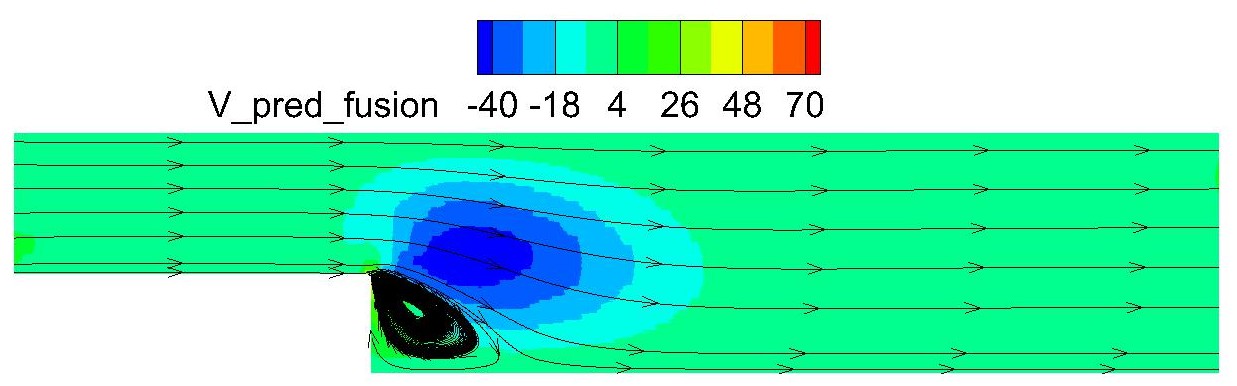}
        \caption{Fusion-DeepONet, V-velocity}
    \end{subfigure}

    \caption{Qualitative comparison of U-velocity (left column) and V-velocity (right column) contours for DSMC (top), DeepONet (middle), and Fusion-DeepONet (bottom).}
    \label{fig:comparison_3Methods}
\end{figure}

\begin{figure}[htbp]
    \centering
    \begin{subfigure}{0.48\textwidth}
        \centering
        \includegraphics[width=\linewidth]{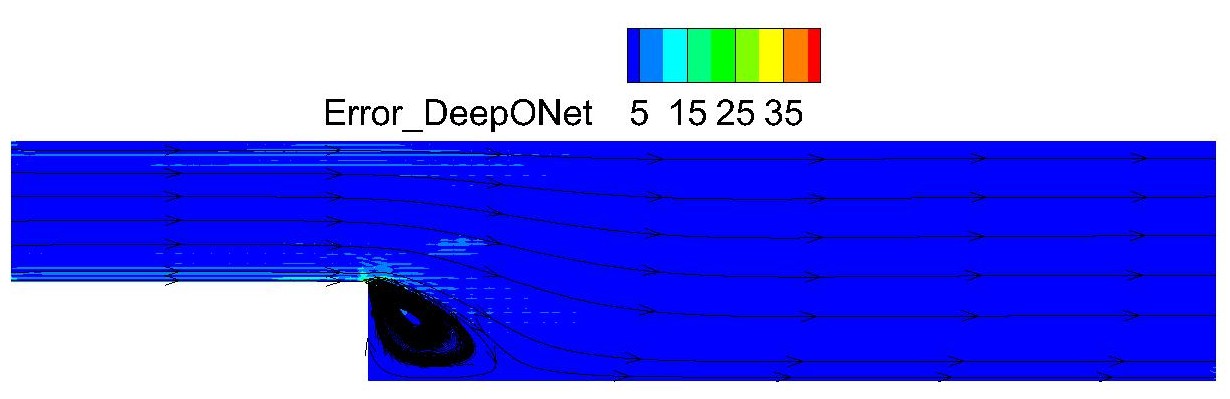}
        \caption{DeepONet error field}
        \label{fig:error_deeponet}
    \end{subfigure}
    \hfill
    \begin{subfigure}{0.48\textwidth}
        \centering
        \includegraphics[width=\linewidth]{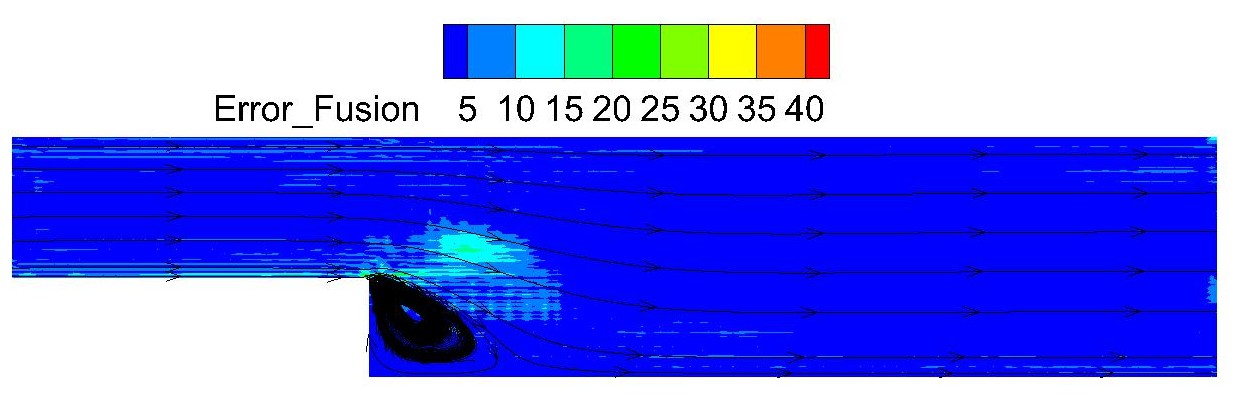}
        \caption{Fusion-DeepONet error field}
        \label{fig:error_fusion}
    \end{subfigure}
    \caption{Comparison of velocity error distributions for the height-variation test case ($h/H=0.44$). The DeepONet approach (a) produces visibly lower error intensity in the shear layer and recirculation region compared to the Fusion-DeepONet (b).}
    \label{fig:error_fusion_vs_deeponet}
\end{figure}

Figures~\ref{fig:comparison_3Methods} and \ref{fig:error_fusion_vs_deeponet} provide a direct comparison between the baseline DSMC results, the DeepONet predictions, and the Fusion-DeepONet approach for the unseen step-height ratio of $h/H=0.44$. 

In Figure~\ref{fig:comparison_3Methods}, the U-velocity and V-velocity contours are presented side by side. The DeepONet predictions (middle row) capture both the streamwise velocity acceleration above the step and the recirculation structure below the step with high fidelity, closely resembling the reference DSMC fields (top row). The Fusion-DeepONet predictions (bottom row) display some deviations in the vortex core and in the shear layer. Slight differences are also observed in the V-velocity component.

The discrepancy is quantified in Figure~\ref{fig:error_fusion_vs_deeponet}, where error maps are shown for DeepONet (left) and Fusion-DeepONet (right). The DeepONet error field exhibits relatively low intensity, with the largest deviations localized near the separation and reattachment zones. On the other hand, the Fusion-DeepONet error field shows broader regions of elevated error, particularly in the shear layer and wake region, indicating that the Fusion approach introduces additional uncertainty in predicting small-scale flow features. These results confirm that DeepONet provides superior generalization in the scarce-data regime, while the Fusion model may require significantly more training samples to achieve comparable accuracy.

\setlength{\floatsep}{0pt}
\setlength{\textfloatsep}{0pt}

\section{Concluding Remarks}

This study presented a convolutional DeepONet surrogate model, augmented with a physics-guided zonal loss, for predicting rarefied flow over a micro backward-facing step. The work establishes both methodological innovations and practical implications, which can be summarized as follows:

\begin{itemize}
    \item \textbf{Effectiveness of the Zonal Loss.}  
    The proposed zonal loss provides superior accuracy in physically critical regions of the flow, particularly within recirculation bubbles where $U<0$. As demonstrated in Table~2, global error metrics alone can be misleading by underestimating deficiencies in vortex resolution. By emphasizing localized errors, the zonal loss achieves a more faithful reconstruction of separated flow structures, ensuring that engineering-critical phenomena are captured with higher fidelity.
    
    \item \textbf{Insights from the Fusion-DeepONet Comparison.}  
    The comparative analysis with Fusion-DeepONet, summarized in Table~3, should not be interpreted as a shortcoming of the hybrid model. Instead, it provides a valuable scientific observation: increasing architectural complexity without sufficient training data can degrade generalization performance. This highlights a fundamental trade-off between model sophistication and data availability in computationally expensive regimes. Such insights are crucial for guiding the design of surrogate models in rarefied gas dynamics.
    
    \item \textbf{Quantification of Computational Savings.}  
    One of the most significant contributions of this work lies in computational efficiency. The trained surrogate is capable of generating predictions for new parametric cases in milliseconds, compared to the hours or days required for full DSMC simulations. This represents several orders-of-magnitude acceleration, thereby enabling many-query tasks such as uncertainty quantification, parametric sweeps, and design optimization that would otherwise be infeasible with DSMC alone.
\end{itemize}

In summary, the framework delivers both scientific and practical contributions. Scientifically, it introduces a physics-guided loss function that redefines error evaluation by aligning it with physically meaningful flow regions, and it sheds light on the interplay between model complexity and data availability in surrogate modeling. Practically, it establishes a pathway toward orders-of-magnitude acceleration in rarefied flow simulations, bridging the gap between the accuracy of DSMC and the speed required for real-world engineering design.  

Beyond the present backward-facing step configuration, the methodology holds promise for a broad class of applications in micro- and hypersonic flows where DSMC remains indispensable yet prohibitively expensive. Future extensions could integrate additional physics-informed constraints, multi-fidelity training strategies, or coupling with uncertainty quantification frameworks, further enhancing the applicability and robustness of surrogate modeling in rarefied gas dynamics.

From a computational perspective, the efficiency gain of the proposed surrogate is striking. For instance, generating a single DSMC solution at low Knudsen numbers in the slip-flow regime typically requires about 24 hours of wall-clock time on a single Intel Core-i7 CPU core. By contrast, training the DeepONet surrogate on a high-end GPU (NVIDIA A100-SXM4-80GB) takes approximately 20--30 minutes. This comparison---``minutes vs. hours''---highlights the orders-of-magnitude acceleration achieved by the surrogate, effectively transforming tasks such as parametric sweeps, optimization, and uncertainty quantification from computationally prohibitive to practically feasible.

Future work will involve exploring adaptive weighting for the zonal loss function, wherein the hyperparameter \(\alpha\) is learned during training rather than being fixed. Furthermore, the application of this framework to three-dimensional geometries will be investigated.
The proposed surrogate modeling approach will be extended to other challenging rarefied flow problems, such as shockwave-boundary layer interactions in supersonic flows and gas-surface chemistry in atmospheric re-entry vehicles, where localized, non-equilibrium phenomena are critical.
Ultimately, this rapid and uncertainty-aware surrogate model can be integrated into a complete multi-fidelity optimization framework for the design of MEMS devices and hypersonic vehicle components, enabling robust design under uncertainty in rarefied gas environments.

\section*{Appendix: Hyperparameter Details}
\label{appendix:hyperparameters}

\section{Hyperparameter Details}
\label{appendix:hyperparameters}

For clarity and reproducibility, this section provides a comprehensive summary of the key hyperparameters used for training and model architecture across the four main experimental setups. The specifications are detailed across two tables for improved readability. Table \ref{tab:hyperparameters_1} outlines the configurations for the Knudsen number study and the primary height ablation study. Table \ref{tab:hyperparameters_2} details the configurations for the simplified height study and the derivative-enhanced Fusion model.

\begin{table}[h!]
    \centering
    \caption{Hyperparameter specifications for the Knudsen and primary Height studies.}
    \label{tab:hyperparameters_1}
    \begin{tabular}{@{}lll@{}}
        \toprule
        \textbf{Hyperparameter} & \textbf{Knudsen Number Study} & \textbf{Height Ablation Study} \\
        \midrule
        \multicolumn{3}{l}{\textit{\textbf{Training Hyperparameters}}} \\
        \quad Epochs & 1500 & 1500 \\
        \quad Batch Size & 512 & 512 \\
        \quad Base Learning Rate & $1 \times 10^{-4}$ & $1 \times 10^{-4}$ \\
        \quad Optimizer & AdamW & AdamW \\
        \quad Weight Decay & $1 \times 10^{-5}$ & $5 \times 10^{-5}$ \\
        \quad Clipnorm & 1.0 & Not specified \\
        \midrule
        \multicolumn{3}{l}{\textit{\textbf{Model Architecture}}} \\
        \quad Branch Width / Depth & 256 / 6 & 384 / 4 \\
        \quad Trunk MLP Width / Depth & 256 / 4 & 384 / 4 \\
        \quad Head Width / Depth & 512 / 4 & 768 / 4 \\
        \quad Activation Functions & ReLU (CNN), Tanh (MLP) & ReLU (CNN), Tanh (MLP) \\
        \midrule
        \multicolumn{3}{l}{\textit{\textbf{Loss Function Details}}} \\
        \quad Primary Loss Function & Zonal (Two-Zone) & MSE, Zonal, GMSE (Ablation) \\
        \quad Zonal Weight ($\alpha$) & 0.7 & 0.6 \\
        \quad GMSE Parameters & N/A & $\sigma=10.0, \gamma=1.0, C_o=0.2$ \\
        \bottomrule
    \end{tabular}
\end{table}

\begin{table}[h!]
    \centering
    \caption{Hyperparameter specifications for the simplified Height and Fusion studies.}
    \label{tab:hyperparameters_2}
    \begin{tabular}{@{}lll@{}}
        \toprule
        \textbf{Hyperparameter} & \textbf{Simplified Height Study} & \textbf{Fusion DEL Study} \\
        \midrule
        \multicolumn{3}{l}{\textit{\textbf{Training Hyperparameters}}} \\
        \quad Epochs & 1500 & 1500 \\
        \quad Batch Size & 512 & 512 \\
        \quad Base Learning Rate & $1 \times 10^{-4}$ & $1 \times 10^{-4}$ \\
        \quad Optimizer & AdamW & AdamW \\
        \quad Weight Decay & $5 \times 10^{-5}$ & $5 \times 10^{-5}$ \\
        \quad Clipnorm & 1.0 & 1.0 \\
        \midrule
        \multicolumn{3}{l}{\textit{\textbf{Model Architecture}}} \\
        \quad Branch Width / Depth & 256 / 4 & 512 / 6 \\
        \quad Trunk MLP Width / Depth & 256 / 3 & 512 / 4 \\
        \quad Head Width / Depth & 512 / 3 & 1024 / 4 \\
        \quad Activation Functions & ReLU (CNN), Tanh (MLP) & ReLU (CNN), Tanh (MLP) \\
        \midrule
        \multicolumn{3}{l}{\textit{\textbf{Loss Function Details}}} \\
        \quad Primary Loss Function & Zonal (Two-Zone) & Fusion \\
        \quad Zonal Weight ($\alpha$) & 0.6 & 0.7 \\
        \quad Gradient Weight ($\lambda$) & N/A & 0.05 \\
        \bottomrule
    \end{tabular}
\end{table}

\clearpage 
\bibliographystyle{unsrt}     
\bibliography{References}

\end{document}